\documentclass[a4,letterpaper,11pt]{article}

\pdfoutput=1 

\usepackage{jheppub} 
\usepackage[T1]{fontenc}
\usepackage{lmodern}
\usepackage{amsmath}
\usepackage{amssymb}
\usepackage{graphicx}
\usepackage{xspace}
\usepackage{multirow}
\usepackage[absolute]{textpos}

\usepackage{epstopdf}

\newcommand{\nn}{\nonumber} 

\newcommand{\be}{\begin{equation}}
\newcommand{\ee}{\end{equation}}
\newcommand{\bea}{\begin{eqnarray}}
\newcommand{\eea}{\end{eqnarray}}

\newcommand{\vect}[1]{\mathbf{#1}}
\newcommand{\abs}[1]{\left\lvert #1\right\rvert}

\newcommand{\minus}{\!-\!}
\newcommand{\plus}{\!+\!}

\newcommand{\df}{\mathrm{d}}
\newcommand{\Lqcd}{\Lambda_{\text{QCD}}}

\newcommand{\run}{\text{run}}
\newcommand{\mur}{\mu_\run}
\newcommand{\nur}{\nu_\run}

\newcommand{\GeV}{\text{ GeV}}

\newcommand{\nkll}{N$^k$LL\ }
\newcommand{\nkllprime}{N$^k$LL$'$\ }
\newcommand{\as}{\alpha_s}
\newcommand{\MSbar}{\overline{\text{MS}}}

\newcommand{\cusp}{\mathrm{cusp}}

\newcommand{\Ga}{\Gamma}
\newcommand{\eg}{\emph{e.g.}~}
\newcommand{\ie}{\emph{i.e.}~}

\newcommand{\cO}{\mathcal{O}}

\newcommand{\cC}{\mathcal{C}}
\newcommand{\cI}{\mathcal{I}}

\newcommand{\cH}{\mathcal{H}}

\newcommand{\cL}{\mathcal{L}}

\newcommand{\hats}{\hat{s}}
\newcommand{\hatt}{\hat{t}}
\newcommand{\hatu}{\hat{u}}

\newcommand{\zed}{\mathbb{Z}}

\newcommand{\wt}{\widetilde}

\newcommand{\eq}[1]{Eq.~\eqref{#1}}
\newcommand{\eqs}[2]{Eqs.~\eqref{#1} and \eqref{#2}}
\newcommand{\eqss}[3]{Eqs.~\eqref{#1}, \eqref{#2}, and \eqref{#3}}

\renewcommand{\sec}[1]{Sec.~\ref{sec:#1}}
\newcommand{\ssec}[1]{Sec.~\ref{ssec:#1}}

\newcommand{\appx}[1]{App.~\ref{app:#1}}
\newcommand{\fig}[1]{Fig.~\ref{fig:#1}}

\newcommand{\tab}[1]{Table~\ref{tab:#1}}

\newcommand{\tft}{\widetilde{f}^\perp}
\newcommand{\tS}{\widetilde{S}}

\allowdisplaybreaks[1]

\DeclareMathOperator{\Imag}{Im}

\DeclareMathOperator{\Res}{Res}

\usepackage{mathtools}
\DeclarePairedDelimiter\ceil{\lceil}{\rceil}
\DeclarePairedDelimiter\floor{\lfloor}{\rfloor}

\begin{document}

\begin{textblock}{2}(12.7,3.1)%
  LA-UR-17-27820
  \end{textblock}%

\title{ \Large A fast and accurate method for perturbative resummation of transverse momentum-dependent observables}

\author[a,b]{Daekyoung Kang,}
\author[a]{Christopher Lee,}
\author[a]{Varun Vaidya}

\affiliation[a]{Theoretical Division, MS B283, Los Alamos National Laboratory, 
Los Alamos, NM 87545, USA }
\affiliation[b]{Key Laboratory of Nuclear Physics and Ion-beam Application (MOE) and Institute of Modern Physics, Fudan University, Shanghai, China 200433}

\emailAdd{dkang@fudan.edu.cn}
\emailAdd{clee@lanl.gov}
\emailAdd{vvaidya@lanl.gov}

\abstract{ 
We propose a novel strategy for the perturbative resummation of transverse momentum-dependent (TMD) observables, using the $q_T$ spectra of gauge bosons ($\gamma^*$, Higgs) in $pp$ collisions in the regime of low (but perturbative) transverse momentum $q_T$ as a specific example. First we introduce a scheme to choose the factorization scale for virtuality in momentum space instead of in impact parameter space, allowing us to avoid integrating over (or cutting off) a Landau pole in the inverse Fourier transform of the latter to the former. The factorization scale for rapidity is still chosen as a function of impact parameter $b$, but in such a way designed to obtain a Gaussian form (in $\ln b$) for the exponentiated rapidity evolution kernel, guaranteeing convergence of the $b$ integral. We then apply this scheme to obtain the $q_T$ spectra for Drell-Yan and Higgs production at NNLL accuracy. In addition, using this scheme we are able to obtain a fast semi-analytic formula for the perturbative resummed cross sections in momentum space: analytic in its dependence on all physical variables at each order of logarithmic accuracy, up to a numerical expansion for the pure mathematical Bessel function in the inverse Fourier transform that needs to be performed just once for all observables and kinematics, to any desired accuracy.
}

\maketitle

\section{Introduction}
\label{sec:Intro}

The transverse momentum spectra of gauge bosons is well trodden territory. They are important for measurements of, \eg Higgs production, as well as the dynamics of QCD in Drell-Yan (DY) processes. There are calculations available at NNLL+NNLO accuracy using a variety of resummation schemes both using the framework of soft collinear effective theory (SCET) \cite{Bauer:2000ew,Bauer:2000yr,Bauer:2001ct,Bauer:2001yt,Bauer:2002nz}, \eg \cite{Gao:2005iu,Idilbi:2005er,Mantry:2010mk,GarciaEchevarria:2011rb}, and Collins-Soper-Sterman (CSS) \cite{Collins:1984kg} formalisms, \eg \cite{deFlorian:2000pr,deFlorian:2001zd,Catani:2010pd,Bozzi:2010xn,Becher:2010tm}, and even N$^3$LL+NNLO \cite{Bizon:2017rah} (see also calculations in, \eg\cite{Li:2016axz,Li:2016ctv,Vladimirov:2016dll,Gehrmann:2014yya,Luebbert:2016itl,Echevarria:2015byo,Echevarria:2016scs}). Joint resummation of threshold and tranverse-momentum logs is even possible to NNLL and beyond (\eg \cite{Li:2016axz,Lustermans:2016nvk,Marzani:2016smx,Muselli:2017bad}).Why, then, do we wish to visit this subject anew? 

\begin{figure}
\centerline{\scalebox{.55}{\includegraphics{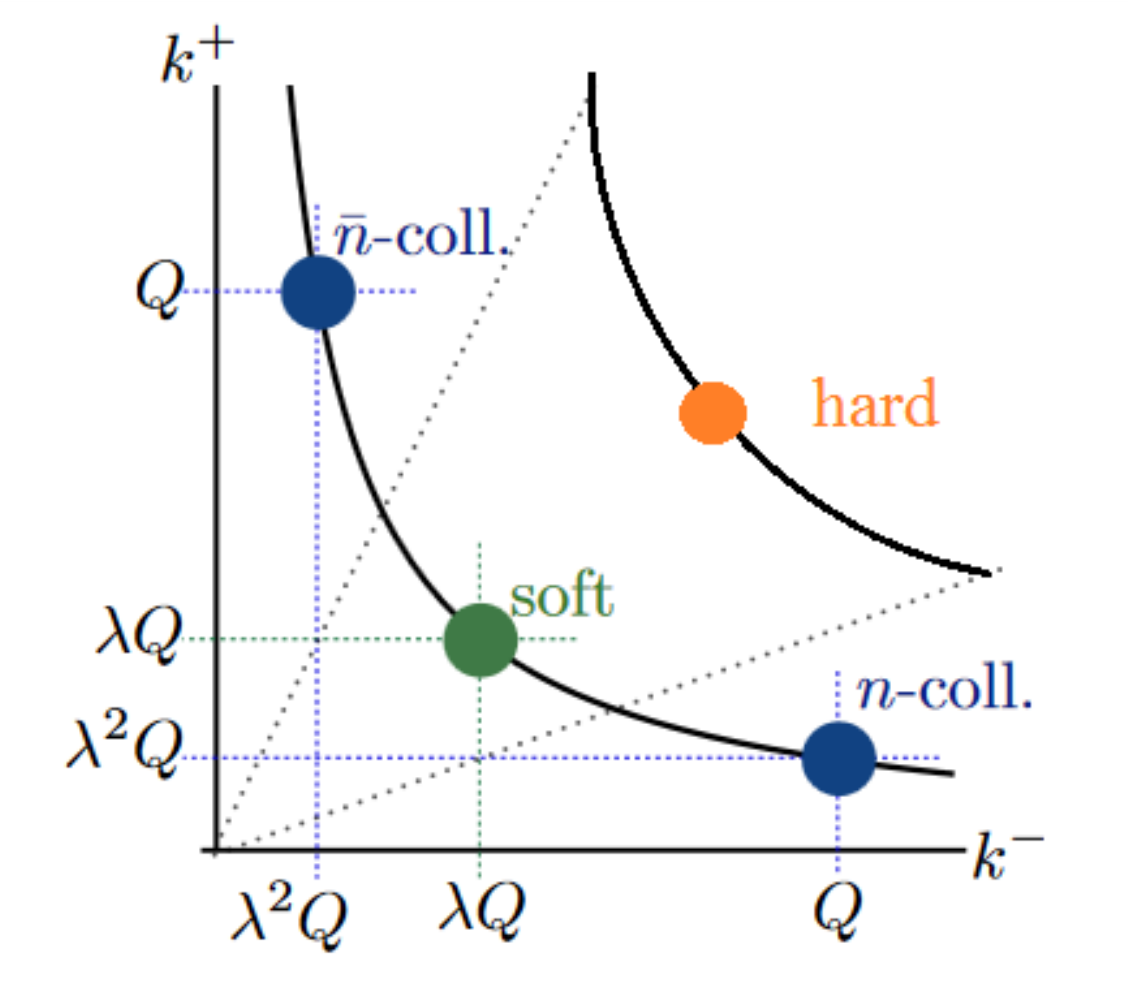}} \scalebox{.6}{\includegraphics{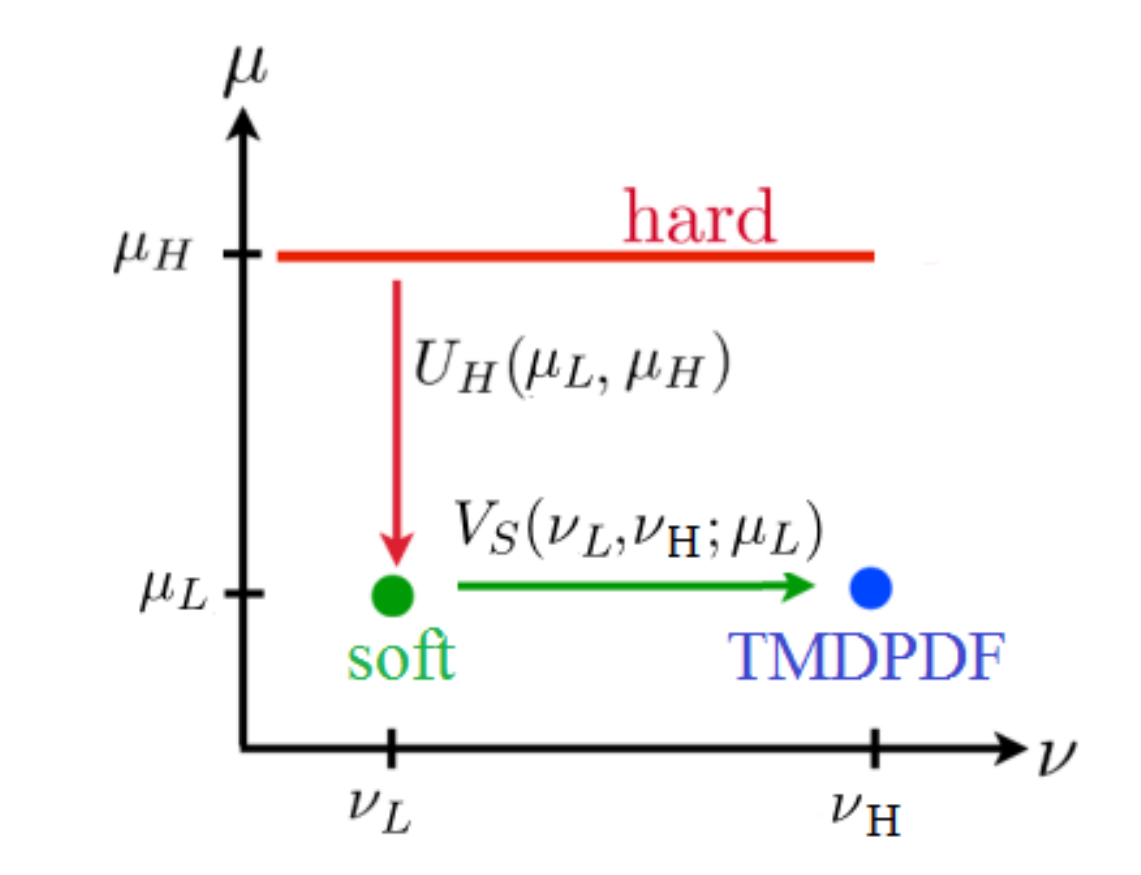}}}
\vskip-0.2cm
\caption[1]{\emph{Left:} EFT modes and their scalings in light-cone momentum $k^\pm$ space. For the TMD cross sections we consider, the small parameter can be taken to be $\lambda \sim q_T/Q$ or $\sim Qb_0$ in impact parameter $b$ space, where $b_0 = be^{\gamma_E}/2$. \emph{Right:} RG and rapidity RG evolution. $\mu$ runs between the hard and soft hyperbolas of virtuality shown in the left-hand figure, while $\nu$ runs between the soft and collinear modes which are separated only by rapidity. The evolution is path independent, one convenient path is shown here.}
\label{fig:modes} 
\end{figure}

This has mostly to do with the peculiar structure of the factorized cross section which makes the resummation of large logarithms an interesting problem. The cross section can be factorized in terms of a hard function, which lives at a virtuality $Q$, the invariant mass of the gauge boson, and soft and the beam functions (or TMDPDFs) which describe the IR physics and live at the virtuality $q_T \ll Q$, which is the transverse momentum of the gauge boson. The soft and collinear emissions are the ones providing the recoil for the transverse momentum of the gauge boson. This automatically means that these functions are convolved with each other in transverse momentum space so that the $q_T$ of the gauge boson is a sum of the $q_T$ contribution from each emission:
\begin{align}
\label{txsec}
\frac{ d \sigma} {d^2 q_T dy} &= \sigma_0 C_t^2(M_t^2,\mu) H (Q^2; \mu ) \int d^2 \vec{q}_{Ts} d^2 \vec{q}_{T1} d^2\vec{q}_{T2}  \delta^2\bigl(\vec{q_T} - (\vec{q_{Ts}} +\vec{q}_{T1}+\vec{q}_{T2}) \bigr)  \\
&\quad\times S( \vec{q}_{Ts} ;\mu,\nu) f_1^{\perp}\Bigl(\vec{q}_{T1}, x_1, p^-;\mu ,\nu\Bigr)f_2^{\perp}\Bigl(\vec{q}_{T2},x_2, p^+;\mu ,\nu\Bigr) \,, \nn
\end{align}
at $s=(P_1+P_2)^2$, with colliding protons of momenta $P_{1,2}$, and gauge boson invariant mass $Q^2$ and rapidity $y$. For the case of the Higgs, we have a Wilson coefficient $C_t$ after integrating out the top quark. (For DY we  just set $C_t=1$ in \eq{txsec}, and consider explicitly only the $\gamma^*$ channel in this paper.)
Here, $S$ is the soft function accounting for the contribution of soft radiation to $\vec{q}_T$, $f_{1,2}^\perp$ are the TMDPDFs (or beam functions) accounting for the contribution of radiation collinear to the incoming protons to $\vec{q}_T$, and they depend on kinematic variables $p^\mp = Qe^{\mp y} = x_{1,2}\sqrt{s}$. The peculiarity of the factorization is that even though the TMDPDFs form a part of the IR physics, they depend on the hard scale $Q$ (\emph{c.f.} \cite{Chiu:2007dg}), which, as we shall see later, will play an important role in our resummation formalism. The hard function $H$ encodes virtual corrections to the hard scattering process, computed by a matching calculation from QCD to SCET. The scale $\mu$ is the renormalization scale normally encountered in the $\MSbar$ scheme and plays the role of separating hard modes (integrated out of SCET) from the soft and collinear modes, by their virtuality. The additional rapidity renormalization scale $\nu$, introduced in \cite{Chiu:2011qc,Chiu:2012ir}, arises from the need to separate soft and collinear modes, which share the same virtuality $\mu$, in their rapidity (\fig{modes}). The cross section itself is independent of these arbitrary virtuality and rapidity boundaries, but the renormalization group (RG) evolution of factorized functions from their natural scales, where they have no large logs, to arbitrary $\mu,\nu$ can be used to resum the large logs in the cross section.

\subsection{RG and RRG Evolution in Impact Parameter vs. Momentum Space}

These functions obey the renormalization group (RG) equations in $\mu$
\bea
\mu \frac{d }{d\mu} F_i =  \gamma_{\mu}^i F_i 
\eea
where $F_i$ can be $C_t^2(M_t^2, \mu)$,  $H(Q^2;\mu)$, $S(\vec{q}_{Ts};\mu,\nu)$ or $f_i^{\perp}(\vec{q}_{Ti}, Q, x_i;\mu ,\nu)$. The RG equations in $\nu$ have a more complicated convolution structure: 
\bea
\nu \frac{ d}{d\nu} G_i (\vec{q}_{T}; \nu) = \gamma_{\nu}^i( \vec{q}_{T}) \otimes G_i (\vec{ q_{T}} ;\nu)
\eea
where $G_i$ can be soft functions or TMDPDFs. The symbol $\otimes$ here indicates convolution defined as
\bea
\gamma_{\nu} (\vec{q_T}) \otimes G( \vec{q_T})= \int \frac{d^2p_T}{(2 \pi)^2} \gamma_{\nu} (\vec{q_T} - \vec{p_T} ) G( \vec{p_T})   
\eea
Apart from the complicated structure of the RG equations, the anomalous dimensions themselves are not simple functions but are usually plus distributions \cite{Chiu:2012ir}  which makes it even harder to solve these equations directly in momentum space. A typical strategy to get around this is to Fourier transform to position (\ie impact parameter) space, defining
\bea
\widehat G(\vec{b}) \equiv \int \frac{d^2 q_T}{(2\pi)^2} e^{i \vec{b} \cdot \vec{q_T}}G(\vec{q_T})  \,,\quad \widehat G(\vec{b}) \equiv \frac{1}{2\pi}\wt G(b)\,,\quad b\equiv \abs{\smash{\vec{b}}}\,,
\eea
the latter definitions accounting for the fact that all the distributions we encounter will have azimuthal symmetry in $\vec{q}_T$ or $\vec{b}$. This then gives ordinary multiplicative differential equations (instead of convolutions), and a closed form solution to the RG equations can be easily obtained. Moreover the cross section now takes the simpler structure,
\begin{align}
\label{xsec}
 \frac{ d \sigma} {dq_T^2 dy} &= \sigma_0 \pi (2\pi)^2 C_t^2(M_t^2, \mu) H (Q^2,\mu ) \int db\, b J_0( b q_T) \\
 &\qquad \times \wt S( b ,\mu,\nu) \wt f_1^{\perp}( b, x_1,p^-; \mu ,\nu) \wt f_2^{\perp}(  b, x_2, p^+;  \mu ,\nu)  \,,\nn
\end{align}
where $J_0$ is the $n=0$ Bessel function of the first kind. Note we have changed variables from $\vect{q}_T$ in \eq{txsec} to $q_T^2$ in \eq{xsec}. The $b$-space soft and beam functions $\wt S$ and $\wt f_i^\perp$  now obey multiplicative rapidity RGEs in $\nu$,
\be
\nu\frac{d}{d\nu}\wt G_i = \gamma_\nu^i \wt G_i\,,
\ee
whose anomalous dimensions and solutions we shall give below. Only the $b$ integration in \eq{xsec} stands in the way of a having a simple product factorization of the momentum-space cross section. Finding a way to carry it out will be one the main focuses of this paper.
For perturbative values of $q_T$, the TMDPDF's can be matched onto the PDF's.
The $b$-space cross section, defined as the following product of factors in the integrand of \eq{xsec}:
\be
\label{bxsec}
\wt\sigma(b,x_1,x_2;\mu,\nu) = H(Q^2,\mu) \wt S(b;\mu,\nu) \wt f_1^\perp(b,x_1,p^-;\mu,\nu) \wt f_2^\perp(b,x_2,p^+;\mu,\nu)\,,
\ee
computed in fixed-order QCD perturbation theory then contains logs of $Q b_0$ where $b_0 = b e^{\gamma_E}/2$ (see \eq{sing}). Schematically, the expansion takes the form
\be
\label{logexpansion}
(2\pi)^3\wt\sigma(b) = f_i(x_1)f_{\bar i}(x_2) \exp\bigg[ \sum_{n=0}^\infty\sum_{m=0}^{n+1}\Bigl(\frac{\as(\mu)}{4\pi}\Bigr)^n G_{nm}\ln^m Qb_0\biggr] \,,
\ee
where $i,\bar i= g$ for Higgs production and $i=q$ for DY, and where we ignore effects of DGLAP evolution for the moment (we include them in \eq{sing} and in all our analysis below). This takes the typical form of a series of Sudakov logs. The number of coefficients $G_{nm}$ that need to be known is determined by the desired order of resummed accuracy. Using the heuristic power counting $\ln Qb_0\sim 1/\as$ in the region of large logs needing resummation, the leading log (LL) series includes the $\cO(1/\as)$ terms $m=n+1$, the next-to-leading log (NLL) series the $\cO(1)$ terms up to $m=n$, at NNLL the $\cO(\as)$ terms up to $m=n-1$, etc. When we later talk about resummation in momentum space, we will define our accuracy by the corresponding terms in the $b$-space integrand that we have successfully inverse Fourier transformed (cf. \cite{Almeida:2014uva}).

For a TMD cross section, the logs in the full QCD expansion \eq{logexpansion} are factored into logs from the hard and soft functions and TMDPDFs of ratios of the arbitrary virtuality and rapidity factorization scales $\mu,\nu$ and the physical virtuality and rapidity scales defining each mode. Each function contains logs:
\be
C_t^2 = C_t^2\Bigl(\ln \frac{\mu^2}{M_t^2}\Bigr)\,\ H = H\Bigl(\ln \frac{\mu^2}{Q^2}\Bigr)\,,\ \wt S = \wt S\Bigl(\ln\mu b_0,\ln\frac{\mu}{\nu}\Bigr)\,,\ \wt f^\perp = \wt f^\perp \Bigl(\ln \mu b_0,\ln\frac{\nu}{p^\pm}\Bigr)\,.
\ee
These logs reflect the natural virtuality and rapidity scales where each function ``lives'' and where logs in each are minimized. For example, at one loop, the logs in the QCD result \eq{sing} split up into individual hard, soft, and collinear logs from \eqss{Cn}{Sn}{Iijcoefficients},
\begin{align}
&-\zed_H \frac{\Gamma_0}{2} \ln^2 Qb_0 - \gamma_H^0 \ln Qb_0- \gamma_{C_t^2}^0 \ln M_tb_0 = - \zed_H \frac{\Gamma_0}{2} \ln^2\frac{\mu}{Q} + \gamma_H^0\ln\frac{\mu}{Q}  + \gamma_{C_t^2}^0\ln\frac{\mu}{M_t} \nn\\
 &\qquad\qquad\qquad   + \zed_S\frac{\Gamma_0}{2}\Bigl(\ln^2 \mu b_0 + 2\ln\mu b_0 \ln\frac{\nu}{\mu}\Bigr)  + \zed_f \Gamma_0 \ln\mu b_0\ln\frac{\nu^2}{Q^2} + 2\gamma_f^0 \ln\mu b_0\,, 
\end{align}
where the individual anomalous dimension coefficients satisfy the constraints $\zed_H + \zed_S + 2\zed_f = 0$ and $\gamma_H^0+\gamma_{C_t^2}^0 + 2\gamma_f^0 = 0$. (For DY, $\gamma_{C_t^2}=0$.) RG evolution of each factor---hard, soft, and collinear---in both virtuality and rapidity space from scales where the logs are minimized, namely, $\mu_H\sim Q, \mu_T \sim M_t$ and, naively, $\mu_{S,f}\sim 1/b_0$ for the virtuality scales, while $\nu_S\sim \mu_S$ and $\nu_f\sim Q$ for the rapidity scales, to the common scales $\mu,\nu$ achieve resummation of the large logs, to an order of accuracy determined by the order to which the anomalous dimensions and boundary conditions for each function are known and included. This will be reviewed in further detail in \sec{resumxsec}.

This, at least, is the procedure one would follow to resum logs in impact parameter space. It corresponds, in SCET language, to how to obtain the result of the standard CSS resummation through traditional \cite{Collins:1984kg} or modern techniques \cite{Collins:2011zzd}, as well as recent EFT treatments like \cite{Neill:2015roa}. Then the resummed $b$-space cross section is Fourier transformed back to momentum space via \eq{xsec}. The main issue with this procedure is that the strong coupling $\as(\mu)$ in the soft function and TMDPDFs is then evaluated at a $b$-dependent scale $\mu_{S,f}\sim 1/b_0$, which enters the nonperturbative regime at sufficiently large $b$ in the integral in \eq{xsec}. So the integrand must be cut off before reaching the Landau pole in $\as$. There are quite a few procedures in the literature to implement precisely such a cutoff by introducing models for nonperturbative physics, see \eg \cite{Qiu:2000hf,Sun:2013dya,DAlesio:2014mrz,Collins:2014jpa,Scimemi:2017etj}.

Motivated by these observations,  in this paper we explore the following main questions:
\begin{itemize}
\item Even though the natural scale for minimizing the logarithms in the soft function and TMDPDFs is a function of the impact parameter $b$, can we actually set scales directly in momentum space, after performing the $b$ integration? (without an arbitrary cutoff of the $b$ integration?)

\item If that is possible, can we obtain a closed-form expression for the cross section which will be accurate to any resummation order and  ultimately save computation time?
\end{itemize}

In \sec{resumxsec} we shall propose a way to answer the first question, and in \sec{analytic} we shall develop a method to answer the second. To aid the reader in quickly grasping the main points of our paper, we offer a more detailed-than-usual summary of these sections here, which is somewhat self-contained and can be used as a substitute for the rest of the paper upon a first reading. Readers interested in the details of our arguments can then delve into the main body of the paper. Except for a brief discussion near the end, we emphasize we address only the perturbative computation of the cross section in this paper.

\subsection{A hybrid set of scale choices for convergence of the $b$ integral}

Regarding the first question, the issue with leaving the $\mu,\nu$ scales for the soft function and TMDPDFs unfixed before integrating over $b$ in \eq{xsec} is that the integral, while avoiding the Landau pole from long-distance/small-energy scales, is then plagued by a spurious divergence from \emph{large}-energy/\emph{short}-distance emissions \cite{Chiu:2012ir}, \eg at NLL accuracy:
\begin{align}
\label{singlelog}
 \frac{d \sigma}{d q_T^2 dy} &=\sigma_0 \pi (2\pi)^2 C_t^2(M_t^2, \mu_T)U^{\rm NLL}(\mu_H, \mu_T, \mu_L) H(Q^2;\mu_H)  \int db\, b J_0(b q_T) \wt S(b;\mu_L,\nu_L) \\ 
 &\qquad \times \wt f_1^\perp(b,x_1,p^-;\mu_L,\nu_H) \wt f_2^\perp(b,x_2,p^+;\mu_L,\nu_H)  \exp\left[ -\Gamma_0 \frac{\alpha(\mu_L)}{\pi}\ln\left(\frac{\nu_H}{\nu_L}\right)\ln(\mu_L b_0) \right]\,, \nn
 \end{align}
 although $ H,\wt S,\wt f^\perp$ are truncated to tree level at NLL while $C_t= \as$. 
 The hard scale $\mu_H$ is usually set to $i Q$ to implement what is called $\pi^2$ resummation to improve perturbative convergence \cite{Ahrens:2008qu,Ahrens:2008nc}. 
 This integral, as we will see below in \eq{softNLL}, is still divergent. At this point, $\mu_L$ and $\nu_{H,L}$ are $b$-independent and cannot help with regulating the integral. What we need in \eq{singlelog} is a factor that damps away the integrand for both large \emph{and} small $b$. 
 In this paper, we adopt the approach that there are already terms in the physical cross section itself that can play the role of this damping factor and that we should use them. Namely, at NLL$'$ order and beyond, the soft function evaluated at the low scales $\mu_L,\nu_L$ in the integrand of \eq{singlelog} contains logs of $\mu_L b_0$ that we can use to regulate the integral, see \eq{Sn}. Since $\wt S(b,\mu_L,\nu_L)$ no longer contains large logs (if $\mu_L,\nu_L$ are chosen near the natural soft scales), it is typically truncated to fixed order (see \tab{NkLL}). However, we know that the logs themselves still exponentiate, being predicted by the solution \eq{Sevolved2} to the RG and $\nu$RG equations. If we could keep the exponentiated one-loop double log in $\wt S$ in \eq{Sn} in the integrand of \eq{singlelog}, $\exp\bigl[\frac{\as(\mu_L)}{8\pi}\zed_S \Gamma_0 \ln^2\mu_L b_0\bigr]$, where $\zed_S = -4$, it would play precisely the role that we desire. Now, as we argue below, if we are going to keep this term exponentiated, we should also include a piece of the 2-loop rapidity evolution kernel $\sim \as^2 \ln^2\mu_L b_0 \ln(\nu_H/\nu_L)$ given by \eqs{UVdef}{gammanuexp} in the exponent, as it is of the same form and same power counting, so that the terms we wish to promote to the exponent of \eq{singlelog} at least at NLL order are:
\be
\label{Sexp}
\wt S_{\text{exp}} = \exp \biggl[ \frac{\as(\mu_L)}{4\pi}\zed_S \frac{\Gamma_0}{2} \ln^2 \mu_L b_0 + \Bigl(\frac{\as(\mu_L)}{4\pi}\Bigr)^2 \zed_S \Gamma_0 \beta_0 \ln^2\mu_L b_0 \ln \frac{\nu_H}{\nu_L}\biggr]\,.
\ee
These are terms that would otherwise be truncated away at strict NLL accuracy. Since they are subleading, we are in fact free to choose to include them (and not other subleading terms that are formally of the same order). While this is admittedly a bit \emph{ad hoc}, we take the view that it is no more arbitrary than any regulator or cutoff we might choose to introduce to \eq{singlelog}, and these are terms that actually exist in the expansion of the physical cross section. We can rephrase this choice of subleading terms in \eq{Sexp} to include in \eq{singlelog} as part of our freedom to choose the precise scale $\nu_L$ in \eq{singlelog} (the variation of which anyway probes theoretical uncertainty due to missing subleading terms).  Namely, if one were otherwise to choose $\nu_L\sim \mu_L$ in $\wt S$ in \eq{singlelog}, we propose then shifting that choice to:
\be
\label{nuLshift}
\nu_L\to \nu_L^* = \nu_L(\mu_L b_0)^{-1+p}\,,\qquad p = \frac{1}{2}\biggl[ 1 - \frac{\as(\mu_L)\beta_0}{2\pi}\ln\frac{\nu_H}{\nu_L}\biggr]\,,
\ee
which we derive in \eqs{nuLstar}{eq:n}. This achieves the shifting of the terms in the exponent of \eq{Sexp} that would otherwise be truncated away into the integrand of \eq{singlelog} where they appear explicitly, and can be used to regulate the $b$ integral. This particular choice of regulator factor in \eq{nuLshift} is motivated, furthermore, by the fact that it will allow us actually to evaluate the $b$ integral \eq{singlelog} (semi-)\emph{analytically}, as we show in \sec{analytic}. Maintaining a Gaussian form for the exponent in $\ln b$ inside the $b$ integral will be crucial to this strategy.

Beyond NLL, we will choose to keep the same shifted scale choice \eq{nuLshift}, but to ensure that we do not introduce higher powers of logs of $\mu_L b_0$ than quadratic into the exponent of the integrand in \eq{singlelog}, we make one additional modification to how we treat the rapidity evolution kernel. Namely, in the all orders form of the rapidity evolution kernel:
\be
\label{Vdef}
V(\nu_L,\nu_H;\mu_L) = \exp\biggl[ \gamma_\nu^S(\mu_L)\ln\frac{\nu_H}{\nu_L}\biggr]\,,
\ee
given in \eq{UVdef}, where the rapidity anomalous dimension takes the form \eq{gammanuexp},
\begin{align}
\gamma_\nu^S(\mu_L) &=  \quad  \frac{\as(\mu_L)}{4\pi} \quad \biggl[\zed_S \Gamma_0 \ln \mu_L b_0  \biggr] \\
&\quad +  \Bigl(\frac{\as(\mu_L)}{4\pi}\Bigr)^2 \biggl[ \zed_S \Gamma_0\beta_0 \ln^2 \mu_L b_0 +( \zed_S  \Gamma_1 + 2\gamma_{RS}^0 \beta_0 )\ln \mu_L b_0 + \gamma_{RS}^1\biggr] \,, \nn
\end{align}
we divide the anomalous dimension into a purely ``conformal'' part containing only the diagonal pure cusp terms with a single log of $\mu_L b_0$ and the same for the non-cusp part $\gamma_{RS}$. We divide the rapidity evolution kernel \eq{Vdef} into corresponding ``conformal'' and ``non-conformal'' parts:
\be
\label{Vdivision}
V(\nu_L,\nu_H;\mu_L) = V_\Gamma(\nu_L,\nu_H;\mu_L) V_\beta(\nu_L,\nu_H;\mu_L)\,,
\ee
where $V_\Gamma$ contains pure anomalous dimension coefficients,
\be
\label{VGammaintro}
V_\Gamma (\nu_L,\nu_H;\mu_L) = \exp\biggl\{ \ln\frac{\nu_H}{\nu_L} \sum_{n=0}^\infty \Bigl(\frac{\as(\mu_L)}{4\pi}\Bigr)^{n+1}(\zed_S \Gamma_n \ln\mu_L b_0 + \gamma_{RS}^n)\biggr\}\,,
\ee
and $V_\beta$ contains all the terms with beta function coefficients, whose expansion is shown in \eq{Vbeta}. We will keep $V_\Gamma$ exponentiated as in \eq{VGammaintro}, and the shift $\nu_L\to \nu_L^*$ in \eq{nuLshift} will turn it into a Gaussian in $\ln\mu_L b_0$ and thus  allow us to carry out the $b$ integral in \eq{singlelog} (updated beyond NLL). However, to keep this Gaussian form of the exponent, we will then choose to truncate $V_\beta$ at fixed order. The logs of $\mu_L b_0$ in $V_\beta$ will give integrals in \eq{singlelog} that we can carry out by differentiating the basic result we obtain in \sec{analytic}.

Admittedly, this expansion and truncation of $V_\beta$ is not part of any usual scheme for N$^k$LL resummation, but is our addition. In particular, $V_\beta$ still contains large logs of $\nu_H/\nu_L$ as seen in \eq{Vbeta}. This means that starting at NNLL order, we will not actually exponentiate \emph{all} the logs that appear at this accuracy, as usual log counting schemes in the exponent require. This is the price we choose to pay for the (semi-)analytic solution we obtain in \sec{analytic}, which requires a Gaussian exponent in $\ln b$ in the $b$ integrand. This is essentially an implementation of Laplace's method for evaluating the $b$ integral. As we argue in \ssec{NNLLbeyond}, our truncation of $V_\beta$ in fixed order is not as bad as failing to exponentiate large logs involving $\mu$ (which we do exponentiate) would be. There is never more than a single large log of $\nu_H/\nu_L$ appearing in the exponent of the rapidity evolution. Thus, the series of terms in the fixed-order expansion of the exponentiated $V_\beta$ are suppressed at every order by another power of $\as$.\footnote{In the $\mu$ evolution kernels, the exponents, \eg \eqs{KGammaexp}{etaexp}, themselves contain higher and higher powers of large logs of $\mu_H/\mu_L$, and truncating any part of it to fixed order would not be sensible. Truncating $V_\beta$ in \eq{Vbeta} to the same order as other corresponding genuinely fixed-order terms at N$^k$LL accuracy makes more sense. Loosely speaking, we maintain counting of logs in the exponent for most of the cross section, except for $V_\beta$, in which we revert to older log counting in the fixed-order expansion (N$^k$LL$_\text{E}$ vs. N$^k$LL$_\text{F}$ in \cite{Chiu:2007dg}.)} Our expansion of $V_\beta$ should be viewed as asymptotic expansion, which, indeed, we find truncating at a finite order yields a good numerical approximation to the resummed cross section (within the theoretical uncertainties otherwise present in the resummed cross section at N$^k$LL accuracy) in the perturbative region.\footnote{We expect that the way in which it breaks down for small $q_T$ will yield clues to the behavior of the nonperturbative contributions to the cross section, which however are not the subject of this paper.} Note, furthermore, that in the conformal limit, $V_\beta =1$, and the exponentiated part $V_\Gamma$ of the rapidity evolution would be exact.

We should point out that, through the shift \eq{nuLshift}, we do introduce $b$ dependence into our choice of scale $\nu_L$, so we would not call our resummation scheme \emph{entirely} a momentum-space scheme. (See \cite{Ebert:2016gcn,Monni:2016ktx}  for such proposed methods.) We do, however, leave the $\mu_L$ scale unfixed until \emph{after} the $b$ integration, and this still allows us to avoid integrating over a Landau pole in $\as(\mu_L)$ in \eq{xsec}. 

In \sec{resumxsec} we also use our freedom to determine exactly where $\mu_L\sim q_T$ should be in order to improve the convergence of the resummed perturbative series. We argue it should be set at a value such that other unresummed fixed-order logs make a minimal contribution to the final momentum-space cross section. For small values of $q_T$, this scale turns out to be shifted to slightly higher values $\mu_L\sim q_T + \Delta q_T$. Without making such a shift we find instabilities in the evaluation of the cross section. This is similar in spirit to the shift $\mu_L\to q_T+q_T^*$ proposed in \cite{Becher:2011xn,Becher:2012yn}, though not identical in motivation, implementation, or interpretation in terms of nonperturbative screening.

\subsection{A semi-analytic result for the $b$ integral with full analytic dependence on momentum-space parameters}
\label{ssec:intro3}

If we stopped there, our choice of scale \eq{nuLshift} might be no more than just another in a long series of proposed schemes to avoid the Landau pole in \eq{xsec}, and, in addition, our division of the rapidity evolution kernel \eq{Vdef} into an exponentiated and a fixed-order truncated part in \eq{Vdivision} would be quite unnecessary and inexplicable. However, what we find in \sec{analytic} is that all of these scheme choices together yield a form of the $b$-space integrand \eq{bxsec} that is Gaussian in $\ln b$ so that we can integrate it analytically into a fairly simple form, modulo a numerical approximation for the pure Bessel function in \eq{xsec}. The dependence on all physical parameters and scales such as $q_T,\mu_{H,L},\nu_{H,L}$, is obtained analytically. We now briefly summarize our procedure and results.

With the division \eq{Vdef} of the rapidity evolution kernel and the scale choice $\nu_L^*$ in \eq{nuLshift}, the momentum-space cross section \eq{xsec} can be written in the form, given in \eq{resummedIb}, 
\be
\label{qTcs}
\frac{d\sigma}{dq_T^2 dy} = \frac{\sigma_0}{2} C_t^2(M_t^2, \mu_T) H(Q^2,\mu_H) U(\mu_L,\mu_H, \mu_T) I_b(q_T,Q;\mu_L,\nu_L^*,\nu_H)\,,
\ee
where we isolated the $b$ integral,
\be
\label{Ibintro}
I_b(q_T,Q;\mu_L,\nu_L^*,\nu_H) \equiv \int_0^\infty db\,b J_0(bq_T) \wt F(b,x_1,x_2,Q;\mu_L,\nu_L^*,\nu_H) V_\Gamma(\nu_L^*,\nu_H;\mu_L)\,,
\ee
$\wt F$ contains the fixed-order terms, including powers of logs of $\mu_L b_0$, contained in the soft function, TMDPDFs, and the part $V_\beta$ in \eq{Vdivision} of the rapidity evolution kernel that we choose to truncate at fixed order. As we will show in \sec{analytic}, the exponentiated part of the rapidity evolution kernel $V_\Gamma$ in \eq{VGamma}, with the scale choice $\nu_L^*$ in \eq{nuLshift}, can be written in the form of a pure Gaussian in $\ln b$,
\be
\label{VGauss}
V_\Gamma = C e^{-A \ln^2(\mu_L b_0 \chi)}\,,
\ee
where $C,A,\chi$ are functions of the scales $\mu_L,\nu_L,\nu_H$ and the rapidity anomalous dimension, given explicitly in \eq{ACUeta}. In particular $A\sim \Gamma[\as(\mu_L)]$. If we could figure out how to integrate this Gaussian against the Bessel function in \eq{Ibdef}, we would be done. Now, the presence of terms in $\wt F$ in \eq{Ibdef} with nonzero powers of $\ln \mu_L b_0$ can be obtained from the basic result by differentiation, as we will derive in \ssec{fixedorderlimit}, so we really only need to figure out how to evaluate the basic integral,
\be
\label{basicintegral}
I_b^0= \int_0^\infty db\, b J_0(b q_T)\, e^{-A\ln^2(\Omega b)}\,,
\ee
where $\Omega \equiv \mu_L e^{\gamma_E}\chi/2$.

Now, our mathematical achievements in this paper do not reach so far as to evaluate \eq{basicintegral} analytically in its precise form. We will, however, develop a procedure to evaluate it in a closed form, with analytic dependence on $q_T,A,\Omega$ (and thus all scales and anomalous dimensions), to arbitrary numerical accuracy determined by the goodness of an approximation we use for the Bessel function. We find a basis in which to expand the pure Bessel function, in which just a few terms are sufficient to reach a precision better than needed for NNLL accuracy in the resummed cross section, and which can be systematically improved as needed. The details of this derivation are in \sec{analytic}, but we summarize the key steps here.

The first step is to use a Mellin-Barnes representation for the Bessel function,
\be
\label{Jzero}
J_0(z) = \frac{1}{2 \pi i}\int_{c-i \infty}^{c+i \infty} dt \frac{\Gamma[-t]}{\Gamma[1+t]} \left(\frac{1}{2} z \right)^{2t}  \,,
\ee
where the contour lies to the left the poles of the gamma function $\Gamma(-t)$, so $c<0$. The choice $c=-1$ turns out to be well behaved, and useful as it is closely related to the fixed-order limit of \eq{Ibintro} (see \ssec{largeqT}). This trades the $b$ integral in \eq{basicintegral} for the $t$ integral, and we obtain
\be
\label{Ib0f}
I_b^0 = -\frac{2}{\pi q_T^2}\frac{e^{-AL^2}}{\sqrt{\pi A}}\int_{- \infty}^{\infty} dx\, \Gamma(-c-ix)^2 \sin[\pi (c+ix)] e^{-\frac{1}{2}[x - i(c-t_0)]^2}\,,
\ee
where we parametrized the contour in \eq{Jzero} as $t=c+ix$, and where $t_0 = -1+AL$, where $L = \ln (2\Omega/q_T)$. We also used the reflection formula $\Gamma(-t)\Gamma(1+t) = -\pi\csc(\pi t)$.

It may appear that we are no farther along than when we started with \eq{basicintegral}---we still have to do the $x$ integral. However, we now observe that thanks to the form of the Gaussian with a width $\sim \sqrt{A}$, which vanishes in the limit $\as\to 0$, we only need to know the rest of the integrand, in particular 
\be
\label{f}
f(t)\equiv \Gamma(-c-ix)^2\,,
\ee 
in a fairly small region of $x$. In fact we shall not need it out to more than $\abs{x}\sim 1.5$ for any of our applications. Thus if we can find a good basis in which to expand $f$ where every term gives an analytically evaluable integral in \eq{Ib0f}, we shall be in good shape. 

Now, this would not have been a good strategy in \eq{Ibintro} for the Bessel function itself, as it is highly oscillatory out to fairly large $b$, and the Gaussian does not damp the integrand away quickly---its width only grows as $\as$ (i.e. $A$) goes to zero. However, inside \eq{Ib0f}, we find an expansion of $f$ (\eq{f}) in terms of Hermite polynomials $H_n$ to work very well:
\be
\label{seriesexpansion}
\Gamma(1-ix)^2  =  e^{-a_0 x^2} \sum_{n=0}^\infty c_{2n} H_{2n}(\alpha x) 
	+ \frac{i \gamma_E}{\beta} e^{-b_0 x^2} \sum_{n=0}^\infty c_{2n+1} H_{2n+1}(\beta x) \,,
\ee
where we now pick $c=-1$ and factor out Gaussians with widths set by $a_0,b_0$ which closely (but not exactly) resemble the real and imaginary parts of $\Gamma(1-ix)^2$ itself, near $x=0$. Their departures from an exact Gaussian are accounted for by the remaining series of Hermite polynomials. It would be natural to choose the scaling factors $\alpha,\beta$ for the Hermite polynomials to be $\alpha^2 = a_0$ and $\beta^2 = b_0$, but instead we leave them free, to be determined empirically to optimize fast convergence of the series. We find we can get sufficient numerical accuracy acceptable for NNLL accuracy in the final cross section with just a few (3 or 4) terms in each series, real and imaginary. The coefficients $c_n$ in \eq{seriesexpansion} still have to be determined by the numerical integrals \eq{HermiteCoefficients}, which unfortunately prevents us from having a fully analytic result for the momentum-space cross section. However, the series \eq{seriesexpansion} with these numerical coefficients depends only on properties of the pure mathematical function $\Gamma(1-ix)^2$ itself---not on any physical parameters. The dependence on these we keep analytically. All that is left is to evaluate analytically the integral of each Hermite polynomial against the Gaussian in \eq{Ib0f}, leading to the result we derive in \eq{IbHermite},
\be
\label{Ibseriesintro}
I_b^0  = \frac{2}{\pi q_T^2}  \sum_{n=0}^{\infty} 
\Imag \biggl\{    c_{2n} \cH_{2n}(\alpha,a_0) + \frac{i\gamma_E}{\beta} c_{2n+1} \cH_{2n+1}(\beta,b_0)  \biggr\}\,,
\ee
where each term $\cH_{n}$ is defined by the integral,
\be
\label{Hnintro}
\cH_{n}(\alpha,a_0) =  \frac{1}{\sqrt{\pi A}} e^{-A(L-i\pi/2)^2}\int_{-\infty}^\infty dx\, H_{n}(\alpha x) e^{-a_0 x^2-\frac{1}{A}( x + z_0)^2}\,,
\ee
each of which has the closed form result,
\be
\label{Hnanalytic}
\cH_n(\alpha,a_0) =  \frac{(-1)^n n! \, e^{\frac{-A(L-i\pi/2)^2}{1+a_0 A}}}{(1+a_0 A)^{n+\frac{1}{2}}}\sum_{m=0}^{\floor*{n/2}}\frac{1}{m!}\frac{1}{(n-2m)!} \Bigl\{ [ A(\alpha^2 - a_0)-1] (1+a_0 A)\Bigr\}^m (2\alpha z_0)^{n-2m} \,,
\ee
the first several of which are written out explicitly in \eq{Hnexplicit}. In \eqs{Hnintro}{Hnanalytic}, $z_0 = A(\pi/2 + iL)$, in terms of which the integral \eq{Ib0f} can be written, the shifted exponents arising from absorbing the sine function in \eq{Ib0f}. 

The results \eq{Hnanalytic} for the integrals \eq{Hnintro} are the primary mathematical result of our paper. The final and primary physics result of our paper, \eq{thefinalfinalresult}, the resummed cross section in momentum space, is given in terms of the analytic result \eq{Ibseriesintro} for $I_b^0$ above.

While a first glance at these formulas may not be particuarly illuminating, we would like to emphasize that the results \eq{Hnanalytic} of the integrals \eq{Hnintro} in terms of which the final result is written contain within them explicit dependence on all the physical parameters such as $q_T$ and the scales $\mu_{L,H},\nu_{L,H}$ that one would want to vary not only to evaluate the cross section but estimate its theoretical uncertainties. This is made very fast to compute by our explicit analytic formula, modulo only the numerically computed coefficients in \eq{seriesexpansion}, but that can be done once and for all, for any TMD observable or kinematics.

It is important to emphasize that the result \eq{thefinalfinalresult} we give for the resummed momentum-space cross section represents, then, a triple expansion:
\begin{itemize}
\item \emph{Perturbative expansion:} usual expansions in $\alpha_s$ of matching coefficients and resummed exponents in \eq{qTcs}, counting $\as \ln(\mu_H/\mu_L)\sim 1$ or $\as \ln(\nu_H/\nu_L)\sim 1$, and fixed-order tails (not shown in \eq{qTcs}).

\item \emph{$V_\beta$ expansion:} The additional truncation of the $V_\beta$ part of the rapidity evolution kernel in \eqs{Vdivision}{VGammaVbeta} to a fixed order in $\as$, according to \tab{NkLL}, makes possible the integration of a rapidity exponential \eq{VGauss} Gaussian in $\ln b$, and behaves as an asymptotic expansion. This expansion becomes exact in the conformal limit of QCD.

\item \emph{Hermite expansion:} The integral of the Gaussian in \eq{VGauss} against the Bessel function in \eq{basicintegral} is performed in terms of analytic integrals, by expanding $J_0$ through the representation \eq{Jzero} and the series of Gaussian-weighted Hermite polynomials \eq{seriesexpansion}, truncated to a finite number of terms, as needed to achieve a numerical accuracy in the cross section better than the perturbative uncertainty already present.
\end{itemize}
These are the expansions we find necessary to obtain the analytical (up to the numerical Hermite coefficients) result for the cross section in \eq{thefinalfinalresult}. Each expansion is systematically and straightforwardly improvable. The last two expansions could be avoided if one is satisfied with a fully numerical evaluation of the $b$ integral in \eq{Ibintro}. We find the expansions worthwhile as they yield the faster and similarly accurate formula \eq{thefinalfinalresult}.\footnote{In our calculations, we found a factor of 5 improvement in speed with our formula for the $q_T$ distribution vs. numerically integrating \eq{Ibintro} at every $q_T$.} In the rest of the paper, we will do our best to make clear which expansion(s) are being used at each stage.

\bigskip

The remainder of our paper is organized as follows. 
Before concluding \sec{analytic}  we match our resummed result onto fixed-order perturbation theory in \ssec{matching}  and obtain and illustrate our results resummed to NNLL accuracy and matched to $\cO(\as)$ fixed order. In \sec{non_pert} we offer some comments about expected nonperturbative corrections to our perturbative predictions, and in \sec{compare} we survey other methods to resum TMD cross sections in the literature as compared to ours. We conclude in \sec{conclusions}. In the Appendices we offer an array of technical results we need to evaluate the integrals and  cross sections in the rest of the paper, as well as some alternatives to particular choices of schemes or methods we made in the main body of the paper.

\section{Resummed cross section}
\label{sec:resumxsec}

In this section, we first review RG and rapidity RG methods to resum logs of separated hard and soft/collinear virtuality scales and collinear and soft rapidity scales in TMD cross sections. We review a standard procedure to set scales in impact parameter space, and then inverse Fourier transforming to momentum space. Then we propose a hybrid scale setting scheme where the soft rapidity scale is chosen to depend on $b$, but the virtuality scales are chosen only \emph{after} we transform back to momentum space, allowing evaluation of the $b$ integral without encountering a Landau pole. We also organize the rapidity evolution kernel in a way that anticipates making use of it to perform the $b$ integral semi-analytically in \sec{analytic}. We also address the choice of the soft virtuality scale itself in momentum space to ensure stable power counting of logs.

\subsection{RGE and $\nu$RGE solutions}

We defined the $b$-space cross section in \eq{bxsec}.
The cross section is independent of the virtuality and rapidity factorization scales $\mu,\nu$, but each factor $H,\wt S,\wt f^\perp$ does depend on them, and contains logs of ratios of the scales $\mu,\nu$ to their ``natural'' virtuality or rapidity scales $\mu_H$, $(\mu_L,\nu_L)$ and $(\mu_L,\nu_H)$, at which no large logarithms exist.  Thus we would like to evaluate each factor at these separate scales, and then use RG and $\nu$RG evolution to take them to the common scales $(\mu,\nu)$ at which the cross section is evaluated. The solutions to these evolution equations are in a form where the large logs of ratios of separated scales are resummed or exponentiated.

\subsubsection{Hard function}

The hard function $H=\abs{C}^2$ depends only on the virtuality scale $\mu$, and obeys the RGE,
\be
\label{hardRGE}
\mu\frac{d}{d\mu}C(Q^2,\mu) = \gamma_C(\mu)C(Q^2,\mu) \Rightarrow \mu\frac{d}{d\mu}H(Q^2,\mu) = \gamma_H(\mu)H(Q^2,\mu)\,,
\ee
where the anomalous dimension takes the form,
\be
\gamma_C(\mu) = -\frac{\zed_H}{2}\Gamma_{\text{cusp}}[\as(\mu)]\ln\frac{\mu}{Q} + \gamma_C[\as(\mu)] \,,\qquad \gamma_H = \gamma_C + \gamma_C^*\,,
\ee
where $\Gamma_{\text{cusp}}$ is known as the cusp anomalous dimension, the proportionality constant $\zed_H = 4$, and $\gamma_C[\as]$ is the non-cusp part of the anomalous dimension. The cusp anomalous dimension can be written as an expansion in the strong coupling $\alpha_s(\mu)$. 
\be
\Gamma_{\text{cusp}}[\as(\mu)] = \sum_{i=0}^{\infty}{\left(\frac{\alpha_s(\mu)}{4\pi}\right)^{i+1} \Gamma_{i}}
\ee 

The RGE \eq{hardRGE} has the solution
\be
\label{Hevolved}
C(Q^2,\mu) = C(Q^2,\mu_H)U_C(\mu_H,\mu) \Rightarrow H(Q^2,\mu) = H(Q^2,\mu_H) U_H(\mu_H,\mu)\,,
\ee
where the evolution kernel is
\begin{align}
\label{UHdef}
U_C(\mu_H,\mu) &= \exp\biggl\{ \int_{\mu_H}^\mu \frac{d\mu'}{\mu'} \gamma_C(\mu')\biggr\} \\
&= \exp\biggl\{ -\frac{\zed_H}{2} K_\Gamma(\mu_H,\mu) - \frac{\zed_H}{2}\eta_\Gamma(\mu_H,\mu)\ln\frac{\mu_H}{Q} + K_{\gamma_C}(\mu_H,\mu)\biggr\}\,, \nn
\end{align}
and $U_H = \abs{U_C}^2$. The pieces $K_\gamma,\eta_\Gamma,K_\gamma$ of the evolution kernel are given by:
\begin{subequations}
\label{Ketadef}
\begin{align}
K_\Gamma(\mu_0,\mu) &= \int_{\mu_0}^\mu \frac{d\mu'}{\mu'} \Gamma_{\text{cusp}}[\as(\mu')]\ln\frac{\mu'}{\mu_0} \\
\eta_\Gamma(\mu_0,\mu) &=  \int_{\mu_0}^\mu \frac{d\mu'}{\mu'} \Gamma_{\text{cusp}}[\as(\mu')] \\
K_\gamma(\mu_0,\mu) &= \int_{\mu_0}^\mu \frac{d\mu'}{\mu'} \gamma[\as(\mu')] \,.
\end{align}
\end{subequations}
Explicit expressions for these kernels up to NNLL accuracy are given in \appx{def-K}.

For the case of the Higgs production, we have another Wilson coefficient ($C_t^2$) obtained from integrating out the top quark. So in addition to the hard function, we also have a running for this coefficient.
\bea
\mu \frac{d}{d\mu} C_t^2(M_t^2,\mu) = \gamma_{C_t^2} C_t^2(M_t^2,\mu)
\eea
The anomalous dimension takes the general form 
\bea
 \gamma_{C_t^2} = \sum_{i=0}^{\infty} \left(\frac{\alpha_s(\mu)}{4\pi}\right)^{i+1} \gamma_i
\eea
where $\gamma_i$ is a number. The RGE has the solution 
\bea
C_t^2(M_t^2,\mu) =  C_t^2(M_t^2,\mu_T)U_{C_t^2}(\mu_T,\mu)
\eea
where the evolution kernel is 
\bea
\label{UTdef}
U_{C_t^2}(\mu_T,\mu)= \text{exp} \Bigg \{\int_{\mu_T}^{\mu}\frac{d\mu'}{\mu'}\gamma_{C_t^2}\Bigg \}= \text{exp} \Bigg \{K_{\gamma_{C_t^2}}(\mu_T, \mu) \Bigg\}
\eea
where 
\be
K_{\gamma_{C_t^2}}(\mu_0,\mu) = \int_{\mu_0}^\mu \frac{d\mu'}{\mu'} \gamma_{C_t^2}[\as(\mu')] 
\ee
Explicit expressions for these kernels up to NNLL accuracy are given in \appx{def-K}.

\subsubsection{Soft function and TMDPDFs}
The soft function in $b$ space obeys the $\mu$- and $\nu$-RGEs,
\begin{align}
\mu\frac{d}{d\mu}\wt S(b;\mu,\nu) = \gamma_\mu^S(\mu,\nu) \wt S(b;\mu,\nu)\,, \  \nu\frac{d}{d\nu}\wt S(b;\mu,\nu) = \gamma_\nu^S (\mu,\nu) \wt S(b;\mu,\nu) 
\end{align}
while the TMDPDFs/beam functions obey
\begin{align}
\mu\frac{d}{d\mu}\wt f_i^\perp(b,x_i,p^\pm;\mu,\nu) &= \gamma_\mu^f(\mu,\nu) \wt f_i^\perp(b,x_i,p^\pm;\mu,\nu)\,, \\
 \nu\frac{d}{d\nu}\wt f_i^\perp(b,x_i,p^\pm;\mu,\nu) &= \gamma_\nu^f (\mu,\nu) \wt f_i^\perp(b,x_i,p^\pm;\mu,\nu)\,.\nn
\end{align}
The $\mu$ anomalous dimensions take the form:
\begin{subequations}
\label{gammamu}
\begin{align}
\label{gammamuS}
\gamma_\mu^S(\mu,\nu) &= -\zed_S \Gamma_{\text{cusp}}[\as(\mu)] \ln\frac{\mu}{\nu} + \gamma_\mu^S[\as(\mu)]\,, \\
\label{gammamuf}
\gamma_\mu^f(\mu,\nu) &= \zed_f \Gamma_{\text{cusp}}[\as(\mu)] \ln\frac{\nu}{p^\pm} + \gamma_\mu^f[\as(\mu)]\,,
\end{align}
\end{subequations}
where $\mu$ and $\nu$ independence of the cross section require $\zed_H = 2\zed_f = -\zed_S = 4$, and $\gamma_H[\as] = -\gamma_\mu^S[\as] - 2\gamma_\mu^f[\as]$. In $\gamma_\mu^f$ we recall the large rapidity scales are given by $p^\pm = Qe^{\pm y} = x_{1,2}\sqrt{s}$  for the two colliding hard partons.
Note $p^+ p^- = Q^2$. As for the form of the $\nu$ anomalous dimensions, at one-loop fixed order in perturbation theory, they take the values
\be
\label{gammanu}
\gamma_\nu^S = \zed_S \frac{\alpha_s(\mu)}{4 \pi}\Gamma_0 \ln \mu b_0\,,\quad \gamma_\nu^f = \zed_f \frac{\alpha_s(\mu)}{4 \pi}\Gamma_0 \ln\mu b_0\,,
\ee
where 
\be
\label{mub}
b_0 = \frac{be^{\gamma_E}}{2}\,,
\ee
the $\mu$-scale at which rapidity logs are minimized in $b$ space. Beyond $\cO(\as)$, the form of the $\nu$ anomalous dimensions can be deduced from the consistency relation:
\be
\frac{d}{d\ln\mu}\gamma_\nu^i(\mu,\nu) = \frac{d}{d\ln\nu} \gamma_\mu^i(\mu,\nu) = \zed_i \Gamma_{\text{cusp}}[\as(\mu)]\,.
\ee
Solving this equation in $\mu$, we obtain
\be
\label{gammanurun}
\gamma_\nu^i (\mu,\nu) = \zed_i\int_{1/b_0}^\mu d\ln\mu' \Gamma_{\text{cusp}}[\as(\mu')] + \gamma_{Ri}[\as(1/b_0)]= \zed_i\eta_\Gamma(1/b_0,\mu) + \gamma_{Ri}[\as(1/b_0)]\,,
\ee
where the boundary condition of the evolution at $1/b_0$ determines the non-cusp part $\gamma_{Ri}[\as]$ of the $\nu$ anomalous dimension. The independence of the cross section \eq{bxsec} on $\nu$ requires, again, $\zed_S = -2\zed_f$, and $\gamma_{RS} = -2\gamma_{Rf}$.

The solutions of the $\mu$ and $\nu$ RGEs for $\wt S$ and $\wt f^\perp$ are:
\begin{subequations}
\begin{align}
\label{Sevolved}
\wt S(b;\mu,\nu) &= \wt S(b;\mu_L,\nu_L)U_S(\mu_L,\mu;\nu)V_S(\nu_L,\nu;\mu_L)  \\
&=  \wt S(b;\mu_L,\nu_L)V_S(\nu_L,\nu;\mu)U_S(\mu_L,\mu;\nu_L) \nn \\
\label{fevolved}
\wt f_i^\perp(b,x_i,p^\pm;\mu,\nu) &= \wt f_i^\perp(b,x_i,p^\pm;\mu_L,\nu_H)U_f(\mu_L,\mu;\nu)V_f(\nu_H,\nu;\mu_L) \\
&=  \wt f_i^\perp(b,x_i,p^\pm;\mu_L,\nu_H)V_f(\nu_H,\nu;\mu)U_f(\mu_L,\mu;\nu_H)\,,  \nn
\end{align}
\end{subequations}
where each pair of equalities accounts for two, equivalent paths for RG evolution in the two-dimensional $\mu,\nu$-space (see \fig{modes}). The evolution kernels $U_{S,f}$ in the $\mu$ direction are:
\begin{subequations}
\label{USUf}
\begin{align}
\label{US}
U_S(\mu_L,\mu;\nu) &= \exp\biggl\{-\zed_S K_\Gamma(\mu_L,\mu) - \zed_S \eta_\Gamma(\mu_L,\mu)\ln \frac{\mu_L}{\nu} + K_{\gamma_S}(\mu_L,\mu)\biggr\} \\
\label{Uf}
U_f(\mu_L,\mu;\nu) &= \exp\biggl\{\zed_f \eta_\Gamma(\mu_L,\mu)\ln\frac{\nu}{p^\pm} + K_{\gamma_f}(\mu_L,\mu)\biggr\} \,.
\end{align}
\end{subequations}
Note that the $\mu$ anomalous dimension for $\wt f^\perp$ in \eq{gammamuf} does not have a log of $\mu$ in its cusp anomalous dimension term, so no $K_\Gamma$ term appears in its evolution kernel $U_f$ in \eq{Uf}. Meanwhile, the $\nu$ evolution kernels $V_{S,f}$ are given by integrals over $\nu$ of \eq{gammanurun},
\begin{subequations}
\label{VSVf}
\begin{align}
\label{VS}
V_S(\nu_L,\nu;\mu) &= \exp\biggl\{\biggl[\zed_S \eta_\Gamma(1/b_0,\mu) + \gamma_{RS}[\as(1/b_0)]\biggr]\ln\frac{\nu}{\nu_L}\biggr\} \\
\label{Vf}
V_f(\nu_H,\nu;\mu) &=\exp\biggl\{\biggl[\zed_f \eta_\Gamma(1/b_0,\mu) + \gamma_{Rf}[\as(1/b_0)]\biggr]\ln\frac{\nu}{\nu_H}\biggr\} \,.
\end{align}
\end{subequations}

\subsubsection{RG evolved cross section}

We can now put these pieces together to express the cross section \eq{bxsec} in terms of the hard, soft, and beam functions evolved from their natural scales where logs in each are minimized, and thus logs in the whole cross section are resummed:
\begin{align} 
\label{sigrun}
\wt\sigma(b,x_1,x_2,Q;\mu_i,\nu_i;\mu,\nu) &=U_{\text{tot}}(\mu_i,\nu_i;\mu,\nu) \, C_t^2(M_t^2, \mu_T) H(Q^2;\mu_H)\wt S(b;\mu_L,\nu_L) \\
&\quad\times \tft_1(b,x_1,p^-;\mu_L,\nu_H) \tft_2(b,x_2,p^+;\mu_L,\nu_H)
\,,\nn 
\end{align}
where
\begin{align}
\label{Utotmunu}
&U_{\text{tot}}(\mu_i,\nu_i;\mu,\nu) \equiv U_{C_t^2}(\mu_T, \mu)U_H(\mu_H, \mu) U_S(\mu_L, \mu; \nu)V_S(\nu_L,\nu;\mu_L)\,U_f^2(\mu_L, \mu; \nu) V_f^2(\nu_H,\nu;\mu_L)\Big]
\nn\\ 
&\quad =
\exp\biggl\{-\zed_H K_\Gamma(\mu_H,\mu) - \zed_S K_\Gamma(\mu_L, \mu) - \zed_H \eta_\Gamma(\mu_H,\mu)\ln\frac{\mu_H}{Q} +K_{\gamma_H}(\mu_H,\mu) + K_{\gamma_{C_t^2}}(\mu_T,\mu)
\nn \\ 
& \qquad\quad +\eta_\Gamma(\mu_L,\mu)\Bigl[-\zed_S \ln\frac{\mu_L}{\nu} + 2\zed_f \ln\frac{\nu}{Q}\Bigr]
+ K_{\gamma_S}(\mu_L,\mu) + 2K_{\gamma_f}(\mu_L,\mu)  \\ 
& \qquad\quad
+ \Bigl[ \zed_S \eta_\Gamma(1/b_0,\mu_L) + \gamma_{RS}[\as(1/b_0)]\Bigr] \ln\frac{\nu}{\nu_L} + 2 \Bigl[ \zed_f \eta_\Gamma(1/b_0,\mu_L) + \gamma_{Rf}[\as(1/b_0)]\Bigr] \ln\frac{\nu}{\nu_H} \biggr\} \nn
. 
\end{align}
Using the relations $\zed_H = -\zed_S = 2\zed_f = 4$, $\gamma_H +\gamma_{C_t^2}= -\gamma_\mu^S - 2\gamma_\mu^f$, and $\gamma_{RS} = -2\gamma_{Rf}$, we obtain the simpler expression,
\begin{align} 
\label{Utot}
U_{\text{tot}}(\mu_i,\nu_i;\mu,\nu) & =
\exp\biggl\{4 K_\Gamma(\mu_L,\mu_H) - 4 \eta_\Gamma(\mu_L,\mu_H) \ln\frac{Q}{\mu_L} - K_{\gamma_H}(\mu_L,\mu_H) - K_{\gamma_{C_t^2}}(\mu_L,\mu_T)
\nn\\ & \qquad
+\Big[ -4\, \eta_\Gamma (1/b_0,\mu_L)+ \gamma_{R\, S}\big[\alpha_s(1/b_0) \big] \Big] \ln \frac{\nu_H}{\nu_L}\biggr\} \,,\end{align}
in which we observe that the explicit dependence on the arbitrary scales $\mu$ and $\nu$ has exactly canceled out, leaving only the dependence on the natural scales $\mu_{H,T,L,b}$ and $\nu_{L,H}$ where the hard, soft, and beam functions live. Note $U_{C_t^2},K_{\gamma_{C_t^2}}$ are present only in the case of the Higgs.

In \eq{Utot}, we envision that the rapidity evolution takes place at (or around) the scale $1/b_0$ (see \fig{modes}). Then we can actually just expand the evolution factor $\eta_\Gamma(1/b_0,\mu_L)$ and the rapidity anomalous dimension $\gamma_{RS}$ in a fixed-order expansion in $\as(\mu_L)$, to the order required for N$^k$LL accuracy. This is in fact what we will do below. Then it becomes useful to split up $U_{\text{tot}}$ in \eq{Utot} into two factors, 
\be
\label{UVsplit}
U_{\text{tot}}(\mu_i,\nu_i;\mu,\nu) = U(\mu_L,\mu_H) V(\nu_L,\nu_H;\mu_L)\,,
\ee
where
\begin{align}
\label{UVdef}
U(\mu_L,\mu_H, \mu_T) &= \exp\biggl\{4 K_\Gamma(\mu_L,\mu_H) - 4 \eta_\Gamma(\mu_L,\mu_H) \ln\frac{Q}{\mu_L} - K_{\gamma_H}(\mu_L,\mu_H)- K_{\gamma_{C_t^2}}(\mu_L,\mu_T)\biggr\} \nn \\
V(\nu_L,\nu_H;\mu_L) &= \exp\biggl\{ \gamma_\nu^S (\mu_L) \ln\frac{\nu_H}{\nu_L}\biggr\} \,,
\end{align}
which are of course just $U_H(\mu_L,\mu_H)U_{C_t^2}(\mu_L,\mu_T)$ and $V_S(\nu_L,\nu_H;\mu_L)$ as given by \eqss{UHdef}{UTdef}{VS}. For brevity in the rest of the paper we will just use $U,V$ in \eq{UVdef}.

Inside $V$ in \eq{UVdef}, we use the fixed-order expansion of $\gamma_\nu^S(\mu_L)$ given in \eq{gammanurun} using the expansions \eq{etaexp2} for $\eta_\Gamma$ and  \eq{gammaRSexp} for $\gamma_{RS}$:
\begin{align}
\label{gammanuexp}
\gamma_\nu^S(\mu_L) &=  \quad \frac{\as(\mu_L)}{4\pi} \quad \biggl[\zed_S \Gamma_0 \ln \mu_L b_0 + \gamma_{RS}^0 \biggr] \\
&\quad +  \Bigl(\frac{\as(\mu_L)}{4\pi}\Bigr)^2 \biggl[ \zed_S \Gamma_0\beta_0 \ln^2 \mu_L b_0 +( \zed_S  \Gamma_1 + 2\gamma_{RS}^0 \beta_0 )\ln \mu_L b_0 + \gamma_{RS}^1\biggr] \nn \\
&\quad + \Bigl(\frac{\as(\mu_L)}{4\pi}\Bigr)^3 \biggl\{ \frac{4}{3}\zed_S \Gamma_0\beta_0^2 \ln^3 \mu_L b_0 + [ \zed_S (\Gamma_0 \beta_1 + 2\Gamma_1\beta_0) + 4\gamma_{RS}^0\beta_0^2]\ln^2 \mu_L b_0 \nn \\
&\qquad\qquad\qquad  + [\zed_S \Gamma_2 + 2\gamma_{RS}^0 \beta_1 + 4\gamma_{RS}^1\beta_0]\ln \mu_L b_0 + \gamma_{RS}^2\biggr\} + \cdots \,, \nn
\end{align}
In practice we truncate this expansion at the appropriate order of logarithmic accuracy. We will always pick $\mu_L$ in such a way that none of these generate large logs (either $\mu_L\sim 1/b_0$ in $b$ space, or in momentum space in such a way that they remain small after inverse transformation---see \ssec{muLscale}), except the factor of $\ln(\nu_H/\nu_L)$ in \eq{UVdef}. This is an observation that will become key below, when we split $\gamma_\nu$ into two separate parts in \ssec{NNLLbeyond}.

\subsubsection{How to choose the scales?}

To evaluate the cross section \eq{sigrun} (and its inverse Fourier transform back to momentum space \eq{xsec}) explicitly, we need to make explicit choices for the scales $\mu_{H,L}$ and $\nu_{L,H}$ between which to run in \eq{Utot}. Choosing these near the scales at which the logs in each individual function are minimized in principle achieves resummation of all large logarithms. However, these natural choices are different in impact parameter and momentum space.

There are various possible ways in which this resummation can be handled. In this paper, we envision, in \eqs{UVsplit}{UVdef} and \fig{modes}, running the hard function in $\mu$ to the natural low scale of the soft and collinear functions, and the soft function in $\nu$ to the natural rapidity scale of the TMDPDFs.
The high scales $\mu_H$ and $\nu_H$ for the running of the hard and soft functions are unambiguously best chosen near the invariant mass $\sim Q$ of the gauge boson. The choices of the low scales $\mu_L$ and $\nu_L$ are under debate since we can choose those scales either in $b$ space or in momentum space.

The $\mu$ scales, like in a usual EFT, are a measure of virtuality of the modes that contribute to that function. For the hard function, this virtuality scale, not surprisingly, is the hard scale $Q$, which also happens to be the scale  choice for which the logarithms in the hard function are minimized.
The virtuality for soft and beam functions is of the order of the transverse momentum that the function contributes to the total transverse momentum. This can be seen in momentum space where the product of these functions in impact parameter space turns into a convolution over transverse momentum, \eq{txsec}.  Since the total transverse momentum is a sum over the transverse momenta contributed by each function,  for a given total $q_T$, the  contribution of any one of these functions traverses a range of scales. While this situation is not unique to this observable, what is different is the dependence of the TMDPDF on the hard scale $Q$.  As we will see, due to this $Q$ dependence, the conjugate natural scale to $b_0$ in the resummed result is no longer  $q_T$ but is shifted away from $q_T$ towards $Q$.

However, the final aim of any resummation is to have a well behaved perturbative series. Whenever the fixed order logs become too large, the expansion in $\alpha_s$ does not converge, and it becomes necessary to reorganize the series in terms of resummed exponents. A successful resummation is then one in which the fixed order terms that are left behind form a rapidly converging series in $\alpha_s$. Since the large logarithms are, in fact, the terms that spoil the convergence of the fixed order perturbative series, the general strategy is then to minimize the effect of these logarithms in the residual fixed order series.

Keeping these issues in mind, we explore two possible sets of scale choices for $\mu_L,\nu_L$ for resummation: the standard choices in impact parameter space in \ssec{CSS}, and a new proposed set of choices in \ssec{mom} allowing evaluation of the resummed cross section in momentum space.

\subsection{Scale choice in impact parameter space}
\label{ssec:CSS}

To choose scales for resummation, we need some idea about the natural scales at which each of the three functions (hard, soft and TMDPDF) live. This is easily seen by looking at their behavior up to one loop. From the results given in \appx{fixedorder}, we find that each of these functions are function of the logs,
\be
H = H\Bigl(\ln\frac\mu Q\Bigr)\,,\quad \wt S = \wt S\Bigl( \ln \mu b_0,\ln \frac{\nu}{\mu}\Bigr)\,,\quad \wt f_\perp = \wt f_\perp\Bigl( \ln \mu b_0,\ln \frac{\nu}{p^\pm}\Bigr)\,,
\ee
given in impact parameter space for $\wt S,\wt f_\perp$. In this space, it is perfectly evident from the fixed-order calculation that the natural scale which minimizes all the logs in $\mu$ for the soft and beam functions is $\mu =\mu_L \sim 1/b_0$. Since the final cross section at a given $q_T$ involves an integral over a range of $b$, the scale choice is in fact spread over a range of  scales. This is to be expected from the earlier discussion of their being no unique physical scale for the soft and beam functions. The natural scales for the various functions then are $\mu=\mu_H$ for the hard function, $(\mu,\nu) = (\mu_L,\nu_L)$ for the soft function and $(\mu,\nu) = (\mu_L,\nu_H)$ for the beam functions, where $\mu_L, \nu_L \sim 1/b_0$ and $\nu_H , \mu_H \sim Q$ (recall $p^+p^- = Q^2$).

All the logs can then be resummed by running the hard function from the scale $Q$ to $1/b_0$ and the soft function in $\nu$ from $Q$ to $1/b_0$. This will produce the result (for the central values, not counting scale variations) of the CSS formalism. Therefore, this scheme resums logarithms of the form $\ln( Q b_0)$. The power counting adopted for this resummation then is straightforward since there is only one type of log. It is usually chosen as $\alpha_s \ln(Q b_0) \sim 1$. Leading log (LL) accuracy  then resums $\alpha_s^n \ln^{n+1}(Q b_0)$, with NLL and NNLL down by one and two powers of the logarithm respectively.

Since the lower scales $\mu_L$ are chosen in $b$ space, the cross section involves an inverse Fourier transform over arbitrarily large values of $b$, so  eventually we hit the nonperturbative scale which manifests itself in the form of the Landau pole: $\alpha_s(1/b_0)$. This corresponds to the fact that the beam and soft functions can contribute arbitrarily small values of transverse momentum even when the total transverse momentum is perturbative. This is usually handled by putting a sharp or smooth cutoff in $b$ space which provides a way to model nonperturbative physics \cite{Qiu:2000hf,Sun:2013dya,DAlesio:2014mrz,Collins:2014jpa,Scimemi:2017etj}. The impact of these nonperturbative effects will be discussed in \sec{non_pert}.

The obvious advantage of this scheme is that the power counting is unambiguous and we can guarantee that with the central values of scale choices in $b$ space, all the logs in the residual fixed order series are set exactly to zero.
As far as the choice of central values is concerned, the terms that are resummed are exactly equal to the CSS resummation formalism \cite{Collins:1984kg}. However, due to the introduction of the new rapidity renormalization scale $\nu$, there is much better control over which terms can be included in the exponent and which terms remain in the fixed order \cite{Chiu:2012ir,Chiu:2011qc}. This directly translates into a much better estimates of error due to missing higher order terms.

\begin{figure}
\centerline{\scalebox{.55}{\includegraphics{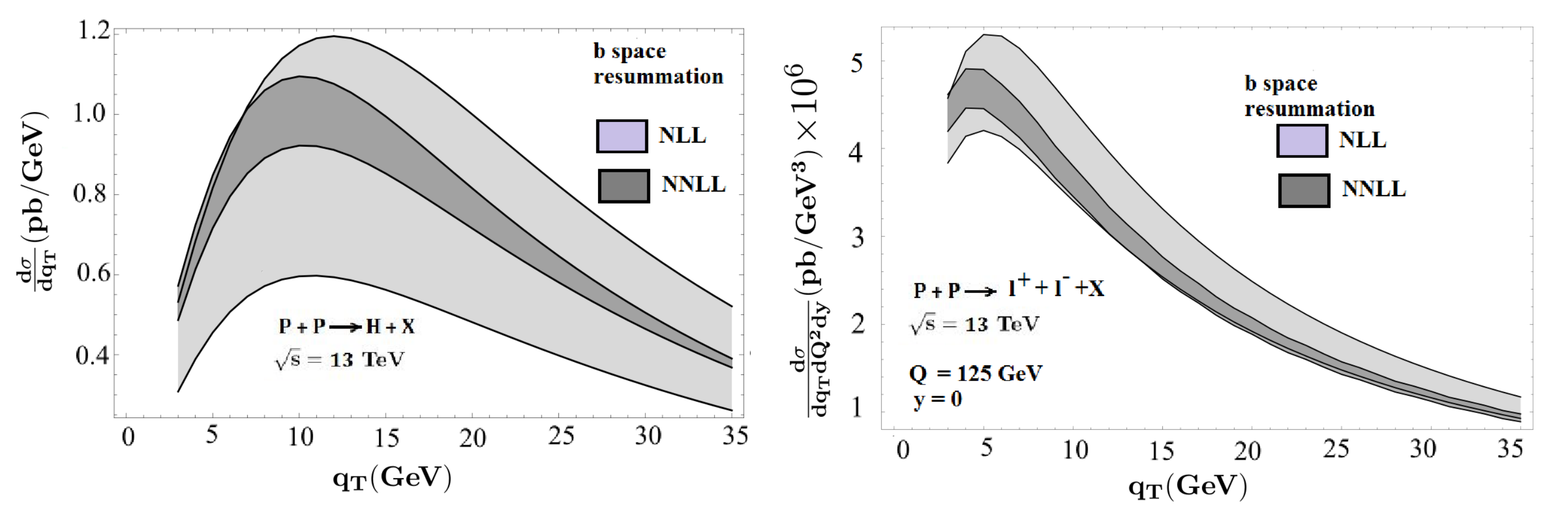}}}
\vskip-0.2cm
\caption[1]{Result of $b$-space resummation for Higgs production and DY, setting the central values for the low scales for resummation at $\mu_L,\nu_L = 1/b_0$ and varying around these by factors of 2 to obtain the uncertainty bands. The plots in momentum space are obtained by an inverse Fourier transform., which hits a Landau pole for large $b$, which must thus be cut off, see \fig{bspace}. The uncertainty for the DY curve is actually thus underestimated due to an ambiguity in this cutoff, see also \fig{bspace}.}
\label{fig:bspaceresum} 
\end{figure}

Another advantage of having control over what exactly goes in the exponent is when we match the resummed cross section to the fixed order cross section at large $q_T$. 
To maintain accuracy over the full (perturbative) range of $q_T$ ($Q \geq q_T \gg  \Lambda_{QCD}$), we need to turn off resummation at the value $q_T$ where the nonsingular contribution is the same order as that of the singular one. Due to the two independent scales $\mu,\nu$ available, this can be done very easily by using profiles in these scales to smoothly turn off the resummation and simultaneously match onto the full (including non-singular pieces) fixed order cross-section. This technique was implemented for the Higgs transverse spectrum in \cite{Neill:2015roa} to obtain the cross section to NNLL+NNLO accuracy. In this paper, for the purposes of comparison with other resummation schemes, we present the results for the cross section at NNLL accuracy for both the Higgs and DY using this scheme (\fig{bspaceresum}).

\begin{figure}
\centerline{\scalebox{.55}{\includegraphics{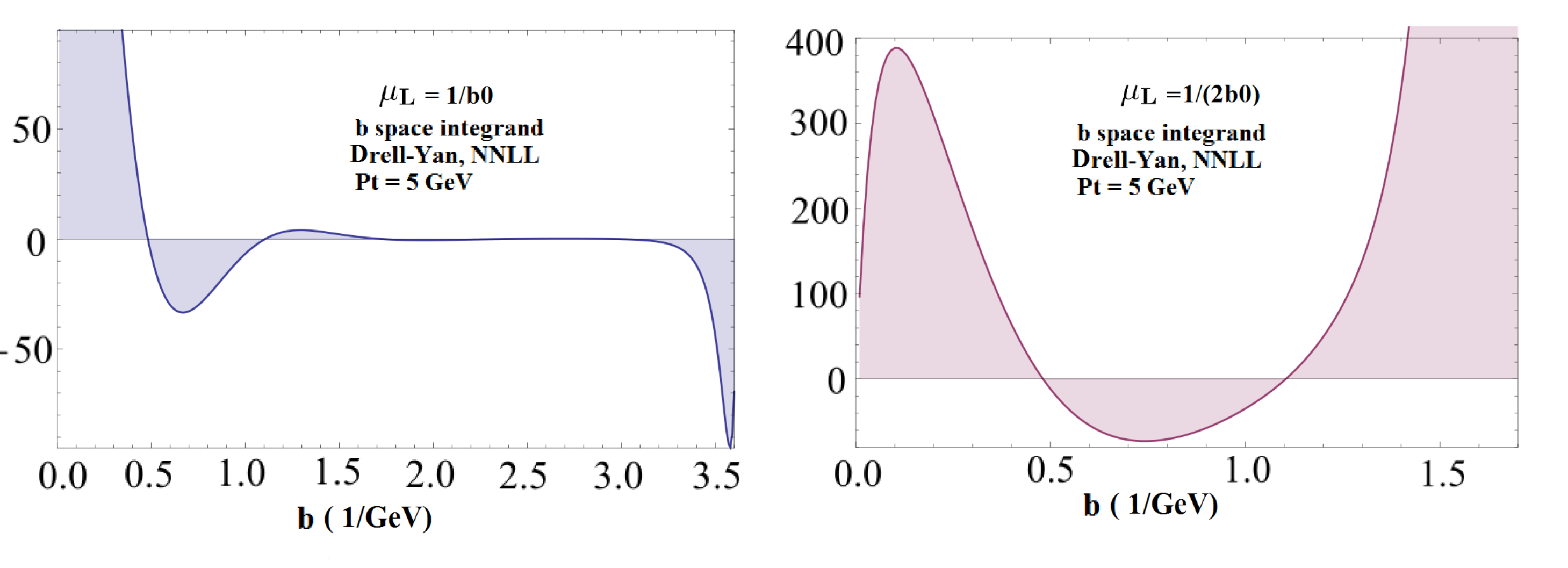}}}
\vskip-0.4cm
\caption[1]{$b$-space integrand $K(b)$ given in \eq{inverseFT} for $b$-space resummation. On the left is the result of the central value $\mu_L=1/b_0$ for the low virtuality scale, for which a somewhat stable plateau exists before hitting the Landau pole above $b\sim 3\text{ GeV}$, allowing imposition of a cutoff to which the result is insensitive. On the right is shown the result of varying this scale down by a factor of 2, in which case the pole moves to lower $b$, and no stable region for a reasonable cutoff exists.}
\label{fig:bspace} 
\end{figure}

After having decided on the central values, we next need to estimate the size of higher-order perturbative corrections we have missed by using scale variations. This is accomplished by varying the two renormalization scales $\mu $ and $\nu$ independently as detailed in \cite{Neill:2015roa}. 
Since we resummed and chose scales in $b$ space, our final result involves an inverse Fourier transform over the resummed $b$ space result:
\begin{align}
\label{inverseFT}
\frac{d\sigma}{dq_T^2 dy} &= \frac{\sigma_0}{2} \int db\, bJ_0(bq_T)  U\Bigl(\mu_H \!\sim\! i Q, \mu_T \!\sim\! M_t, \mu_L \!\sim\! \frac{1}{b_0}\Bigr)  V\Bigl(\nu_L \!\sim \!\frac{1}{b_0},\nu \!\sim\! Q;\mu_L\! \sim\! \frac{1}{b_0}\Bigr)\nn\\
&\qquad \times f(x_1,\mu_L\sim1/b_0)f(x_2,\mu_L\sim 1/b_0) \equiv \frac{\sigma_0}{2} \int db\,   K(b)
\end{align}
where $K(b)$ is the complete $b$-space integrand. It turns out the cusp anomalous dimension for DY ($\Gamma_0 = 4C_F$) is much lower in magnitude than that for Higgs ($\Gamma_0 = 4C_A$). This results in a much lower damping effect at large values of $b$, see \fig{bspace}. The plot shows the $b$ space integrand $K(b)$ for $q_T= 5$ GeV.

The divergence beyond $b= 3$ in the left-hand plot in \fig{bspace} is the Landau pole. It is clear that there is only a narrow window of stability before we hit  the Landau pole. The situation worsens when try and do a scale variation about the central choice $ \mu_L \sim 1/b_0 $, specifically $\mu \sim \mu_L/2$, as shown in the right-hand plot of \fig{bspace}. What this does is to bring the Landau pole closer by factor of 2,  in which case there is no clear separation between the perturbative and nonperturbative regimes. It is then unclear how a hard cut-off or even a smooth one would give an accurate estimation of the perturbative uncertainty band.
The error bands in \fig{bspaceresum} for the case of DY at NNLL, therefore cannot be generated meaningfully via this scale variation. For the purposes of error estimation then we were forced to make an educated guess for the upper boundary of the error band at NNLL. Clearly, we would like to find a procedure that does better.

\subsection{Resummation in momentum space} \label{ssec:mom}

The Landau pole in the $b$-space resummation scheme above comes about due to the running of the strong coupling $\alpha_s$ all the way down to $1/b_0$ which goes all the way down to $\Lqcd$.
A natural question to ask, then, is if we can avoid the Landau pole by choosing the $\mu$ scale in momentum space. It is clear from the above discussion that we cannot choose a single scale in momentum space which will put all the logs to 0. However, what we can ask is whether it is possible to make an appropriate choice for $\mu,\nu$ directly in momentum space such that the resummed exponent, on an average, minimizes the contribution from the residual fixed order logs. These small logs, though nonzero, can then be included order by order ensuring that they only contribute to the same order as the error band.

\subsubsection{Leading logs}

As before let us assume the power counting $ \alpha_s \ln( \mu_H/\mu_L)$ , $\alpha_s \ln( \nu_H/\nu_L) \sim 1$, where $\mu_L,\nu_L $ will be our choices of the renormalization scales in momentum space.
According to this power counting then, at leading log (LL) we wish to resum terms of the form $\as^n \ln^{n+1}(\frac{\mu_H}{\mu_L})$ in the exponent of the cross section. We also assume that the residual logs of the form $\ln( \mu_L b_0)$ are small (of $\cO(1)$) and need not be resummed. The resummation then involves running just the hard function from $\mu_H \sim Q$ to $\mu_L $. All the residual fixed order logs as well as logs in $\nu$ are then subleading at this order.
The cross section we get is
\bea
\frac{d \sigma}{d q_T^2 dy} =  \frac{(2\pi)\sigma_0}{2} U^{\rm LL}(\mu_H, \mu_T, \mu_L)  \delta^2(\vec{q_T})f(x_1,q_T)f(x_2,q_T)
\label{LLc}
\eea
where we can obtain $U_H^{\rm LL}$ from \eq{UHdef}.
This is highly singular and gives a trivial result at nonzero $q_T$.  This suggests that the supposedly higher order pieces that we are ignoring are not unimportant. Let's look at the fixed order pieces left over at one loop:
\be
\label{ffS}
(2\pi)^3 \tilde f^{\perp}_1 \tilde f^{\perp}_2 \tilde S =  1+  \zed_S \Gamma_0\frac{\as(\mu_L)}{8\pi}\Bigl( 2 \ln\frac{Q}{\mu_L}\ln(\mu_L b_0) + \ln^2(\mu_L b_0) \Bigr) + O(\alpha_s \ln(\mu_L b_0), \alpha_s^2)
\ee
The biggest term here appears to be $\ln(Q/\mu_L)\ln(\mu_L b_0)$. According to our naive power counting, this term is subleading at LL and hence should be ignored.

 Going to momentum space, this term gives us $\ln(\frac{Q}{\mu_L})/q_T$, which is in fact the leading logarithmic term in the fixed order cross section at nonzero $q_T$. So then it appears that we haven't resummed any of the logs which contribute at nonzero $q_T$, which makes sense since we get a trivial result at nonzero $q_T$. We can then conclude that the LL cross section in Eq.~(\ref{LLc}) with this power counting only gives us the result at zero $q_T$. We will have to go to NLL to have any handles to ``fix'' it. 
 
\subsubsection{Next-to-leading logs}
\label{ssec:NLL}

The next logical step is to go to NLL. We first update the hard function to resum all logs of the form $\alpha_s^n \ln^n(Q/\mu_L)$. However this by itself is not useful since it will lead to the same problem as for the LL case in that it will only contribute at zero $q_T$. But power counting demands that we also run the soft function in $\nu$ from the scale $\nu_H \sim Q$ to $\nu_L \sim \mu_L$.  Using \eq{gammanurun}, 
\bea \label{gamS}
\gamma_{\nu}^S (\mu_L)= -4\Gamma_0\frac{\as(\mu_L)}{4\pi}\ln(\mu_L b_0)
\left[1+ \frac{\as(\mu_L)}{4\pi}  \beta_0 \ln(\mu_L b_0) + O\left(\as^2\ln^2(\mu_L b_0)\right) \right ]
\eea
We can then use the leading term of this anomalous dimension to resum the soft function:
\bea
\label{softNLL}
V^{\rm NLL}(\nu_H,\nu_L;\mu_L) &&= \exp\left[ -\Gamma_0 \frac{\alpha(\mu_L)}{\pi}\ln\left(\frac{\nu_H}{\nu_L}\right)\ln(\mu_L b_0) \right] \\
 \frac{d \sigma}{d q_T^2 dy} &&=\frac{\sigma_0}{2}  U^{\rm NLL}(\mu_H, \mu_T, \mu_L)\int db\, b J_0(b q_T) \,V^{\rm NLL}(\nu_H,\nu_L;\mu_L)f(x_1,\mu_L)f(x_2,\mu_L) \nn\\
&&= \sigma_0 U^{\rm NLL}(\mu_H, \mu_T, \mu_L) e^{-2 \omega_s \gamma_E} \frac{ \Gamma[1- \omega_s]}{\Gamma[\omega_s]} \frac{1}{\mu_L^2} \left(\frac{\mu_L^2}{q_T^2}\right)^{1-\omega_s}f(x_1,\mu_L)f(x_2,\mu_L) \nn
\eea
where $\omega_s= \Gamma_0\frac{\as}{2\pi}\ln\left(\frac{\nu_H}{\nu_L}\right)$ .  This result works for $q_T >0 $.  Clearly then we still have a singularity in the cross section at $\omega_s = 1$. As was noted in earlier papers \cite{Chiu:2012ir}, this is a divergent series. The reason for this is that the logarithms in $b$ space of the form $\ln^n( \mu b_0)$ do not translate directly to logarithms of $\ln^n(\mu q_T)$. The simplest example of this is the inverse Fourier transform of $\ln(\mu b_0)$. This gives us a term proportional to the plus distribution function $ \cL_0 =\tfrac{1}{\mu^2}\left[\mu ^2/q_T^2\right]_+$ as defined in \cite{Chiu:2012ir}. For nonzero values of $q_T$, this is simply $1/q_T^2$. The same thing happens for higher powers of logarithms. For example $\ln^2(\mu b_0)$ also gives a term proportional to   $\cL_0$  along with terms which go like $\ln(\mu /q_T)$. So an all order summation of the logarithms in $b$ space eventually adds up all of these pieces (whose coefficient is controlled by $\omega_s$) which leads to a divergence for sufficiently large $\omega_s$. 

So this step softens the singularity somewhat  moving it away from $q_T =0$, so that we at least have a nonzero cross section for nonzero $q_T$. However, the result is still singular. The singularity results from the single logarithm of $\ln(\mu_L b_0)$ in the exponent which diverges at very small values of $b$. We would have expected that the integrand would receive its dominant contribution from the region $b \sim 1/q_T$. Instead, as it was noted in \cite{Chiu:2012ir} it is the UV region $b\sim 1/Q$ which appears to dominate the integral.

Since our resummation kernel in $b$ space is unstable, we cannot yet assign a power counting to the residual logs. This is because the fixed order logs (of the form $\as^n \ln^m(\mu_L b_0)$ are weighted by this exponent in the inverse Fourier transform. So before we can talk of power counting for these logs, we must stabilize this kernel such that the region $b \sim 1/q_T$ dominates.
 Clearly, to cure this singularity (which is a UV singularity) we need an even powered logarithm $\ln^{2m}(\mu b_0)$ with a negative coefficient in the soft resummation exponent.  
 So we move ahead and attempt to see if we can resum the next biggest term (which would technically be subleading at this order) $\sim\as \ln^2(\mu_L b_0)$.  There are two places we can find such a double logarithm, one from the second term in \eq{gamS}, which is part of the 2-loop rapidity anomalous dimension, and another from the fixed-order soft function at 1-loop, see \eq{Sn}, or \eq{ffS}, which includes the beam function contribution. Since the term in \eq{gamS} is already part of the RRG exponent, including it just means tacking on a subleading term in the exponent, which we are free to do at NLL order. As for the fixed-order term in \eq{Sn}, while standard resummation schemes tell us to truncate the logs in the soft function $\wt S(b,\mu_L,\nu_L)$ at fixed order, these logs in the soft function, though not large, do actually exponentiate, see \eq{Sevolved2}.  The higher order logs are subleading at NLL, but again, we are free to include them, either in fixed-order or exponentiated form.

Specifically, together with standard soft exponent at NLL, the terms we want to put in the total soft exponent are:
\bea
 &&\ln V^\text{NLL}(\nu_H,\nu_L; \mu_L)
  + \ln^2(\mu_L b_0)\left[ -\Gamma_0\frac{\as}{2\pi}-2\Gamma_0\frac{\as^2}{2\pi} \frac{\beta_0}{4\pi} \ln\left(\frac{\nu_H}{\nu_L}\right)\right]   
\nn \\
 &&=-2\Gamma_0\frac{\as}{2\pi}
 \left[\ln\left(\frac{\nu_H}{\nu_L}\right)\ln(\mu_L b_0)+\ln^2(\mu_L b_0)\left(\frac{1}{2}+ \as \frac{\beta_0}{4\pi} \ln\left(\frac{\nu_H}{\nu_L}\right)\right) \right]\,.
\label{Ustable}
\eea
The typical default choice for the low rapidity scale is $\nu_L = \mu_L$. We imagine making a choice near this scale. But we will rescale it so that the extra $\ln^2(\mu_L b_0)$ terms in \eq{Ustable} get automatically included in the standard soft NLL exponent. 

We now attempt to reproduce the above terms in the pure NLL soft exponent, with a modified scale choice $\nu_L^*$: 
\be
\label{VSNLLstar}
V_S^{\text{NLL}}(\nu_H,\nu_L^*;\mu_L)  =\exp\biggl\{ - \frac{\as(\mu_L)}{4\pi}4\Gamma_0 \ln(\mu_L b_0) \ln\frac{\nu_H}{\nu_L^*}\biggr\}\,.
\ee
This can be done with the scale choice:
\be\label{nuLstar}
  \nu_L^* = \nu_L (\mu_L b_0)^{-1+p} 
\ee 
The value of $p$ that allows us to obtain the double log terms in \eq{Ustable} is determined by comparing them with the exponent in \eq{VSNLLstar}: 
\bea \label{VSNLL}
\ln V_S^\text{NLL}(\nu_H,\nu_L^*;\mu_L) &= & \gamma_\nu^{S (0)} \ln\left(\frac{\nu_H}{\nu_L^*}\right)
\nn\\
&=& -4\Gamma_0 \frac{\as(\mu_L)}{4\pi}\ln(\mu_L b_0) \left[
\ln\left(\frac{\nu_H}{\nu_L}\right)+(1-p)\ln(\mu_L b_0)
\right]
\eea
By comparing above equation to \eq{Ustable} we find
\bea
p&=& 
\frac{1}{2}\left[1-  \frac{\as(\mu_L) \beta_0}{2\pi}\ln\left(\frac{\nu_H}{\nu_L}\right)\right]
\label{eq:n}
\eea
This ensures that we have now resummed all the terms of the form $\as \ln^2(\mu_L b_0)$. (We assume one did not choose the default $\nu_L$ scale on the right-hand side of \eq{nuLstar} with nontrivial $b$ dependence.)
Since $\Gamma_0>0$, the double log now provides the necessary stability to the exponential kernel in $b$ space (at both large and small values of $b$). With this term in place, we can now talk of a systematic power counting for the fixed order logs, which, hitherto, was not meaningful.

So we now  adopt the usual power counting that $\ln(\mu_L b_0) \sim 1$, i.e., this log is small. We still need to confirm numerically, that the fixed order logs that remain ($\cO(\as\ln(\mu_L b_0))$, $\cO(\as^2 \ln^3(\mu_L b_0))$, $\cO(\as^2 \ln^2(\mu_L b_0))$, etc.), when integrated against our exponent in $b$ space, are actually small so that our series is well behaved perturbatively. With this power counting in place, our result for the resummation at NLL then looks like 
\bea
\label{nustarcs}
\frac{d \sigma}{d q_T^2 dy} =\frac{\sigma_0}{2}\,  U^{\text{NLL}}(\mu_H, \mu_T, \mu_L) \int db \, b J_0(b q_T)\, V^{\text{NLL}}(\nu_H,\nu_L^*;\mu_L)f(x_1,\mu_L)f(x_2,\mu_L)
\eea
At NLL, all fixed order logs are subleading and hence not included. 

We will find below that not only does the quadratic $\ln^2\mu_L b_0$ term in the exponent of the $b$ integrand in \eq{nustarcs} introduced by the scale choice \eq{nuLstar} make the integral converge for both small and large $b$, it actually makes it integrable analytically (after a very good numerical approximation for the Bessel function). We will describe this in detail in \sec{analytic}. First, we explore how to generalize the above-described procedure beyond NLL.

\subsubsection{NNLL and beyond}
\label{ssec:NNLLbeyond}

At NNLL and higher order, we have some freedom in choosing how to update the accuracy of the rapidity evolution kernel in \eq{UVdef}. In this subsection, we will use this freedom to look for a way to preserve both the stable power counting of fixed-order logs of $\mu_L b_0$ after integrating over $b$ as well as our ability, in \sec{analytic}, to evaluate that $b$ integral (semi-)analytically.

A simple, standard, choice would simply be to include the next order terms in the rapidity anomalous dimension \eq{gammanuexp}, \eg the $\cO(\as^2)$ terms at NNLL, while keeping the scale choice $\nu_L^*$ in \eq{nuLstar} for the soft rapidity scale. There are two somewhat undesirable consequences of this choice, however. First, as we noted at the end of \ssec{NLL}, an exponent in the $b$ integrand that is quadratic in $\ln\mu_L b_0$ will make the integral analytically computable, to a very good approximation to be described in \sec{analytic}---but not higher powers of $\ln\mu_L b_0$, which will begin to enter starting at N$^3$LL order in \eq{gammanuexp}. Second, starting at NNLL order, maintaining the scale choice $\nu_L^*$ would put some terms into the exponent twice, namely, the $\zed_S \Gamma_0\beta_0$ term in $\cO(\as^2)$, once from the shift from $\nu_L$ to $\nu_L^*$ \eq{VSNLL} in the $\cO(\as)$ term of the exponent, and once from updating the anomalous dimension appearing in \eq{UVdef} with the two-loop value in \eq{gammanuexp}. Of course, this is compensated by the shift from $\nu_L$ to $\nu_L^*$ in the fixed-order soft function.  From \eq{Sn}:
\begin{align}
\label{Snustar}
\wt S^{*(1)}(\mu_L,\nu_L^*) &= \zed_S \frac{\Gamma_0}{2}\Bigl(\ln^2 \mu_L b_0 + 2\ln \mu_L b_0 \ln\frac{\nu_L^*}{\mu_L}\Bigr) + c_{\wt S}^1 \\
&=  \zed_S \frac{\Gamma_0}{2}\Bigl[\ln^2 \mu_L b_0 + 2\ln \mu_L b_0 \Bigl( \ln\!\frac{\nu_L}{\mu_L} - \frac{1}{2} \ln\mu_L b_0 - \frac{\as(\mu_L)\beta_0}{4\pi} \ln\frac{\nu_H}{\nu_L}\ln \mu_L b_0\Bigr)\!\Bigr] + c_{\wt S}^1 \nn \\
&= \zed_S \frac{\Gamma_0}{2} \Bigl( 2\ln \mu_L b_0\ln\frac{\nu_L}{\mu_L} - \frac{\as(\mu_L)\beta_0}{2\pi}\ln\! \frac{\nu_H}{\nu_L} \ln^2\mu_L b_0\Bigr) + c_{\wt S}^1\,,\nn
\end{align}
where we see that the one-loop double log of $\mu_L b_0$ has canceled---the scale choice $\nu_L^*$ has promoted this term to the exponent in \eqs{UVdef}{gammanuexp}. The added $\cO(\as^2)$ term subtracts off (at fixed order) the corresponding $\Gamma_0\beta_0$ term that was double-added to the exponent at NNLL. While this is acceptable as far as power counting goes, it seems awkward to have this term double-counted in the exponent by itself. Similar observations apply to additional terms at higher orders.

There are a number of ways to avoid these problems, while maintaining the desirable quadratic $\ln^2\mu_Lb_0$ terms in the exponent of the $b$ integrand of the cross section. One possibility would be just to revert back to the ordinary scale choice $\nu_L$ from $\nu_L^*$ beyond NLL, but for meaningful comparisons between orders of accuracy, we should maintain the same scale choice as we increase accuracy. Moreover this solution would not prevent higher than quadratic power terms in $\ln\mu_L b_0$ from entering the exponent, which will spoil our analytic integration below. Since the choice $\nu_L^*$ is needed at NLL to stabilize the $b$ integral in \eq{nustarcs} and restore a meaningful resummed power counting, we will go ahead and keep it beyond NLL as well. We will then need a prescription for how to update the rapidity evolution kernel in \eq{VSNLL} that respects the stabilization of the $b$ integral at NLL, without introducing unwanted double counting of terms or higher powers of $\ln^{n>2}\mu_L b_0$ as discussed in the previous paragraph. 

Another possibility, then,  is to just keep the NLL part of the exponent in the rapidity evolution kernel \eq{VSNLLstar}, and expand out the NNLL and higher-order parts in $\as$, \ie 
\bea
V_S(\nu_H,\nu_L^*;\mu_L)&=&e^{\gamma_\nu^{S(0)} \ln \frac{\nu_H}{\nu_L^*}}e^{(\gamma_\nu^{S(1)}+\gamma_\nu^{S(2)}+\cdots) \ln \frac{\nu_H}{\nu_L^*}}
\nn \\
&=&e^{\gamma_\nu^{S(0)} \ln \frac{\nu_H}{\nu_L^*}} \left[1+ \gamma_\nu^{S(1)} \ln \frac{\nu_H}{\nu_L^*}  +\cdots \right]
\label{VStruncated}
\eea
At NNLL we would keep the second term of order $\as$ in the bracket, and at N$^3$LL two more terms of order $\as^2$ would be included, etc. This is not insensible, as the rapidity kernel in \eq{UVdef} contains only a single large log ($\ln \nu_H/\nu_L$) multiplying the whole $\nu$-anomalous dimension. Thus, while the NLL terms are all order $\as \times 1/\as\sim 1$ in log counting and should be exponentiated, the NNLL $\cO(\as^2)$ part of the $\nu$-anomalous dimension and higher-order terms are all truly suppressed by additional powers of $\as$. This is in contrast to the $\mu$ evolution kernels, \eg \eqs{KGammaexp}{etaexp}, where there are infinite towers of terms of the same order in log counting, because terms at higher powers in $\as$ are multiplied by large logs of $\mu/\mu_0$. This is not the case in \eq{gammanuexp}, since higher order terms in $\as$ generated by $\mu$ running do not come with large logs---we are doing the rapidity evolution at a scale $\mu_L\sim 1/b_0$, generating only small logs of $\mu_L b_0$. All the effects of $\mu$ running between widely separated scales and their associated large logs are contained in $U$ in \eq{UVdef}. 

However, we do not need to be so draconian in truncating the terms we could resum using the RRG kernel in \eq{UVdef}. The terms in \eq{VStruncated} that we either exponentiate or truncate at fixed order are basically the same order as terms in the fixed-order expansion of the soft function $\wt S(b,\mu_L,\nu_L^*)$ given by \eqs{Sfixedorder}{Sn} that are kept at each order of logarithmic accuracy, so there is a freedom in choosing which terms in the rapidity anomalous dimension \eq{gammanuexp} and the soft function \eq{Sfixedorder} we will exponentiate or leave in a fixed-order expansion, to obtain desirable behavior of the $b$-space integrand in \eq{xsec}.

Let us then use this freedom to divide the terms in the rapidity anomalous dimension \eq{gammanuexp} into two parts, those that we will exponentiate and those that we will expand out in fixed order. Namely, we will exponentiate all the pure $\Gamma_n$ and $\gamma_{RS}^n$ anomalous dimension terms; these are at most single logarithmic in $\mu_L b_0$, and the shift from $\nu_L$ to $\nu_L^*$ introduces the double logs  $\frac{1}{2}\zed_S\Gamma_n \ln^2\mu_L b_0$ in the fixed-order soft function $\wt S$ associated with $\Gamma_n$, see \eq{Sn}; as well as (part of) the $\sim \zed_S\Gamma_n \beta_0\ln^2\mu_L b_0$ term in the rapidity anomalous dimension \eq{gammanuexp}, which stabilize the $b$-space integrand. The remaining terms will be expanded out in $\as$, and these are all associated with all the beta function terms coming from $\as$ running of the ``pure'' $\Gamma_n$ and $\gamma_{RS}^n$ terms. Concretely, we split the rapidity evolution kernel in \eq{UVdef} into:
\be
\label{Vsplit}
V(\nu_L,\nu_H;\mu_L) = V_\Gamma(\nu_L,\nu_H;\mu_L) V_\beta(\nu_L,\nu_H;\mu_L)\,,
\ee
where
\be
\label{VGamma}
V_\Gamma (\nu_L,\nu_H;\mu_L) = \exp\biggl\{ \ln\frac{\nu_H}{\nu_L} \sum_{n=0}^\infty \Bigl(\frac{\as(\mu_L)}{4\pi}\Bigr)^{n+1}(\zed_S \Gamma_n \ln\mu_L b_0 + \gamma_{RS}^n)\biggr\}\,,
\ee
which remains exponentiated and contains all the ``pure'' anomalous dimension terms, and
\begin{align}
\label{Vbeta}
V_\beta (\nu_L,\nu_H;\mu_L) &= 1 + \Bigl(\frac{\as(\mu_L)}{4\pi}\Bigr)^2 \Bigl( \zed_S \Gamma_0\beta_0 \ln^2\mu_L b_0 + 2\gamma_{RS}^0\beta_0 \ln \mu_L b_0\Bigr) \ln\frac{\nu_H}{\nu_L} + \cdots\,,
\end{align}
which is the fixed-order expansion of the part of the rapidity evolution kernel \eq{UVdef} coming from all the remaining ($\beta_n$) terms in \eq{gammanuexp} that are not included in \eq{VGamma}. This division \eq{Vsplit} ensures that the exponentiated part of the rapidity evolution kernel contains at most double logs of $\mu_L b_0$ after the shift from $\nu_L$ to $\nu_L^*$, and also avoids double counting of the $\beta_0$ induced terms in \eq{gammanuexp} in the exponent as described above.

We can give a more formal definition of the two factors $V_\Gamma$ and $V_\beta$ in \eqs{VGamma}{Vbeta}. Since they are built out of pieces of the anomalous dimension $\gamma_\nu^S(\mu_L)$ in \eq{gammanuexp}, we go back to its all orders expression given by \eq{gammanurun}:
\begin{align}
\gamma_\nu^{S}(\mu_L) = \zed_S \eta_\Gamma(1/b_0,\mu_L) + \gamma_{RS}[\as(1/b_0)]\,,
\end{align}
and divide up each term into pieces containing just the ``pure'' anomalous dimension coefficients and those generated by beta function terms. For $\gamma_{RS}$ this is easy:
\begin{align}
\label{gammaRSsplit}
\gamma_{RS}[\as(1/b_0)] &= \sum_{n=0}^\infty \Bigl(\frac{\as(1/b_0)}{4\pi}\Bigr)^{n+1} \gamma_{RS}^n \\
&= \sum_{n=0}^\infty \Bigl(\frac{\as(\mu_L)}{4\pi}\Bigr)^{n+1} \gamma_{RS}^n + \sum_{n=0}^\infty \Bigl(\frac{\as(\mu_L)}{4\pi}\Bigr)^{n+1}\biggl(\frac{1}{r^{n+1}} - 1\biggr) \gamma_{RS}^n \nn \\
&\equiv \gamma_{RS}^{\text{conf.}}(\mu_L) + \Delta\gamma_{RS}(\mu_L)\,, \nn
\end{align}
where 
\be
r\equiv \frac{\as(\mu_L)}{\as(1/b_0)}\,.
\ee
The first piece in the last line of \eq{gammaRSsplit} contains those terms in the anomalous dimension that would survive in the conformal limit of QCD, where $\as$ does not run. The second set of terms, in $\Delta\gamma_{RS}$ contain all the beta function induced terms. For example, the ratio $1/r$ has the fixed order expansion up to $\cO(\as^2)$:
\be
\frac{1}{r} = 1 + \frac{\as(\mu_L)}{2\pi}\beta_0 \ln \mu_L b_0 + \frac{\as(\mu_L)^2}{8\pi^2} \Bigl( \beta_1 \ln\mu_L b_0 + 2\beta_0^2 \ln^2\mu_L b_0\Bigr) + \cdots\,.
\ee
and the ``1'' term is subtracted off in \eq{gammaRSsplit}, leaving just the $\beta_i$ terms. The similar thing happens for $1/r^{n+1}$.

We can similarly split up $\eta_\Gamma$ into two pieces. To all orders in $\as$, $\eta_\Gamma(1/b_0,\mu_L)$ is given by \eq{Keta}, and is expanded out in fixed orders in \eq{etaexp2}. We want to split up $\eta_\Gamma$ into the ``pure'' $\Gamma_n \ln\mu_L b_0$ pieces along the diagonal of \eq{etaexp2}, and the remaining $\beta_i$ induced terms. We do this by very straightforwardly defining:
\be
\label{etasplit}
\eta_\Gamma(1/b_0,\mu_L) = \eta_\Gamma^{\text{conf.}}(1/b_0,\mu_L) + \Delta\eta_\Gamma(1/b_0,\mu_L)\,,
\ee
where
\be
\eta_\Gamma^{\text{conf.}}(1/b_0,\mu_L) \equiv \int_{1/b_0}^{\mu_L} \frac{d\mu}{\mu} \sum_{n=0}^\infty \Bigl(\frac{\as(\mu_L)}{4\pi}\Bigr)^{n+1}\Gamma_n = \sum_{n=0}^\infty \Bigl(\frac{\as(\mu_L)}{4\pi}\Bigr)^{n+1}\Gamma_n \ln\mu_L b_0\,
\ee
contains the ``pure'' anomalous dimension terms in $\eta_\Gamma$ (the ones which would survive in the conformal limit), and $\Delta\eta_\Gamma$ is given simply by
\be
\Delta\eta_\Gamma(1/b_0,\mu_L) = \eta_\Gamma(1/b_0,\mu_L) - \eta_\Gamma^{\text{conf.}}(1/b_0,\mu_L)\,.
\ee
We will keep $\eta_\Gamma^{\text{conf.}}$ exponentiated part of \eq{Vsplit}, while $\Delta\eta_\Gamma$ will go into the part expanded in fixed orders in $\as$. The explicit expansion for $\Delta\eta_\Gamma$ is of course given by \eq{etaexp2} with the diagonal terms deleted, or can be worked out to all orders in $\as$ (up to NNLL accuracy) from the expression in \eq{Keta}. The corresponding expression for $\Delta\eta_\Gamma$ up to terms of NNLL accuracy is then:
\begin{align}
\label{Deltaetaexp}
\Delta\eta_\Gamma(1/b_0, \mu_L) &=
 - \frac{\Gamma_0}{2\beta_0}\, \biggl[ \ln r + \frac{\as(\mu_L)}{2\pi} \beta_0 \ln \mu_L b_0 \\ 
 &\qquad\qquad + \frac{\as(\mu_L)}{4\pi}\, \biggl(\frac{\Gamma_1 }{\Gamma_0 }
 \minus \frac{\beta_1}{\beta_0}\biggr)\Bigl(1- \frac{1}{r}\Bigr) + \Bigl(\frac{\as(\mu_L)}{4\pi}\Bigr)^2 \frac{\Gamma_1}{\Gamma_0}2\beta_0\ln \mu_L b_0 \nn \\
 &\qquad \qquad + \frac{\as^2(\mu_L)}{16\pi^2} \biggl(
    \frac{\Gamma_2 }{\Gamma_0 } \minus \frac{\beta_1\Gamma_1 }{\beta_0 \Gamma_0 }
      + \frac{\beta_1^2}{\beta_0^2} -\frac{\beta_2}{\beta_0} \biggr) \frac{1 - \frac{1}{r^2}}{2} + \Bigl(\frac{\as(\mu_L)}{4\pi}\Bigr)^3 \frac{\Gamma_2}{\Gamma_0}2\beta_0\ln \mu_L b_0     \biggr] \,, \nn 
\end{align}
where we notice the subtraction terms at the end of each line just modify the pure anomalous dimension $\Gamma_i/\Gamma_0$ pieces, as designed. Note the following properties of the expanded functions of $r = \as(\mu_L)/\as(1/b_0)$ that appear in each line of \eq{Deltaetaexp}:
\begin{subequations}
\begin{align}
\ln r + \frac{\as(\mu_L)}{2\pi} \beta_0 \ln \mu_L b_0 = - \frac{\as(\mu_L)^2}{8\pi^2} (\beta_1 \ln \mu_L b_0 + \beta_0^2 \ln^2\mu_L b_0) + \cdots \\
1-\frac{1}{r} + \frac{\as(\mu_L)}{2\pi} \beta_0 \ln \mu_L b_0 = - \frac{\as(\mu_L)^2}{8\pi^2} (\beta_1 \ln \mu_L b_0 + 2 \beta_0^2 \ln^2\mu_L b_0) + \cdots \\
\frac{1}{2}\Bigl(1-\frac{1}{r^2}\Bigr) + \frac{\as(\mu_L)}{2\pi} \beta_0 \ln \mu_L b_0 = - \frac{\as(\mu_L)^2}{8\pi^2} (\beta_1 \ln \mu_L b_0 + 3 \beta_0^2 \ln^2\mu_L b_0) + \cdots 
\end{align}
\end{subequations}
etc. So the remaining terms in the expansion of $\Delta\eta_\Gamma$ in \eq{Deltaetaexp} contain only the $\beta_i$ induced terms, that we want to put in to $V_\beta$ in \eq{Vsplit}, as designed.

With the splitting up of terms in \eqs{gammaRSsplit}{etasplit}, we can formally define the two pieces $V_\Gamma,V_\beta$ into which we have split the rapidity evolution kernel in \eq{Vsplit}:
\begin{subequations}
\label{VGammaVbeta}
\begin{align}
V_\Gamma(\nu_L,\nu_H;\mu_L) &= \exp\biggl\{ \ln\frac{\nu_H}{\nu_L} \bigl[ \zed_S \eta_\Gamma^{\text{conf.}}(1/b_0,\mu_L) + \gamma_{RS}^{\text{conf.}}(\mu_L)\bigr]\biggr\} \\
V_\beta(\nu_L,\nu_H;\mu_L) &=  \exp\biggl\{ \ln\frac{\nu_H}{\nu_L} \bigl[ \zed_S \Delta\eta_\Gamma(1/b_0,\mu_L) + \Delta\gamma_{RS}(\mu_L)\bigr]\biggr\}\biggr\rvert_{\text{F.O.}}\,,
\end{align}
\end{subequations}
indicating that $V_\beta$ is to be truncated to fixed order in $\as$ according to \tab{NkLL}. This defines the ``$V_\beta$-expansion'' we first mentioned in \ssec{intro3}. In this paper we shall not need it beyond the $\cO(\as^2)$ terms given in \eq{Vbeta}.

Our master expression for the resummed cross section, using the scale choices and prescriptions we have explained above, is then:
\begin{align}
\label{finalresummedcs}
 \frac{ d \sigma} {dq_T^2 dy} &= \sigma_0 \pi (2\pi)^2 C_t^2(M_t^2, \mu_T) H (Q^2,\mu_H ) U(\mu_L,\mu_H, \mu_T) \int db\, b J_0( b q_T) V_\Gamma(\nu_L^*,\nu_H;\mu_L)  \nn\\
 &\quad\times V_\beta(\nu_L^*,\nu_H;\mu_L)  \wt S( b ;\mu_L,\nu_L^*) \wt f_1^{\perp}( b, x_1,p^-; \mu_L ,\nu_H) \wt f_2^{\perp}(  b, x_2, p^+;  \mu_L ,\nu_H) \,, 
\end{align}
where $U$ in \eq{UVdef} and $V_\Gamma$ given by \eq{VGamma} or \eq{VGammaVbeta} are exponentiated:
\begin{align}
U(\mu_L,\mu_H, \mu_T) &= \exp\biggl\{4 K_\Gamma(\mu_L,\mu_H) - 4 \eta_\Gamma(\mu_L,\mu_H) \ln\frac{Q}{\mu_L} - K_{\gamma_H}(\mu_L,\mu_H)- K_{\gamma_{C_t^2}}(\mu_L,\mu_T)\biggr\}\nn \\
\label{Vstar}
V_\Gamma (\nu_L^*,\nu_H;\mu_L) &= \exp\biggl\{ \ln\frac{\nu_H}{\nu_L^*} \sum_{n=0}^\infty \Bigl(\frac{\as(\mu_L)}{4\pi}\Bigr)^{n+1}(\zed_S \Gamma_n \ln\mu_L b_0 + \gamma_{RS}^n)\biggr\}\,,
\end{align}
and the other objects are truncated at fixed order in $\as$, $H$ being given by \eqs{Hexp}{Cn}, $\wt S$ by \eqs{Sfixedorder}{Sn}, $\wt f^\perp$ being given by \eqss{beammatching}{Iijexpansion}{Iijcoefficients}. The $\beta$-function dependent part of the RRG kernel $V_\beta$ is also truncated at fixed order in our scheme, and is defined in \eq{VGammaVbeta} and given by \eq{Vbeta} up to $\cO(\as^2)$, the highest order we shall need in this paper.

In order for \eq{finalresummedcs} to successfully resum large logs of scale ratios, we recall that the scales $\mu_H,\mu_L$ and $\nu_H,\nu_L$ should be chosen near the natural scales of the respective hard, soft, and beam functions. $\mu_H$ and $\nu_H$ should be chosen $\sim Q$. We expect $\nu_L$ to be chosen $\sim \mu_L$, and then shifted to $\nu_L^*$ according to \eq{nuLstar} to introduce the quadratic damping factor in the $b$ exponent. As for $\mu_L$ itself, it should be chosen $\sim q_T$ in momentum space, although we will explore in \ssec{muLscale} a modified choice for this scale that better preserves stable power counting. For now, it remains a free scale.

\subsubsection{Truncation and resummed accuracy}

Here we summarize the rules for how to truncate the various objects in the full resummed cross section \eq{finalresummedcs}. These rules are mostly standard and well known, but with our introduction of the division of the RRG kernel into exponentiated and fixed order pieces $V_\Gamma \times V_\beta$, it seems prudent to restate these rules here. These are given in \tab{NkLL}.
\begin{table}[t]
\begin{center}
$
\begin{array}{ | c | c | c | c | c | c |}
\hline
\text{accuracy} & \Gamma_n,\beta_n & \gamma_{H,S,f}^n ,\, \gamma_{RS}^n & V_\beta &  H,\wt S,\wt f \\  \hline
\text{LL} & \as & 1 &  1 & 1  \\ \hline
\text{NLL} & \as^2 & \as  & \as & 1  \\ \hline
\text{NNLL} & \as^3 & \as^2  & \as^2 & \as \\ \hline
\text{N$^3$LL} & \as^4 & \as^3  & \as^3 & \as^2 \\ \hline 
\end{array}
$
\end{center}
\vspace{-1em}
\caption{Order of anomalous dimensions, beta function, and fixed-order functions (hard, soft, TMDPDF, and $V_\beta$ in \eq{VGammaVbeta}) required to achieve \nkll and \nkllprime accuracy in the resummed cross section \eq{finalresummedcs}.}
\label{tab:NkLL} 
\end{table}

We should note here that our choice of terms to group into the exponentiated part $V_\Gamma$ and those expanded in fixed order in $V_\beta$ in \eq{finalresummedcs} is not unique. Terms in $V_\Gamma$ and $V_\beta$ in \eqs{VGamma}{Vbeta} can be shifted back and forth by a different choice of prescription, and different choices of scales (such as $\nu_L^*$ in \eq{nuLstar}). Indeed, with our choice of the split between $V_\Gamma$ and $V_\beta$ and the scale $\nu_L^*$, the rapidity evolution exponent contains only a subset of the terms in the full rapidity anomalous dimension \eq{gammanuexp}---but the subset that allows an analytic evaluation of the $b$ integral, as we will show in \sec{analytic}. One may very well make a different set of choices that put a different set of terms in the exponent, based on a different set of desired criteria. This freedom is allowed by the presence of only a single large log in the RRG kernel in \eq{UVdef} at the virtuality scale $\mu_L$. 
We present our particular choice as just one such example. We advocate that readers make use of the freedom to choose the scales $\nu_{L,f}$ and $\mu_{L,H}$ in \eq{finalresummedcs}, either before \emph{or after} $b$ integration, along with the organization of terms in the RRG kernel \eq{Vsplit} into exponentiated and fixed-order parts (beyond NLL) to achieve their desired properties and results for the resummed momentum-space cross section.\footnote{We explore one such scheme in \appx{Cparameter} which allows us to also include single logarithmic terms of the form $\ln(\mu_L b_0)$ at each order of resummation using a simple modification of the choice for $\nu_L^*$.}

 The way we have organized the resummed cross section \eq{finalresummedcs}, all logs of $\mu_L b_0$ are contained in the exponent of $V_\Gamma$, and in fixed-order terms in $V_\beta,\wt S,\wt f_i^\perp$. Furthermore, the power of $\ln\mu_L b_0$ in the exponent \eq{Vstar} with the scale choice $\nu_L^*$ is at most quadratic to all orders in $\as$.  
Our choices to arrange this property are motivated, as we will see later, by the fact that it enables us, using some approximations, to obtain an analytical expression for the $b$ space integral. This is only possible as long as the quadratic nature of the exponent holds. 
At N$^3$LL (or NNLL$'$) and higher order, the fixed-order coefficients $V_\beta,\wt S,\wt f_i^\perp$ contains other powers of $\ln\mu_Lb_0$, but the contributions of these fixed-order logs can be dealt with analytically as well, as long as the exponent is no more than quadratic in $\ln\mu_L b_0$.

\subsubsection{Stable power counting in momentum space and the scale $\mu_L$}
\label{ssec:muLscale}

We have not yet specified exactly what we will choose for the scale $\mu_L$.  All of the above arguments are contingent upon the fact that our power counting $\ln(\mu_L b_0) \sim 1$ holds. What this means in momentum space is that after performing the $b$ integral in \eq{finalresummedcs}, fixed-order logs in the integrand of the form $\ln^n \mu_L b_0$ do not generate parametrically large terms after integration. This, it turns out, depends on what we pick for the scale $\mu_L$.  Let us see what $\mu_L$ we should choose for this power counting to remain true. It turns out that the ``obvious'' choice in momentum space $\mu_L =q_T$ is not always a very good choice for minimizing the contributions of the fixed order logs. This is not too surprising, since the kernel against which they are integrated in transforming back to momentum space is no longer a simple inverse Fourier transform with a single scale $q_T$. Instead it involves an exponent which is also a function of the high scale. It is then natural that the scale at which the logarithms are minimized (if not put exactly to 0) is shifted towards the high scale. This effect is particularly pronounced at low values of $q_T$.

\begin{figure}
\centerline{\scalebox{.55}{\includegraphics{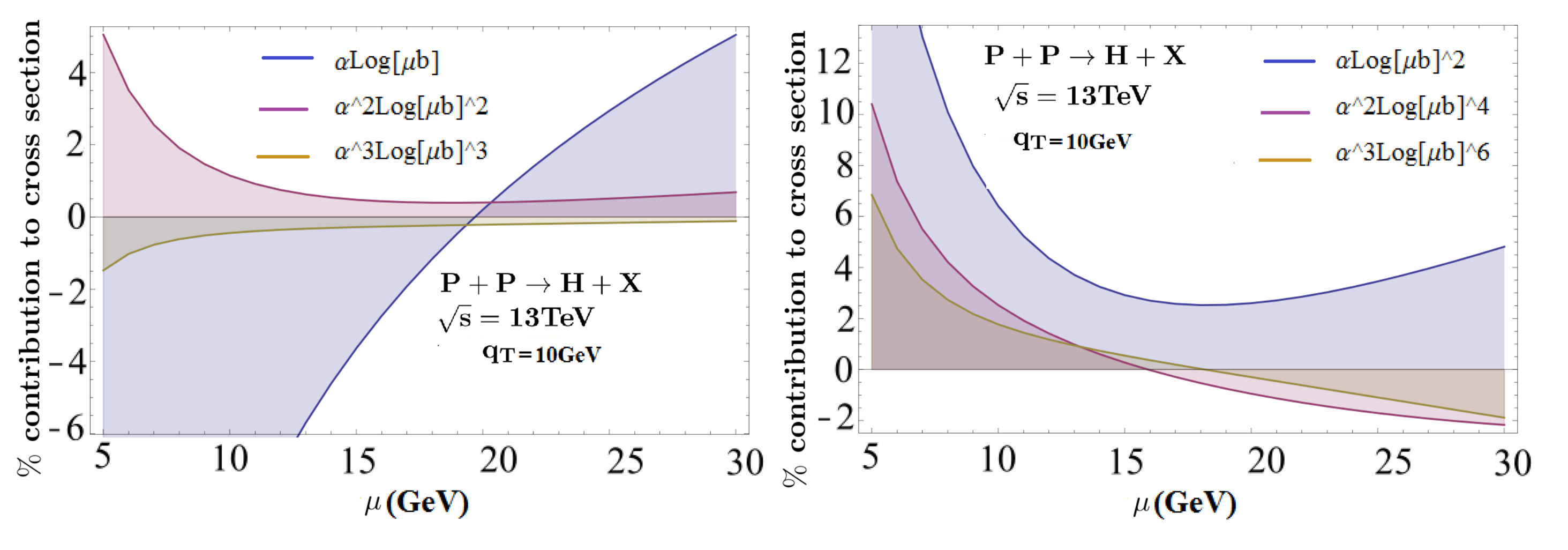}}}
\vskip-0.5cm
\caption[1]{Percentage contribution of fixed order logs to the $b$ integral in \eq{finalresummedcs} as a function of the scale choice $\mu$, assuming a coefficient of $1$ in front of each plotted log, as an estimate of the scale $\mu$ where the fixed-order logs in $b$ space translate to small logs in momentum space.}
\label{fig:fixed} 
\end{figure}

\begin{figure}
\centerline{\scalebox{.55}{\includegraphics{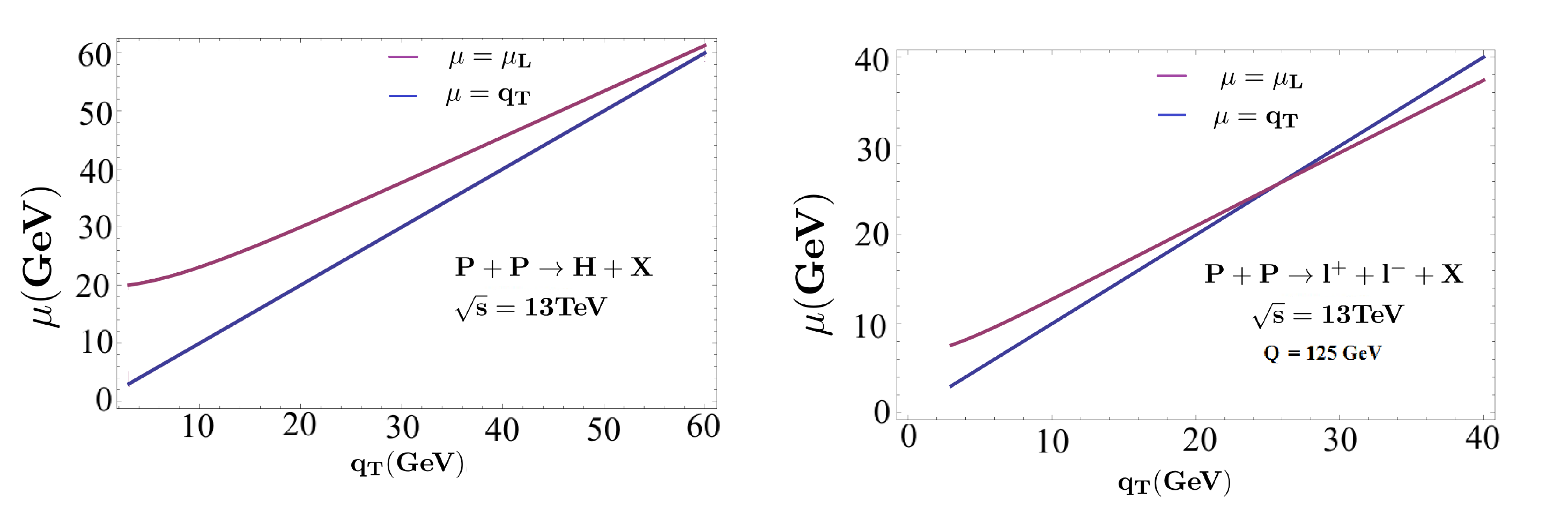}}}
\vskip-0.5cm
\caption[1]{Optimal $\mu$ scale choice in momentum space, corresponding to the solution of the condition \eq{scale} for the peak of the resummation exponent in the $b$ integral, which also turns out to correspond roughly to the location where all the logs in \fig{fixed} make the numerically smallest contributions.}
\label{fig:scaleL} 
\end{figure}

For the rest of the fixed order logarithmic pieces, we then check what value of scale  will minimize the contribution to the cross section ($\lesssim 10 \% $). 
Here, the nature of the resummed kernel can help us. 
The soft exponent in $b$-space provides damping at both small and large values of $b$ which, combined with the Bessel function, gives an integrand which has the form of damped oscillations. Then it is reasonable to assume that the most of the contribution to the integral is coming from the around the region of the first peak.
 A ballpark choice for the scale $\mu$ then is $1/b^*$, where $b^*$ is the scale at which the resummation kernel has the first peak. Since the hard kernel is independent of $b$, we only need to consider the soft resummation.  A straightforward analysis of the $b$-space integrand 
then gives the following condition for the peak 
\bea
q_T b^* \, J_1(q_T b^*) 
&=& J_0( q_T b^*)\left \{ 1- 2\Gamma_0 \frac{\as}{2\pi} \Bigg[\ln \frac{\nu_H}{\mu_L} 
 +2(1-p)\ln(\mu_L b_0^*) \Bigg]\right \}  
\eea
$J_{0,1}(x)$ are the zeroth and first order Bessel Functions of the first kind respectively. 
We now set $b^*=1/\mu_L$ and we can further simplify the expression above by keeping the dominant terms.
\bea
 \frac{q_T}{\mu_L}J_1\left(\frac{q_T}{\mu_L}\right) = J_0\left(\frac{q_T}{\mu_L}\right)\left \{ 1-2\Gamma_0 \frac{\as}{2\pi}
 \ln \frac{\nu_H}{\mu_L} \right \}
\label{scale}
\eea
This can be solved numerically to obtain the scale $\mu_L$.
We  can also confirm this by checking the contribution of the leading term $\as \ln(\mu_L b_0)$ at $O(\as)$  and cross checking it against the next biggest piece $ \as^2 \ln^2(\mu b_0),  \as^2 \ln^3(\mu b_0)$ at $O(\as^2)$. 
While we are currently keeping only one-loop fixed order pieces in our NNLL cross section, we can always include the two-loop pieces $\as^2 \ln^n(\mu b_0)$ which are known to us at NNLL (this would be part of the NNLL$'$ cross section).
\fig{fixed} looks at the percentage contribution of the fixed order logs when integrated against the NNLL exponent in $b$ space. This particular plot is for the Higgs $q_T$ distribution for $q_T = 10 \GeV$. It is pretty clear that $\mu =q_T$ is a poor choice for $\mu$ and that a good choice would be somewhere between $\mu \sim 20 \GeV$ for the contribution from the fixed order logs to be $\sim 2 \%$. In comparison, the value predicted by \eq{scale} is $22 \GeV$. \fig{scaleL} gives the scale choice for Higgs and DY (Choosing a common value of $Q=125 \GeV$ and $\sqrt{s}= 13$ TeV. For low values of $q_T$, the scale is shifted away from $q_T$ toward $Q$ as expected. The shift is far more pronounced for Higgs than for DY since the cusp anomalous dimension for Higgs is much larger so that the soft exponent has far more impact on the shape of the 
$b$-space integrand. At large values of $q_T$, the scale is more or less $q_T$.

The key point to notice here is that apart from the single log $\alpha_s \ln(\mu_L b_0)$, which we will include in the fixed order cross section at NNLL, the contribution from the higher order logs shows a flat behavior at at values of $\mu \geq 15\text{ GeV}$ for $q_T= 10\text{ GeV}$  in \fig{fixed}. So the result is insensitive to the choice of $\mu_L$ as long it is chosen greater than this threshold value.

\begin{figure}
\centerline{\scalebox{.55}{\includegraphics{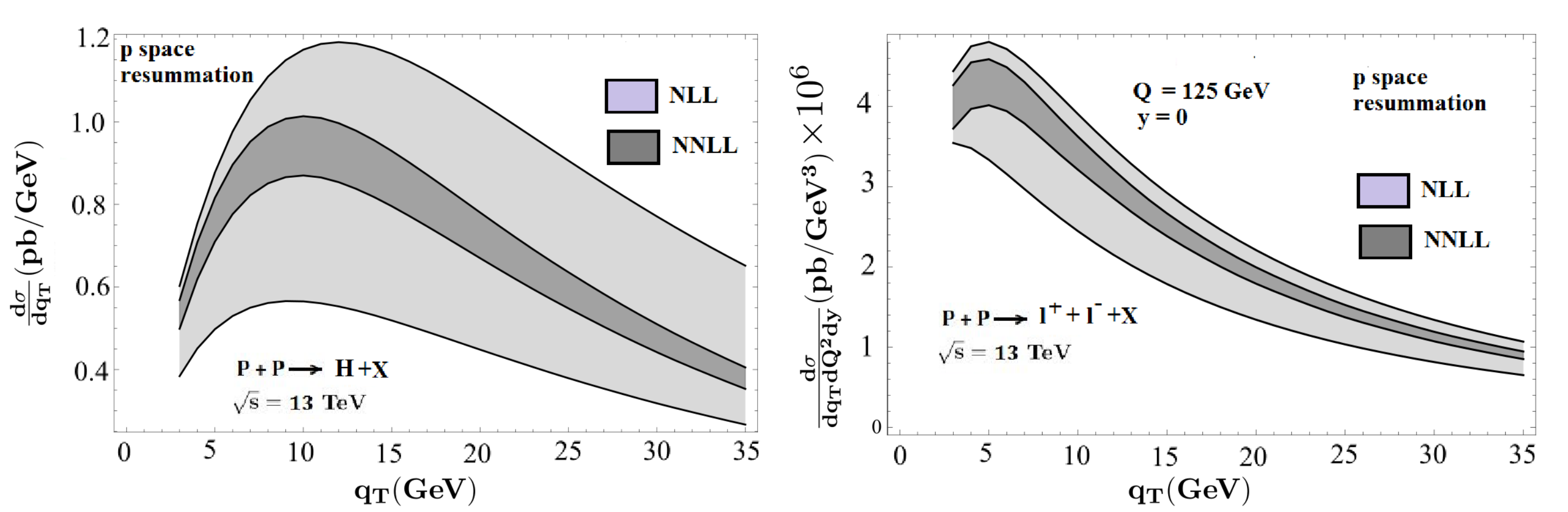}}}
\vskip-0.2cm
\caption[1]{Result of the resummed cross section \eq{finalresummedcs} in momentum space, with the rapidty and virtuality scales $\nu_L^*$ and $\mu_L$ chosen as in \eq{nuLstar} and \fig{scaleL}. No cutoff of the $b$ integral is required, and the $\mu_L$ scale can be varied in its full range from $\mu_L/2$ to $2\mu_L$ to obtain the uncertainty, without hitting a pole as in \fig{bspace}. These plots are obtained by performing the $b$ integral numerically. We stop plotting at low $q_T$ where true nonperturbative effects will enter.}
\label{fig:NNLL} 
\end{figure}

Using these scale choices, we now obtain the transverse spectrum, again at NNLL (\fig{NNLL}). The uncertainties are obtained by scale variations described in \eq{scalevariations}, obtained reliably by varying both $\mu$ and $\nu$ scales, without cutoffs in the $b$ integral.

\section{Explicit formula for resummed transverse momentum spectrum}
\label{sec:analytic}

In this section we will provide an expression for the transverse momentum spectra of gauge bosons that is analytic in terms of its dependence on all the kinematic variables. It requires a numerical but efficient approximation for the Bessel function in the $b$ integral in \eq{xsec} or \eq{finalresummedcs}. 

We can write the resummed cross section \eq{finalresummedcs} in the form
\be
\label{resummedIb}
\frac{d\sigma}{dq_T^2 dy} = \frac{1}{2}\sigma_0 C_t^2(M_t^2, \mu_T) H(Q^2,\mu_H) U(\mu_L,\mu_H, \mu_T) I_b(q_T,Q;\mu_L,\nu_L^*,\nu_H)\,,
\ee
where the we have isolated the terms in the $b$ integral,
\be
\label{Ibdef}
I_b(q_T,x_{1,2},Q;\mu_L,\nu_L^*,\nu_H) \equiv \int_0^\infty db\,b J_0(bq_T) \wt F(b,x_{1,2},Q;\mu_L,\nu_L^*,\nu_H) V_\Gamma(\nu_L^*,\nu_H;\mu_L)\,,
\ee
grouping together the terms in the integrand that are to be expanded in fixed order in $\as(\mu_L)$,
\begin{align}
\label{fixedorderfactor}
\wt F(b,x_{1,2},Q;\mu_L,\nu_L^*,\nu_H) & \equiv (2\pi)^3\wt S( b ,\mu_L,\nu_L^*) \wt f_1^{\perp}( b, x_1, p^-,\mu_L ,\nu_H) \wt f_2^{\perp}(  b,  x_2,p^+, \mu_L ,\nu_H) \nn \\
&\qquad \times V_\beta(\nu_L^*,\nu_H;\mu_L) \,,
\end{align}
separating them from the exponentiated $V_\Gamma$ factor. The $V_\Gamma$ factor is explicitly to all orders in $\as$, plugging in \eq{nuLstar} for $\nu_L^*$ in \eq{Vstar},
\begin{align}
V_\Gamma(\nu_L^*,\nu_H;\mu_L) &= \exp\biggl\{\! \biggl[ \ln \! \frac{\nu_H}{\nu_L} + \Bigl( \frac{1}{2} \plus \frac{\as(\mu_L)\beta_0}{4\pi} \ln\frac{\nu_H}{\nu_L}\Bigr)\!  \ln\mu_L b_0\biggr]  \\
&\qquad\qquad\times \Bigl( \zed_s \Gamma[\as(\mu_L)] \! \ln\mu_L b_0 \plus  \gamma_{RS}[\as(\mu_L)]\Bigr) \! \biggr\} \, ,  \nn
\end{align}
which can always be written in the form 
\bea \label{VGaussian}
V_\Gamma  =  C e^{ -A \ln^2( \Omega b)}
\eea
where $C$, $A$, and $\Omega$ are independent of $b$ and thus constants as far as the integral $I_b$ in \eq{Ibdef} is concerned. 
Explicitly,
\bea \label{ACUeta}
A(\mu_L,\nu_L,\nu_H) &=& -\zed_S\Gamma[\as(\mu_L)] (1-p)=-\zed_S \Gamma[\as(\mu_L)] \left(\frac12+\frac{\as(\mu_L)\beta_0}{4\pi}\ln \frac{\nu_H}{\nu_L} \right)
\\
\Omega&\equiv& \frac{\mu_L e^{\gamma_E}}{2}\chi\,,\quad \chi(\mu_L,\nu_L,\nu_H) = \exp\left\{ \frac{ \ln \nu_H / \nu_L }{1+\frac{\as(\mu_L)\beta_0}{2\pi}\ln \nu_H / \nu_L }+\frac{\gamma_{RS}[\as(\mu_L)]}{2\zed_S\Gamma[\as(\mu_L)]}\right\}
 \nn \\
C(\mu_L,\nu_L,\nu_H) &=& \exp\left\{ A \ln^2\chi + \gamma_{RS} [\as(\mu_L)] \ln \frac{\nu_H}{\nu_L}\right\}  \nn
\eea
These show all the dependence on the scales and on the anomalous dimensions contained in $V_\Gamma$. They are to be truncated to the order in resummed accuracy we intend to work (which we show at NLL and NNLL in \appx{fixedorder} in \eqs{ACUetaNLL}{ACUetaNNLL}). 

Thus we now just have to figure out how to integrate the Gaussian $V_\Gamma$ in $\ln b$ in \eq{VGaussian} against the Bessel function in the integral $I_b$ in \eq{Ibdef}. We will encounter integrals of the form
\be
\label{Ibk}
I_b^k\equiv \int_0^\infty db\, b J_0(b q_T) \ln^k(\mu_L b_0) e^{-A \ln^2\Omega b}\,,
\ee
where the factors $\ln^k(\mu_L b_0)$ come from the fixed-order factor $\wt F$ in \eq{fixedorderfactor}. 
We will first focus on the case where $\wt F = 1$ (\eg at NLL) and compute $I_b^0$, and then discuss below how to generate the terms $I_b^{k>0}$ resulting from integrating the fixed-order terms containing logs of $\mu_L b_0$ against the rest of the integrand.

In the next two subsections we shall develop a method to evaluate $I_b^0$ semi-analytically---with a numerical series expansion for the Bessel function but deriving the analytic dependence of \eq{Ibk} on all relevant physical parameters including $q_T$. Then in \ssec{fixedorderlimit} we shall show how to obtain arbitrary $I_b^k$ from derivatives on $I_b^0$.

\subsection{Representing the Bessel Function}

The main issue in doing the integrals in \eq{Ibk} analytically is the presence of the Bessel function inside the integrand.
The exponent, in our scheme, has terms only up to the quadratic powers of $\ln(\mu_L b_0)$.  Analytic integration is then possible if we can approximate the Bessel function using a series of simpler functions, such as polynomials, which can be easily integrated given the quadratic nature of our exponent. However, a simple power series expansion fails to reproduce the Bessel function in the region, where it contributes to the integrand. This is basically the region up to $b \sim 2 \GeV^{-1}$.  This is because the argument of the Bessel function is  $b q_T$ and at larger values of $q_T$ ($\geq$ 10 GeV),
the power series expansion is no longer useful. We can possibly switch to the large $b\, q_T$ asymptotic form in terms of the cosine function, but then analytic integration is again not possible.

An alternative way then is to rewrite the Bessel function in terms of an integral representation, so that the $b$ space integrand can them be done exactly. This automatically rules out any representations in terms of trigonometric functions, and given the discussion earlier, the most expeditious choice is if a polynomial representation can be used. One choice that we will find amenable to a polynomial expansion is the Mellin-Barnes type representation of the Bessel function, 
\bea
\label{Mellin-Barnes}
J_0(z) = \frac{1}{2 \pi i}\int_{c-i \infty}^{c+i \infty} dt \frac{\Gamma[-t]}{\Gamma[1+t]} \left(\frac{1}{2} z \right)^{2t}  
\,,\eea
where the contour is to the left of all the non-negative poles of the Gamma function, i.e., $c<0$. We give a proof of this identity in \appx{proof}.

Going back to our all-orders $b$ space integral $I_b^0$ given by \eq{Ibk}, it now looks like 
\bea \label{Ib}
I_b^0= \int_0^\infty db\, b J_0(b q_T)\, e^{-A\ln^2(\Omega b)}
= \frac{1}{2 \pi i}  \int_{c-i \infty}^{c+i \infty} dt \, \frac{\Gamma[-t]}{\Gamma[1+t]} 
 \int_0^\infty db\, b \left(\frac{b q_T}{2} \right)^{2t} e^{ -A \ln^2( \Omega b)}
\eea
Since we do not have a Landau pole,  we can extend the limit of integration in $b$ space all the way to infinity, in which case, the $b$ integral is now in the form of a simple Gaussian integral and admits the analytical result:
\bea
I_b^0 &=&  \frac{1}{2\pi i}  \int_{c -i \infty}^{c+i \infty} dt\, \frac{\Gamma[-t]}{\Gamma[1+t]} \left(\frac{q_T}{2}\right)^{2t}\sqrt{\frac{\pi}{A}} \,e^{ (1+t)^2/A- 2(1+t) \ln\Omega}
\nn\\
&=& \frac{1}{i \Omega^2} \frac{1}{\sqrt{4 \pi A}} \int_{c-i \infty}^{c+i \infty} dt\, \frac{\Gamma[-t]}{\Gamma[1+t]}e^{ (1+t)^2/A- 2t L}
\label{Ib-L}\eea
where we have defined 
\be\label{L}
L= \ln \frac{2 \Omega}{q_T}\,.
\ee
Let's examine the Gaussian exponent in this integrand. A simple rearrangement gives us 
\bea\label{Ib-t}
I_b^0 = \frac{ 2 }{i q_T^2} \frac{e^{-AL^2 }}{\sqrt{\pi A}} \int_{c-i \infty}^{c+i \infty} dt\, \frac{\Gamma[-t]}{\Gamma[1+t]}e^{ \frac{1}{A}(t-t_0)^2 }\nn
\eea
where $t_0 = -1+ A L$ , which is a saddle point for this Gaussian and lies on the real line.
In some sense, the $t$ space integral is a dual of the $b$ space integral since the degree of suppression inverts itself from one space to another. So then in a region of large $A$, we should stick to the $b$ space integral, fit a polynomial to the Bessel function (which will now work since the integrand is highly suppressed). On the other hand for small $A$ (which would be relevant for most perturbative series), we should go to $t$ space.

If we parametrize the contour as $t= c+ix$, we have 
\bea
 I_b^0 = \frac{ 2 }{q_T^2} \frac{e^{-AL^2 }}{\sqrt{\pi A}} \int_{-\infty}^{\infty} dx \, \frac{\Gamma[-c-ix]}{\Gamma[1+c+ix]}
 \,e^{ -\frac{1}{A}[x-i(c-t_0)]^2}
\,.\label{saddle}
\eea
It is clear that the path of steepest descent passes through this saddle point ($c=t_0$) and is parallel to the imaginary axis. 
One can then consider doing a Taylor series expansion of the rest of the integrand 
\be\label{ft}
f(t)\equiv \frac{\Gamma[-t]}{\Gamma[1+t]}
\ee
 along this path around the saddle point, truncating the series after a finite number of terms. 

The primary difficulty with a polynomial expansion is that to have  percent level accuracy that we desire for this integral, we need to have a good description of $f(t)$ out to $x_l \sim \sqrt{2 A\ln(10)}$ at which value the exponent drops to 1\% of its value. The factor of $A$ is essentially the cusp anomalous dimension, ( for e.g., it is $ 2 \alpha_s C_A /\pi $ for the Higgs ) which is a small number. Considering the worst case scenario $ A \sim 0.5$, we would need $x_l \sim 1.5$, which clearly cannot be accomplished using a Taylor series expansion because the radius of convergence is around 1. We will, instead, find an expansion for $f(t)$ in the next section in terms of (Gaussian-weighted) Hermite polynomials that performs quite well with just a few terms. There a few customizations of this expansion we will make to optimize fast convergence. Our particular procedure presented here is by no means unique, and we give a couple of alternative methods for expanding and approximating the integrand of $I_b^0$ in \appx{alternatives}. There are, undoubtedly, others that would work also.

\subsection{Expansion in Hermite polynomials}
\label{HStrategy}

One of the difficulties with finding a series expansion of $f(t)$ in \eq{ft} is that it grows exponentially for large $\abs{x}$, with $t=c+ix$. We can factor out this exponential growth by using Euler's reflection formula:
\be
\label{Euler}
\Gamma(z)\Gamma(1-z) = \frac{\pi}{\sin(\pi z)}\,.
\ee
Then
\be
\label{ftfactored}
f(t) = -\frac{\sin(\pi t)}{\pi}\Gamma(-t)^2\,.
\ee
The exponential behavior for large $\abs{x}$ is now factored out in the sine function in front, and we can focus on finding a good expansion for $\Gamma(-t)^2$. The integral \eq{saddle} is then:
\be
I_b^0 = -\frac{2}{q_T^2}\frac{e^{-AL^2}}{\sqrt{\pi A}} \int_{-\infty}^\infty dx\,\Gamma(-c-ix)^2 \frac{\sin[\pi(c+i x)]}{\pi} e^{-\frac{1}{A}[x - i(c-t_0)]^2}\,.
\ee
The sine can be written in terms of exponentials, which shift the linear and constant terms in the Gaussian exponential. It is straightforward to work out that the result is
\begin{align}
I_b^0 = \frac{1}{i\pi q_T^2\sqrt{\pi A}}  \int_{-\infty}^\infty dx\,  \Gamma(-c-ix)^2 &\biggl\{e^{-A(L - i\pi/2)^2} e^{-\frac{1}{A}\bigl[x + \frac{A\pi}{2} - i(c-t_0)\bigr]^2} \\
& -e^{-A(L + i\pi/2)^2}  e^{-\frac{1}{A}\bigl[x - \frac{A\pi}{2} - i(c-t_0)\bigr]^2} \biggr\} \,. \nn
\end{align}
By changing variables in the second line from $x\to -x$, we can write the result compactly as
\be
\label{Ibimag}
I_b^0  = \frac{2}{\pi q_T^2\sqrt{\pi A}} \Imag\biggl\{ e^{-A(L-i\pi/2)^2} \int_{-\infty}^\infty dx\, \Gamma(-c-ix)^2 e^{-\frac{1}{A}\bigl[ x + \frac{A\pi}{2} - i(c-t_0)\bigr]^2}\biggr\}\,.
\ee

The remaining function $\Gamma(-t)^2$ is exponentially damped for large $x$. Indeed, Stirling's formula tells us that
\be
\label{stirling}
\abs{\Gamma(-c-ix)}^2 \to 2\pi e^{-\pi\abs{x}} \abs{x}^{-1-2c}
\ee
as $\abs{x}\to \infty$. The contribution of this exponential tail is further suppressed by the Gaussian factor it multiplies in \eq{Ibimag}.
On the other hand, near $x=0$, the function $\Gamma(-t)^2$ itself closely resembles a Gaussian, times polynomials. To determine the curvature of the Gaussian, we look at the Taylor series expansion of $\Gamma(-t)^2$ near $x=0$:
\bea \label{gexpansion}
\Gamma(-c-ix)^2 &=&\Gamma(-c)^2 \left[(1 - a_0\, x^2)-2i \psi^{(0)}(-c)\, x(1-b_0\, x^2) \right]+\cdots
\,,\\ \nn
a_0&=&2 \psi^{(0)}( -c)^2 + \psi^{(1)}( -c)
\,,\\ \nn
b_0 &=& \tfrac{2}{3} \psi^{(0)}( -c)^2 + \psi^{(1)}( -c) +\tfrac{1}{6} \psi^{(2)}( -c)/\psi^{(0)}( -c)
\,.\eea
It now remains to find a good series expansion for $\Gamma(-c-ix)^2$ that enables us to perform the integral in \eq{Ibimag} analytically and accurately. As noted in \eq{stirling}, $\Gamma(-c-ix)^2$ dies exponentially for large $x$, and the remaining Gaussian in \eq{Ibimag} dies even faster. For $c=-1$, both $\Gamma(-c-ix)^2$ and the Gaussian are significantly nonzero only up to about $\abs{x}\sim 2$. For practical purpose of series expansion, we set $c=-1$ which makes the gamma function less oscillatory than the values $|c|<1$.

This is the saddle point in the limit $ A \rightarrow 0$, i.e., when we are in the fixed order regime with the resummation turned off. This would still induce some imaginary part and hence oscillations in the exponent away from $A=0$, but with a good expansion, this is not an issue.
Thus one can try a series expansion for $\Gamma(1-ix)^2$ in terms of Hermite polynomials $H_n(x)$ which form a complete orthogonal basis. 
They are well known, of course, but we nevertheless remind ourselves of the first several $H_{n}$:
\begin{align}
H_0(x) &= 1 & H_1(x) &= 2x  \\
H_2(x) &= 4x^2 - 2  & H_3(x) &= 8x^3 - 12x \nn \\
H_4(x) &= 16 x^4 - 48 x^2 + 12 & H_5(x) &= 32 x^5 - 160 x^3 + 120 x \nn \\
H_6(x) &= 64x^6 - 480 x^4 + 720 x^2 - 120 & H_7(x) &= 128 x^7 - 1344 x^5 + 3360 x^3 - 1680 x\,, \nn
\end{align}
etc. They satisfy the orthogonality relation
\be
\label{orthogonality}
\int_{-\infty}^\infty dx\,e^{-\alpha^2 x^2} H_m(\alpha x) H_n(\alpha  x) = \alpha^{-1}\sqrt{\pi}\, 2^n n! \delta_{nm}\,.
\ee
 We expand $\Gamma(1-ix)^2$ in terms of these polynomials in the following way:
\be
\label{Hermiteexpansion}
\Gamma(1-ix)^2  =  e^{-a_0 x^2} \sum_{n=0}^\infty c_{2n} H_{2n}(\alpha x) 
	+ \frac{i \gamma_E}{\beta} e^{-b_0 x^2} \sum_{n=0}^\infty c_{2n+1} H_{2n+1}(\beta x) \,,
\ee
We have introduced weighting factors $e^{-a_0 x^2}$ and $i \gamma_E  e^{-b_0 x^2}$ for the real and imaginary parts to help with faster convergence of the series, as they capture the behavior of $\Gamma(1-ix)^2$ near $x=0$, using the values of $a_0$ and $b_0$ obtained from the Taylor series expansion in \eq{gexpansion} at $c=-1$:
\be\label{a0b0}
a_0 =2\gamma_E^2+\frac{\pi^2}{6}\approx 2.31129\, , \quad b_0=\frac{2}{3}\gamma_E^2+\frac{\zeta_3}{3\gamma_E} +\frac{\pi^2}{6}\approx 2.56122
\,. \ee
Note that the real and imaginary parts of the  LHS of \eq{Hermiteexpansion} are  respectively even and odd functions of $x$. Hence on the RHS, we need even polynomials for the real part and odd polynomials for imaginary part. 
Although the relation \eq{orthogonality} would make it seem natural to pick $\alpha^2 = a_0$ and $\beta^2 =b_0$ in the arguments of $H_n$ in the expansion\eq{Hermiteexpansion}, we choose $\alpha,\beta$ to be floating, and will determine their optimal values to ensure the fastest convergence for this expansion.
 
Using \eq{orthogonality}, the coefficients in \eq{Hermiteexpansion} are given by
\bea
\label{HermiteCoefficients}
c_{2n} &=& \frac{\alpha}{\sqrt{\pi}\, 2^{2n} (2n)!} \int_{-\infty}^\infty dx \, \text{Re}\{\Gamma(1-ix)^2\} H_{2n}(\alpha x) e^{-(\alpha^2-a_0) x^2}\,,
\\
c_{2n+1} &=& \frac{\beta^2}{\gamma_E \sqrt{\pi}\, 2^{2n+1} (2n+1)!} 
\int_{-\infty}^\infty dx \, \text{Im}\{\Gamma(1-ix)^2\} H_{2n+1}(\beta x) e^{-(\beta^2-b_0) x^2}\,.\nn
\eea
These integrals still have to computed numerically, as far as we know, but note they are purely mathematical, having no dependence on any of our physical parameters, and need only be computed once.
Thanks to the damped behavior of the integrand and the normalization factors in front, the expansion coefficients fall off fairly rapidly with $n$. 

\begin{figure}
\centerline{
\includegraphics[width=.5\linewidth]{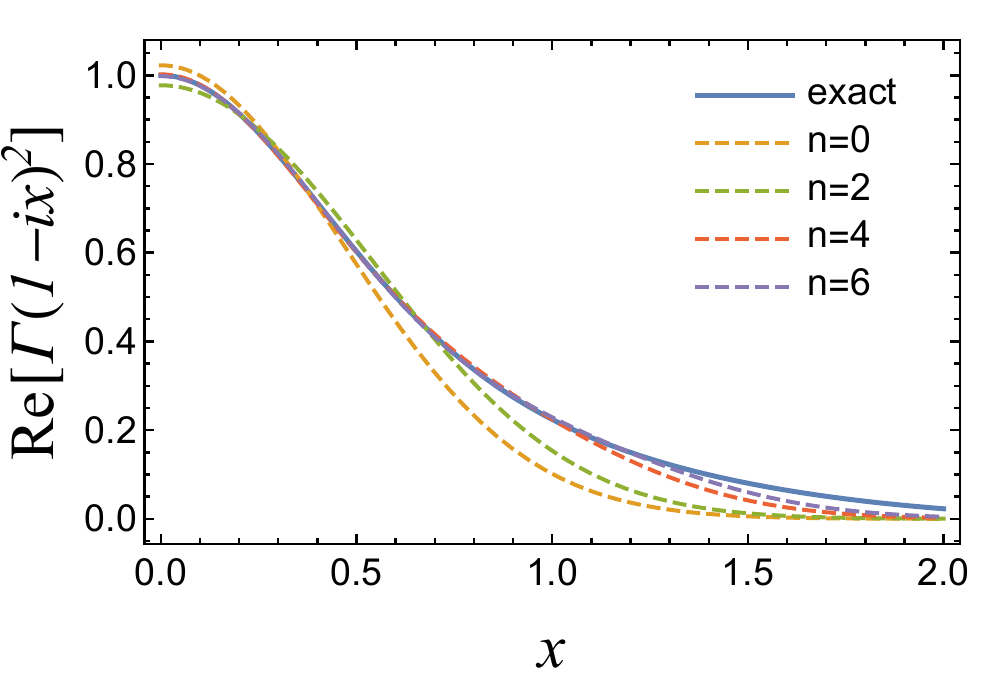}
\hskip 0.2cm
\includegraphics[width=.5\linewidth]{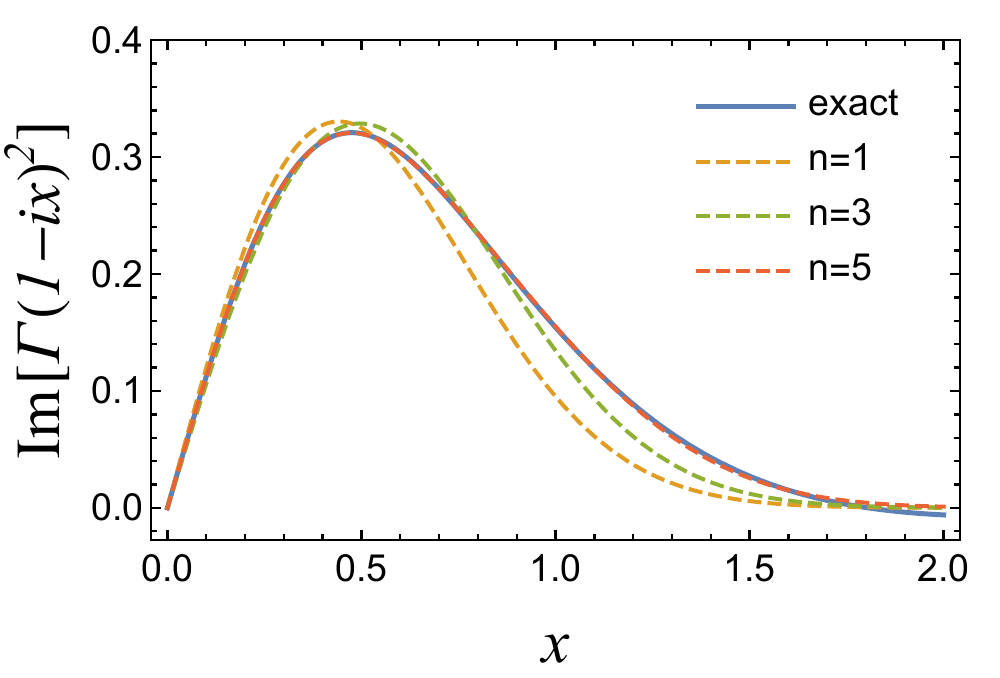}
}
\vskip-0.2cm
\caption[1]{Real and imaginary parts of $\Gamma(1-ix)^2$ compared to series expansion in terms of Hermite polynomials up to 6th (5th) order for the real (imaginary) parts, \ie four terms for the real part and three for the imaginary part.}
\label{fig:Gamma2} 
\end{figure}

To make the series expansion well behaved, the width parameter of Gaussian functions should be positive definite: $\alpha^2- a_0>0$ and $\beta^2- b_0>0$. These widths define the region of $x$ where the function $\Gamma(1-ix)^2$ is expanded in terms of the Hermite bases. For a narrow width, the series converges swiftly with $n$ but is valid only in a narrow region around $x=0$, while for broader width, the convergence is slower but the expansion is valid in a wider region around $x$. The Gaussian function in our integration in \eq{Ibimag} resolves the region $|x-\tfrac{A\pi}{2}| \sim \sqrt{A}$ and for the maximal value we encounter, $A\sim 0.5$, the broadest region is up to $|x| \sim 1.5$. The parameters $\alpha,\beta$ should be chosen so that the Gaussians in \eq{HermiteCoefficients} roughly match the width of this region and resemble $\Gamma(1-ix)^2$ itself as closely as possible. We explored various values of these parameters such that the exponents $\alpha^2 -a_0$ and $\beta^2-b_0$ were between 1 and 10 and found that for small values $\sim 1$ the convergence is slow and hence  more terms are needed for an accurate description. For large values $\sim 10$ the accuracy of the integration in \eq{Ibimag} is very good with a few basis terms but is not further improved by including higher order terms  because the series expansion is resolving only a narrow $x$ region compared to the one dictated by \eq{Ibimag}.
Empirical tests show that the series converges rapidly for $\alpha^2-a_0$ and $\beta^2-b_0$ around $3\sim 5$ while maintaining  required accuracy over the range of $x$ (from 0 to $\pm 1.5 $) desired.
\fig{Gamma2} shows the agreement between the exact result and series expansion up to 3 or 4 basis terms for the real (even) and imaginary (odd) parts, for:
\be\label{alphabeta}
\alpha^2-a_0=4\quad \text{and}\quad \beta^2-b_0=4\,.
\ee  
The coefficients $c_{n}$ for these choices of $\alpha,\beta$ are given by
\begin{align}
\label{cn}
c_0& =1.02248\,,\qquad c_2 = 0.02254\,,\qquad c_4= 0.00206\,, \qquad c_6=3.42\times 10^{-5}
\\ 
c_1&=1.06808\,,\qquad c_3=0.02173\,,\qquad c_5=0.00103\,, \qquad c_7=3.21 \times10^{-6}\, \nn
\end{align}
which indeed show a rapid convergence. In practice we include up to $c_6$ in our numerical results; from $c_7$ onwards the impact is negligible.

From staring at \fig{Gamma2}, one notices a residual deviation in the real part above $x\sim 1$, which thus appears to be the potentially largest source of error from our method. However, the region of larger $x$ in \fig{Gamma2} is suppressed by the remaining Gaussian in \eq{Ibimag}. The remaining deviation can easily be further suppressed if desired by including higher-order polynomials.
In practice, at NNLL accuracy the cross section has $\sim 10$\% uncertainties, and the error due to our series truncation at $n=5$ or $6$ is significantly smaller than this perturbative uncertainty. This is clearly seen in Fig. \ref{fig:Hcompare}, which shows the effect of increasing the total number of terms in the Hermite expansion from 6 to 7.

In terms of this expansion, the integration in \eq{Ibimag} is rewritten in terms of following basis of integrals
\bea \label{cH2n}
\cH_{n}(\alpha,a_0) &=&  \frac{1}{\sqrt{\pi A}} e^{-A(L-i\pi/2)^2}\int_{-\infty}^\infty dx\, H_{n}(\alpha x) e^{-a_0 x^2-\frac{1}{A}( x + z_0)^2}\,,
\eea
where $z_0 = A\pi/2 - i(c-t_0)$. For $c=-1$, $z_0 = A(\pi/2 + iL)$. The integrals for odd $n$ arising from the expansion in \eq{Hermiteexpansion} are obtained from \eq{cH2n} with the substitutions $\alpha\to \beta,a_0\to b_0$. The prefactors in front have been included in the definition of $\cH_n$ for later convenience.

Now we go about evaluating analytically the integrals in \eq{cH2n}. There are a number of ways to do this, we choose one that seems particularly elegant.

\subsubsection{Generating function method to integrate against Hermite polynomials}

Using the generating function for Hermite polynomials,
\be
\label{generating}
e^{2xt -t^2} = \sum_{n=0}^\infty \frac{t^n}{n!} H_n(x)\,,
\ee
we can efficiently evaluate all the integrals $\cH_{n}$ in \eq{cH2n} at the same time. 
By forming the infinite series,
\be
\label{Hnseries}
\cH \equiv \sum_{n=0}^\infty \frac{t^n}{n!} \cH_n(\alpha,a_0) = \frac{e^{-A(L-i\pi/2)^2}}{\sqrt{\pi A}}  \int_{-\infty}^\infty dx\,e^{2\alpha xt - t^2} e^{-a_0 x^2 - \frac{1}{A}( x+ z_0)^2}\,,
\ee
we are able to use the generating function relation \eq{generating} to obtain a Gaussian integral on the RHS. By evaluating the integral on the RHS and expanding the result back out in powers of $t^n$, we will be able to obtain expressions for the individual $\cH_n$.

Rescaling the integration variable and completing the square in the exponent on the RHS of \eq{Hnseries},
\be
\label{HGaussian}
\cH =  e^{-\frac{a_0 z_0^2}{1+a_0 A}} \frac{e^{-A(L-i\pi/2)^2}}{\sqrt{\pi(1+a_0A)}}\int_{-\infty}^\infty dx\, e^{-\bigl[ x + \sqrt{\frac{A}{1+a_0 A}}\,\bigl( \frac{z_0}{A} - \alpha t\bigr )\bigr]^2} e^{\frac{t}{1+a_0 A}\bigl[t(A\alpha^2 - 1 - a_0 A) - 2 \alpha z_0 \bigr] }
\ee
The exponent of the $x$-dependent Gaussian is complex, but the result of integrating it is just $\sqrt{\pi}$ (see \eq{contour}). Thus,
\be
\label{Ht}
\cH = e^{\frac{-A (L-i\pi/2)^2}{1+a_0 A}} \frac{1}{\sqrt{1+a_0 A}} \sum_{m=0}^\infty \frac{t^m}{m!}\frac{1}{(1+a_0 A)^m}\bigl[ t (A\alpha^2 - 1 - a_0 A) - 2\alpha z_0\bigr]^m\,,
\ee
where we have expanded the exponential in $t$ in \eq{HGaussian} in a Taylor series. We cannot directly read off the coefficients of powers of $t$ in \eq{Ht} to obtain $\cH_n$ in \eq{Hnseries} due to the powers of the binomial in $t$, but using the binomial expansion and some reindexing, we can do so. The proof is given in \appx{integral}, with the result:
\be
\label{Hnresult}
\cH_n(\alpha,a_0) = \cH_0(\alpha,a_0) \frac{(-1)^n n!}{(1+a_0 A)^n}\sum_{m=0}^{\floor*{n/2}}\frac{1}{m!}\frac{1}{(n-2m)!} \Bigl\{ [ A(\alpha^2 - a_0)-1] (1+a_0 A)\Bigr\}^m (2\alpha z_0)^{n-2m} \,,
\ee
where $\floor*{\frac{n}{2}}$ is the floor operator, i.e. the integer part of $\frac{n}{2}$, and
\be
\cH_0(\alpha,a_0) =  e^{\frac{-A(L-i\pi/2)^2}{1+a_0 A}} \frac{1}{\sqrt{1+a_0 A}} \,.
\ee
If one desires, \eq{Hnresult} can be written separately for even and odd $n$,
\begin{align}
\cH_{2n} &= \cH_0 \frac{(2n)!}{(1\plus a_0 A)^{2n}} \sum_{m=0}^n \frac{1}{m!}\frac{1}{(2n\minus 2m)!} \Bigl\{ [ A(\alpha^2 \minus a_0)-1] (1\plus a_0 A)\Bigr\}^m (2\alpha z_0)^{2n-2m} \\
\cH_{2n+1} &= \cH_1 \frac{(2n+1)!}{(1+a_0 A)^{2n}} \sum_{m=0}^n \frac{1}{m!}\frac{1}{(2n-2m+1)!} \Bigl\{ [ A(\alpha^2 - a_0)-1] (1+a_0 A)\Bigr\}^m (2\alpha z_0)^{2n-2m}\,, \nn
\end{align}
where
\be
\cH_1 = - \cH_0 \frac{2\alpha z_0}{1+a_0 A}\,.
\ee

\subsubsection{Explicit result of integration}

Explicitly, the first several $\cH_{n}$ given by \eq{Hnresult}, including $\cH_{0,1}$ given above, are given in \eq{Hnexplicit}.
The integral $I_b^0$ in \eq{Ibimag} that we sought to evaluate in the first place is then given explicitly by, for $c=-1$,
\begin{align}
\label{IbHermite}
I_b^0  &= \frac{2}{\pi q_T^2}  \sum_{n=0}^{\infty} 
\Imag \biggl\{    c_{2n} \cH_{2n}(\alpha,a_0) + \frac{i\gamma_E}{\beta} c_{2n+1} \cH_{2n+1}(\beta,b_0)  \biggr\} \\
&= \frac{2}{\pi q_T^2} \Imag \sum_{n=0}^\infty \sum_{m=0}^n \Biggl( c_{2n} \cH_0(\alpha,a_0) \frac{(2n)!}{(1+a_0 A)^{2n}} \frac{1}{m!(2n-2m)!} \nn \\
&\qquad\qquad \qquad \qquad\qquad  \times \Bigl\{ [ A(\alpha^2 - a_0)-1] (1+a_0 A)\Bigr\}^m (2\alpha z_0)^{2n-2m} \nn  \\
&\qquad \qquad + \frac{i\gamma_E}{\beta} c_{2n+1} \cH_1(\beta,b_0) \frac{(2n+1)!}{(1+b_0A)^{2n}} \frac{1}{m!(2n-2m+1)!} \nn \\
&\qquad\qquad \qquad \qquad\qquad  \times \Bigl\{ [ A(\beta^2 - b_0)-1] (1+b_0 A)\Bigr\}^m (2\beta z_0)^{2n-2m}\biggr)\nn
\,,\end{align} 
where the coefficients $c_{n}$ are given by \eq{HermiteCoefficients}. As many terms in the sum over $n$ may be included to achieve the numerical accuracy desired. In practice, we include the first few terms in the sum over $n$ (three or four even $c_{2n}$ and three odd $c_{2n+1}$ coefficients), which gives us percent level accuracy for the cross section in the perturbative resummation region. Although the coefficients $c_{n}$ still need to be evaluated numerically, we note that they depend only on properties of the pure function $\Gamma(1-ix)^2$ and need be determined only once (\eq{cn}). Our Hermite expansion is applied only to this function, which arises solely from the Bessel function $J_0(z)$ which appears in the factorization convolution in \eq{xsec}. The dependence on all the physical variables, such as $q_T,Q$, and the scales $\mu_{L,H},\nu_{L,H}$, enters through the evolution exponent in \eqs{VGaussian}{ACUeta} and is captured analytically by \eq{IbHermite}. For other processes or observables, the same basis and coefficients we used and resultant analytic integration \eq{IbHermite}, should apply, though the number of terms needed to get good convergence (determined by the width of the evolution exponent \eq{VGaussian})  may vary.

\subsection{Fixed order terms}
\label{ssec:fixedorderlimit}

\subsubsection{Fixed-order prefactors in resummed expression}

At NNLL (or NLL$'$) and higher orders, fixed order logarithmic terms of the form $\ln^k(\mu_L b_0)$ appear in the prefactor $\wt F$ multiplying the resummed exponent in the integrand of \eq{Ibdef}. 
Explicitly, $\wt F$ can be expanded:
\be
\label{Fexpansion}
\wt F(b,x_1,x_2,Q;\mu_L,\nu_L^*,\nu_H) = \sum_{n=0}^\infty \sum_{k=0}^{2n} \Bigl(\frac{\as(\mu_L)}{4\pi}\Bigr)^n \wt F^{(n)}_k \ln^k\mu_L b_0\,,
\ee
so that we can isolate integrals of each term in the form \eq{Ibk}. We can build each coefficient $\wt F^{(n)}_k$  out of the coefficients of the soft function and TMDPDFs  in their expansions \eqs{Sn}{Iijexpansion}, reading off the coefficients of each power of $\ln\mu_L b_0$. In doing so we must take into account extra powers of $\ln\mu_L b_0$ in the soft function due to the shifted scale $\nu_L^* = \nu_L(\mu_L b_0)^{-1+p}$ in \eq{nuLstar} that we use. Then we have for the terms we need up to NNLL accuracy,
at tree level:
\begin{align}
\label{Ftree}
\wt F^{(0)}_0 = f_i(x_1,\mu_L) f_{\bar i}(x_2,\mu_L)\,,
\end{align}
where $i,\bar i = g$ for Higgs production and $i=q$ for DY,
and at $\cO(\as)$:
\begin{align}
\label{Foneloop}
\wt F^{(1)}_2 &= f_i(x_1,\mu_L) f_{\bar i}(x_2,\mu_L)  \zed_S \frac{\Gamma_0}{2} (1-1) = 0 \\
\wt F^{(1)}_1 &= f_i(x_1,\mu_L) f_{\bar i}(x_2,\mu_L) \biggl( \zed_S \Gamma_0 \ln\frac{\nu_L}{\mu_L} + \zed_f \Gamma_0 \ln\frac{\nu_H^2}{Q^2} +  2\gamma_f^0\biggr) \nn \\
&\quad - [2P_{ij}^{(0)}\otimes f_j(x_1,\mu_L)] f_{\bar i}(x_2,\mu_L) -  f_i(x_1,\mu_L) [2P_{\bar i j} \otimes f_j(x_2,\mu_L)] \nn \\
\wt F^{(1)}_0 &= f_i(x_1,\mu_L) f_{\bar i}(x_2,\mu_L) c_{\wt S}^1 + [I_{ij}^{(1)}\otimes f_j(x_1,\mu_L)] f_{\bar i}(x_2,\mu_L) + f_i(x_1,\mu_L)[ I_{\bar i j}^{(1)}\otimes f_j(x_2,\mu_L)]\,. \nn
\end{align}
Note that $\wt F_2^{(1)}$ vanishes due to the extra term from the shift $\nu_L\to \nu_L^*$ since the double log term was promoted from the fixed-order soft function to the exponent in \eq{VSNLL}, as designed (though this cancellation will no longer be exact once we implement profile scales in the next subsection). Up to NNLL accuracy, the only $\cO(\as^2)$ terms we need come from the the $\as^2$ piece of the soft function induced by $\nu_L\to\nu_L^*$, and the $\cO(\as^2)$ terms in the $V_\beta$ piece of $\wt F$ in \eq{fixedorderfactor}, which from \eq{Vbeta} we see has the expansion
\be
V_\beta(\nu_L^*,\nu_H;\mu_L) = 1+ \sum_{n=2}^\infty \sum_{k=1}^{n+1} \Bigl(\frac{\as(\mu_L)}{4\pi}\Bigr)^n V^{(n)}_k \ln^k \mu_L b_0 \,,
\ee
where at $\cO(\as^2)$,
\begin{align}
V^{(2)}_3 &= \zed _S\Gamma_0 \beta_0 \Bigl(1-\frac{1}{2}\Bigr) = \frac{1}{2} \zed _S\Gamma_0 \beta_0 \\
V^{(2)}_2 &= \zed_S \Gamma_0 \beta_0 \ln\frac{\nu_H}{\nu_L}  \nn \\
V^{(2)}_1 &= 0\,. \nn
\end{align}
The $V_3^{(2)}$ coefficient actually just multiplies a small log $\ln^2\mu_L b_0$, so strictly at NNLL we can drop it.
Then the only relevant piece of $\wt F^{(2)}$ we would need at NNLL accuracy is
\begin{align}
\label{Ftwoloop}
\wt F^{(2)}_2 &= f_i(x_1,\mu_L) f_{\bar i}(x_2,\mu_L)  \Bigl(V_{2}^{(2)}  - \zed_S\Gamma_0\beta_0 \ln\frac{\nu_H}{\nu_L}\Bigr) = 0\,,
\end{align}
where the second term in $F_2^{(2)}$ came from the shift $\nu_L\to \nu_L^*$ in the one-loop soft function. So at NNLL, all pieces of $\wt F^{(2)}$ (\ie the fixed-order $\cO(\as^2)$ terms) vanish or can be dropped.

The result of integrating each of the powers of $\ln\mu_L b_0$ in \eq{Fexpansion} inside of the integrals $I_b^k$ in \eq{Ibk} can be readily obtained from the analytic resummed result for $I_b^0$ \eq{IbHermite}, by taking derivatives, using:
\begin{align}
\label{logderivatives}
\ln^k(\mu_L b_0) e^{-A \ln^2( \Omega b)} &= \ln^k (\mu_L b_0) e^{-A \ln^2( \mu_L b_0 \chi)} = \left[ \hat \partial_{\chi} \right]^{k} e^{-A \ln^2( \mu_L b_0 \chi)}\,, 
\end{align}
where we used \eq{ACUeta} and we have defined
\bea
\hat \partial_{\chi} = -\frac{1}{2A} \frac{\partial}{\partial \ln \chi}-  \ln \chi\,.
\eea
Using the final expression \eq{IbHermite} for $I_b^0$ we can now write 
\bea\label{Ibk-cH}
I_b^k &=& \left[ \hat \partial_{\chi} \right]^k I_b^0 \nn\\
 &=&  \frac{2}{\pi q_T^2}\sum_{n=0}^{\infty} \text{Im}\Big \{c_{2n} \, \left[ \hat \partial_{\chi} \right]^k\mathcal{H}_{2n}(\alpha,a_0)+ \frac{i\gamma_E}{\beta}c_{2n+1} \,  \left[ \hat \partial_{\chi} \right]^k\mathcal{H}_{2n+1} (\beta,b_0) \Big\}
\eea
In the expression \eq{Hnresult} for $\cH_{n}$, the variable $\chi$ only appears inside of 
\be
\label{Lchi}
L = \ln \left(\frac{2\Omega}{q_T}\right) = \ln\frac{\mu_L e^{\gamma_E}}{q_T} + \ln \chi\,,
\ee
which appears inside $z_0$ and $\cH_0$ in \eq{Hnresult}.
So we can write 
\bea
\label{partialchiL}
 \hat \partial_{\chi} =  -\frac{1}{2A} \partial_{L} - \ln\chi\,.
\eea

To construct fixed order terms up to order $k$, we need up to $k^{\rm th}$ derivatives of the $L$ dependent pieces that make up $I_b^k$. 
There is very simple recursion relation satisfied by the derivatives of $\cH_n$:
\be
\label{recursion}
-\frac{1}{2A}\partial_L \mathcal{H}_{n}(\alpha,a_0) =  -\frac{i z_0}{A(1+a_0 A)} \cH_{n}(\alpha,a_0) + i\frac{n \alpha}{1+a_0 A} \mathcal{H}_{n-1}(\alpha,a_0)\,,
\ee
which can be derived either from the original definition \eq{cH2n} of $\cH_n$, or its explicit all-orders result in \eq{Hnresult}. These proofs are given in \appx{Hderivative}.
Derivatives to arbitrary order $(\hat\partial_\chi)^k\cH_n$ can then be obtained by repeatedly applying this relation \eq{recursion}:
\bea \label{dcH}
\hat \partial_{\chi}  \cH_n &=&  L_1\, \cH_{n}+ \frac{i \alpha}{1+a_0 A} \,n\cH_{n-1}
\,,\\
\left[ \hat \partial_{\chi} \right]^2  \cH_n  
&=& 
\left[L_1^2 +\frac{a_0}{2(1+a_0 A)} \right] \cH_n +2L_1\frac{i \alpha}{1+a_0 A}n \cH_{n-1} +\left[ \frac{i \alpha}{1+a_0 A} \right]^2 n(n-1) \cH_{n-2}
\,,\nn\\
\left[ \hat \partial_{\chi} \right]^3 \cH_n
 &=&L_1 \left[L_1^2 +\frac{3 a_0}{2(1+a_0 A)} \right]\, \cH_n 
	+ 3\left[L_1^2 +\frac{ a_0}{2(1+a_0 A)} \right] \frac{i \alpha}{1+a_0 A}n\cH_{n-1} 
\nn\\ &&
	+3 L_1 \left[ \frac{i \alpha}{1+a_0 A} \right]^2 n(n-1) \cH_{n-2}
	+\left[ \frac{i \alpha}{1+a_0 A} \right]^3 n(n-1)(n-2) \cH_{n-3}  
\,,\nn
\eea
etc., where
\be
L_1 \equiv -\frac{i z_0}{A(1+a_0 A)} - \ln\chi 
		=\frac{1}{1+a_0 A}\left[  \ln\frac{\mu_L e^{\gamma_E}}{q_T}-\frac{i\pi}{2}-a_0 A \ln\chi \right]
\,.
\ee

With these expressions, we can now write the resummed cross section \eq{resummedIb} in $q_T$ space in terms of the above results of integrating $I_b$ and $I_b^k$ in \eqs{Ibdef}{Ibk},
\begin{align}
\label{thefinalresult}
\frac{d\sigma}{dq_T^2 dy} &= \frac{1}{2}\sigma_0 C_t^2(M_t^2,\mu_T)H(Q^2,\mu_H) U(\mu_L,\mu_H, \mu_T) C(\mu_L,\nu_L,\nu_H) \\
&\quad\times \sum_{n=0}^\infty \sum_{k=0}^{2n} \Bigl(\frac{\as(\mu_L)}{4\pi}\Bigr)^n \wt F^{(n)}_k(x_1,x_2,Q;\mu_L,\nu_L,\nu_H)I_b^k(q_T;\mu_L,\nu_L,\nu_H;\alpha,a_0;\beta,b_0)\,. \nn
\end{align}
where the integrals $I_b^k$ are given in final evaluated form in \eq{Ibk-cH} with the first few derivatives $(\partial_\chi)^k \cH_n$ given by \eq{dcH}. The calculation of these integrals have formed the bulk of this \sec{analytic} and constitute one of the main results of this paper.

To turn \eq{thefinalresult} into a final prediction for the perturbative $q_T$ cross section, we still need to match it on to the fixed-order prediction of full QCD for the large $q_T$ tail. We turn our attention now to this task.

\subsubsection{The large $q_T$ limit}
\label{ssec:largeqT}

In the large $q_T\sim Q$ limit the resummation is turned off and this corresponds to the $A\to 0$ limit in the soft evolution of \eq{Ib-L}. In this limit the resummed part of the cross section \eq{thefinalresult} should reduce to a sum of the fixed-order singular terms in the cross section.
Since we evaluate the $I_b^k$'s in \eq{thefinalresult} using an expansion in Hermite polynomials giving \eqs{IbHermite}{Ibk-cH}, it is not obvious whether these expansions, when truncated to our desired number of terms, maintain their accuracy in the fixed order limit. We check the accuracy here.
In the $A\to 0$ limit, the integrals of the even and odd Hermite polynomials reduce to
\bea\label{Ibk-limit}
I_b^0 &\to& 0
\,,\nn\\
I_b^1 &\to& -\frac{1}{q_T^2} f_1  
\,,\nn\\
I_b^2 &\to& -\frac{2}{q_T^2} 
\left[ f_1 \, \ln\frac{\mu_L e^{\gamma_E}}{q_T} -f_2 \gamma_E \right] 
\,,\nn\\
I_b^3 &\to&-\frac{3}{q_T^2}
\left[f_1 \ln^2\frac{\mu_L e^{\gamma_E}}{q_T}  -2 f_2 \gamma_E  \ln\frac{\mu_L e^{\gamma_E}}{q_T} + f_3   \right]
\,,
 \eea
where $f_k$ is a linear combination of the coefficients $c_n$ in the series expansion :
\bea \label{fn}
f_1 &=& \sum_{n=0}^{\infty} \frac{(-1)^n (2n)!}{n!}c_{2n} \approx c_0-2c_2+12 c_4 -120 c_6\approx 0.998051
\,,\\
f_2 &=& \sum_{n=0}^{\infty} \frac{(-1)^n (2n+1)!}{n!}c_{2n+1}\approx c_1-6 c_3 +60 c_5 \approx  0.99950
\,,\nn\\
f_3 &=& \sum_{n=0}^{\infty} \frac{(-1)^n (2n)!}{n!}\left(\gamma_E^2+n \alpha^2\right)  c_{2n}
\approx \gamma_E^2 (c_0-2 c_2+12 c_4-120 c_6)-2\alpha^2( c_2-12 c_4+120c_6)   \nn \\
& & \qquad\qquad\qquad\qquad\qquad\qquad\quad\ \ \approx 0.2828
\nn\,,
\eea
where we used the values in \eqs{alphabeta}{cn}.
We can compare to the result of taking the $A\to 0 $ limit of the exact expression for $I_b^k$ in \eq{Ibk} before expanding the Bessel function in a series representation. Using \eqss{Ibk}{Ib-L}{logderivatives}, we obtain
\be
\label{Ibkexact}
I_{b,\text{exact}}^k
=
\frac{2}{i q_T^2} \frac{1}{\sqrt{\pi A}}  \int  dt\,  f(t)   \left[ \hat \partial_\chi \right]^k   e^{ \frac{(1+t)^2}{A}-2L (1+t) }
\ee
Using \eq{partialchiL}, we can compute
\be
\label{chiderivative}
\hat\partial_\chi e^{ \frac{(1+t)^2}{A}-2L (1+t) } = \Bigl( - \frac{1}{2A} \partial_L - \ln\chi\Bigr) e^{ \frac{(1+t)^2}{A}-2L (1+t) } = \Bigl(\frac{1+t}{A} -\ln\chi\Bigr)e^{ \frac{(1+t)^2}{A}-2L (1+t) } \,,
\ee
which we can re-express as a derivative with respect to $t$,
\be
\label{tderivative}
\Bigl(\frac{1+t}{A} -\ln\chi\Bigr)e^{ \frac{(1+t)^2}{A}-2L (1+t) } = \Bigl(\frac{1}{2}\frac{d}{dt} + \ln\frac{\mu_L e^{\gamma_E}}{q_T}\Bigr)e^{ \frac{(1+t)^2}{A}-2L (1+t) } \,,
\ee
where we also used \eq{Lchi} in the last equality. \eqs{chiderivative}{tderivative} then allow us to make the replacement in \eq{Ibkexact}:
\bea
\left[ \hat \partial_\chi \right]^k  &\to& \left[ \frac{1}{2}\frac{d }{d  t} +\ln\frac{\mu_L e^{\gamma_E} }{q_T}\right]^k 
= \sum_{\ell=0}^{k} \binom{k}{\ell}   \left[\ln\frac{\mu_L e^{\gamma_E} }{q_T}\right]^{k-\ell}  \left[ \frac{1}{2}\frac{d }{d  t}\right]^\ell
\,, \eea
Then, integrating repeatedly by parts in \eq{Ibkexact}, we obtain
\be
I_{b,\text{exact}}^k = 
\frac{2}{iq_T^2} \frac{1}{\sqrt{\pi A}} \sum_{\ell=0}^k {{k}\choose{\ell}} \frac{(-1)^\ell}{2^\ell} \int dt \biggl[\frac{d^\ell}{dt^\ell}  f(t)\biggr]\left[\ln\frac{\mu_L e^{\gamma_E}}{q_T} \right]^{k-\ell}e^{ \frac{(1+t)^2}{A}-2L (1+t) }
\,,\label{Ibkexact-parts} 
\ee
We can now take the $A\to 0$ limit in \eq{Ibkexact-parts}, and in so doing turn the Gaussian into a Dirac delta function:
\be\label{basicI}
 \lim_{A\to 0} \frac{1}{\sqrt{\pi A}}  \int  dt\,    e^{ \frac{(1+t)^2}{A}-2L (1+t) } \left[ \frac{d }{d  t}\right]^\ell f(t)   
 =
i \int  dx\,   \delta (x-i-ic)  \left[ \frac{d }{id  x}\right]^\ell f(c+ix)   
= i  \frac{d ^\ell  }{d  t^\ell} f (-1)\,,
\ee
recalling we pick $c=-1$ in $t=c+ix$, and 
where we used the Dirac delta identity:
\be
\lim_{A\to 0} \exp[-x^2/A]= \sqrt{\pi A} \,\delta (x)\,. 
\ee
Thus, in the $A\to 0$ limit,
\be
I_{b,\text{exact}}^k \to 
\frac{2}{q_T^2} \sum_{\ell=0}^k {{k}\choose{\ell}} \frac{(-1)^\ell}{2^\ell}  \biggl[\frac{d^\ell}{dt^\ell}  f(t)\biggr]_{t=-1}\left[\ln\frac{\mu_L e^{\gamma_E}}{q_T} \right]^{k-\ell}
\,.\label{Ibkexact-limit} 
\ee
We compute the exact derivatives:
\be \label{dfdt}
\frac{d ^\ell }{d  t^\ell} f(-1)  =\left. \frac{d ^\ell }{d  t^\ell} \frac{\Gamma(-t) }{\Gamma(1+t)}\right|_{t=-1}
=
\begin{cases}
0 & \text{for } \ell=0
 \\
1 & \text{for } \ell=1
 \\
4\gamma_E  & \text{for } \ell=2
\\
12\gamma_E ^2 & \text{for } \ell=3
\end{cases}
\ee
Inserting \eq{dfdt} into \eq{Ibkexact-limit},
we have
\bea \label{Ibkexact-final}
I_{b,\text{exact}}^0 &\to& 0
\,,\nn\\
I_{b,\text{exact}}^1 &\to& -\frac{1}{q_T^2}
\,,\nn\\
I_{b,\text{exact}}^2 &\to& -\frac{2}{q_T^2} 
\left[  \ln\frac{\mu_L e^{\gamma_E}}{q_T} - \gamma_E \right] 
\,,\nn\\
I_{b,\text{exact}}^3 &\to&-\frac{3}{q_T^2}
\left[\ln^2\frac{\mu_L e^{\gamma_E}}{q_T}  -2\gamma_E  \ln\frac{\mu_L e^{\gamma_E}}{q_T} + \gamma_E^2   \right]
\,,
\eea
Comparing \eq{Ibkexact-final} to \eq{Ibk-limit}, we find exact values of $f_k$ which are 1,1, $\gamma_E^2\approx 0.33317$ for $k=1,2,3$
and the approximate values in \eq{fn} agree better than 1\% for $k=1,2$ and agree within about 15\% for $k=3$.
Note that the term $I_{b,k}$ at $k=3$ is the $O(\as^2)$ contribution induced by our prescription in \ssec{final} as it multiplies the coefficient $\wt F^{(2)}_3$ in \eq{Ftwoloop},which is suppressed by another order of $\as$ at the cross section level.  
Taking this into account, the error in $f_3$ itself induces only an order of magnitude smaller error in the cross section, much smaller than the total theoretical error at NNLL accuracy, and so we need not be concerned about it. If desired, one can go beyond this accuracy by including higher-order Hermite polynomials in the computation of $I_b^k$, and thus of $f_k$. 

 This means that the fixed order series probes very narrowly the behavior of $f(-1+ix)$ near $x=0$, and the more accurate our expansion is near $x=0$, the higher the order in $\alpha_s$ that we can reproduce accurately in the fixed order (high $q_T$) regime. Practically, we would like to reproduce the one loop fixed order cross section since we would be matching our NNLL predictions to that result in the high $q_T$ region, which means we need our expansion to match the exact result up to the second order in the Taylor expansion of $f$, \ie $\ell=0, 1, 2$.

\subsubsection{Matching to  the fixed order cross section} \label{ssec:matching}

The resummed formula \eq{thefinalresult} accurately predicts the spectrum for relatively low (but not too low) values of $q_T$. At large $q_T\sim Q$, the non-singular terms in $q_T$ become just as big as the logs of $q_T/Q$ themselves, and we should use the fixed-order perturbative expansion for the cross section 
there.  It has been shown in the context of $B\rightarrow X_s +\gamma$ \cite{Ligeti:2008ac}, thrust \cite{Abbate:2010xh} as well as the Higgs jet veto calculation \cite{Berger:2010xi}, that if one does not turn off the resummation then one can over estimate the cross section, in the region where fixed order perturbation theory should suffice,  by an amount which goes beyond the canonical error band in the fixed order result.
This overshoot happens despite the fact that the resummed terms are formally sub-leading
in the expansion.  The reason for this overshoot has  been shown \cite{Ligeti:2008ac, Berger:2010xi, Abbate:2010xh}  to be due to the fact that
there are cancellations between the singular and non-singular terms in the tail region and that
this cancellation will occur only if the proper scale is chosen in the logarithms. We can smoothly combine the resummed and fixed-order formulas by turning off resummation in \eq{thefinalresult} in the high $q_T$ region using profiles in $\mu $ and $\nu$ \cite{Abbate:2010xh}
and match the result to the one loop ($\cO(\alpha_s)$) full theory cross section.

Our resummation as given in \eq{resummedIb} (and evaluated in \eq{thefinalresult}) is neatly divided into two parts, the hard function which runs in $\mu$ and functions $C$ and $I_b$ which implement $\nu$ running.
To turn off resummation at large $q_T$ we implement the following profiles for both $\mu, \nu$ in \eq{resummedIb}:
\begin{subequations}
\begin{align} \label{run}
\mu_L \to \mur (q_T) &= \mu_L(q_T) ^{1-\zeta(q_T)}\, \abs{\mu_H}^{\zeta(q_T)}  \,, \\
\nu_L^* \to \nur (q_T) &= \nu_L^*(b_0;\mur)^{1-\zeta(q_T)}\, \nu_H^{\zeta(q_T)} \\
&=   [\nu_L(\mur b_0)^{-1+p}]^{1-\zeta(q_T)} \nu_H^{\zeta(q_T)} \nn 
\end{align}
\end{subequations}
The function $\zeta(q_T)$ is chosen so that at low values of $q_T$ where resummation is important, its value is 0 while for values near $Q$ it approaches 1. 
$\mu_L(q_T) $ is given by \eq{scale} and illustrated in \fig{scaleL},  and we have indicated $\nur$ so that inside \eq{nuLstar} for $\nu_L^*$ the $\mu_L$ is also set to $\mur$. This is designed so that not only resummation of logs of $\mu_L/\mu_H$ and $\nu_L/\nu_H$ is turned off as $\zeta\to 1$ but also the shifting of logs of $\mu_L b_0$ into the soft rapidity exponent in \eq{Vstar} as we move out of the resummation region. The exponent $p$ defined in \eq{eq:n} is also now evaluated at $\mur$.

\begin{figure}
\centerline{\scalebox{.55}{\includegraphics{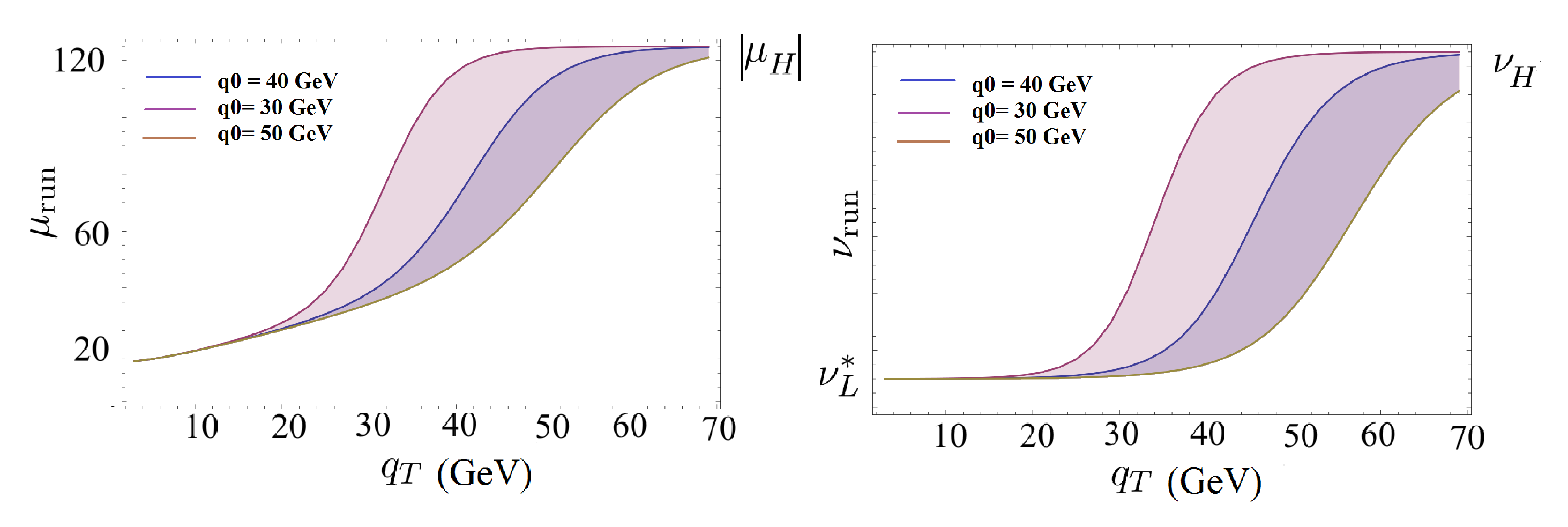}}}
\vskip-0.5cm
\caption[1]{Effect of variation of the transition point $q_0$ in the profile functions for $\mu$, $\nu$  for the case of the Higgs $q_T$ spectrum. $\abs{\mu_H}$ here is $M_h$ = 125 GeV.}
\label{profile_var} 
\end{figure}

\begin{figure}
\centerline{\scalebox{.55}{\includegraphics{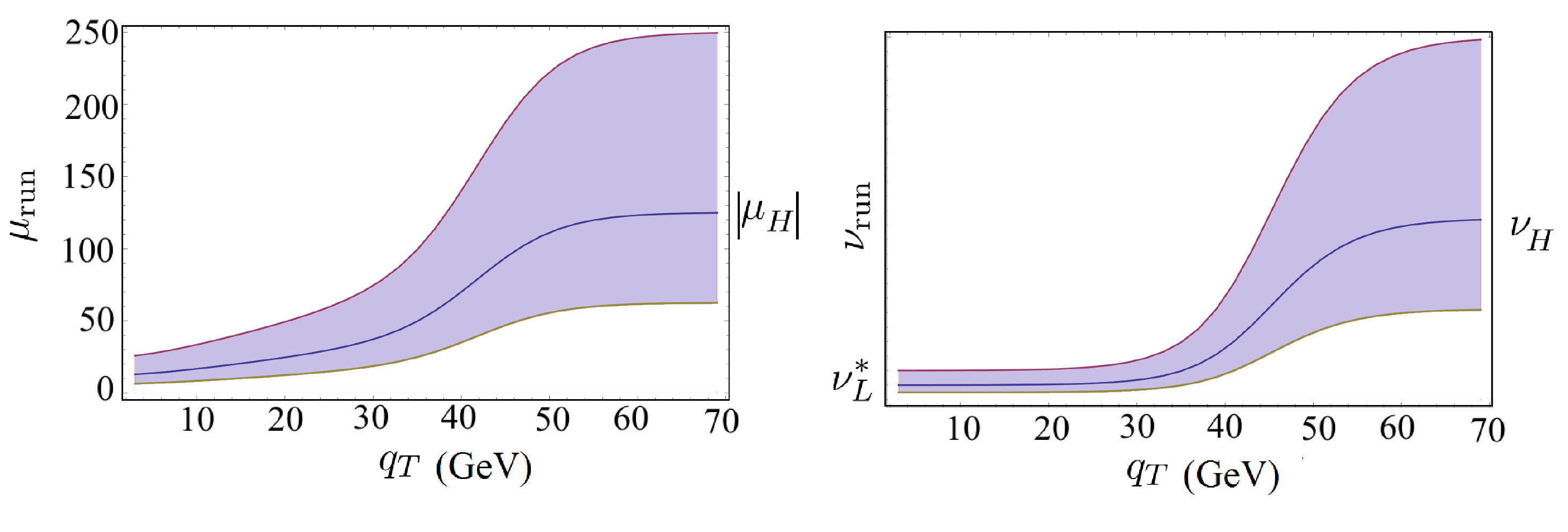}}}
\vskip-0.5cm
\caption[1]{Central profile ($q_0$= 40 GeV) in $\mu$, $\nu$ along with the effect of the scale variations \eq{scalevariations} for the case of the Higgs $q_T$ spectrum. $\abs{\mu_H}$ here is $M_h$ = 125 GeV.}
\label{fig:profile} 
\end{figure}

The choice of  $\zeta(q_T)$ that we make for this function is:
\bea
\zeta(q_T) =  \frac{1}{2} \left(1+\tanh \left[\rho\left(\frac{q_T}{q_0}-1\right)\right]\right)
\eea
Here $q_0$ determines the central value for the transition and $\rho$ determines its rate. 
In practice we use $\rho=3$. The value of $q_0$ is determined by the scale at which the non-singular pieces become as important as the resummed singular cross section. The profiles in $\mu$ and $\nu$ for the case of the Higgs spectrum are shown in \fig{profile}. (Here $q_0$ has been chosen to be 40 GeV for the central profile). We also probe the effect of varying the profiles by varying the value of $q_0$, in our case, between 30 GeV and 50 GeV as shown in Fig. \ref{profile_var}. For each of these profiles, we consider the scale variation by a factor of 2 for $\mu ,\nu$ (Fig. \ref{fig:profile})
For the case of the DY, we use a similar value for $q_0$.

This procedure is straightforward to implement for the hard function running, let's see what effect it has on the $\nu$ running. Going back to the $\nu$ running, our exponent in \eq{UVdef} with the scale choice \eq{nuLstar} for $\nu_L$ looks like 
\bea
V =  \exp\biggl[ \gamma_\nu^S(\mur) \ln\left( \frac{\nu_H}{\nu_L^*} \right) \biggr]  
\eea
Putting in our choices for $\nu_L$ we have 
\bea
 V =  \exp\biggl[ \gamma_{\nu}^S(\mur) \ln\left( \frac{\nu_H}{(\nu_L^*)^{1-\zeta} \nu_H^\zeta} \right) \biggr] = \exp\biggl[(1-\zeta) \gamma_\nu^S(\mur) \ln\left( \frac{\nu_H}{\nu_L^*} \right) \biggr]
\eea
So the effect is to merely multiply the argument of the exponent by a factor of $1-\zeta(q_T)$. Notice that we have also put in $\mur$ in the $\nu$ anomalous dimension since it is a function of $\mu_L$. 
 This is basically the same as $A \rightarrow (1-\zeta)A$ and $\gamma_{RS}\to (1-\zeta)\gamma_{RS}$ in the expressions \eqs{VGaussian}{ACUeta} for the rapidity evolution factor $V_\Gamma$ inside $I_b$ in \eq{Ibdef}. Thus, using these profile scales simply modifies the definitions in \eq{ACUeta} to:
\begin{align} \label{ACprofile}
A(\mu_L,\nu_L,\nu_H;\zeta) &= -(1-\zeta)\zed_S\Gamma[\as(\mur)] (1-p) \\
&= -(1-\zeta)\zed_S \Gamma[\as(\mur)] \frac{1}{2} \Bigl[ 1 + \frac{\as(\mur)\beta_0}{2\pi} \ln\frac{\nu_H}{\nu_L}\Bigr] \nn \\
C(\mu_L,\nu_L,\nu_H;\zeta) &= \exp\left\{ A(\mu_L,\nu_L,\nu_H;\zeta) \ln^2\chi + (1-\zeta)\gamma_{RS} [\as(\mur)] \ln \frac{\nu_H}{\nu_L}\right\}  \,, \nn
\end{align}
and $\Omega,\chi$ in \eq{ACUeta} remain unchanged (with the exception that $\mu_L\to \mur$). The final expressions for $I_b^0$ and $I_b^k$ given by \eqs{IbHermite}{Ibk-cH} are modified accordingly.   So what we are doing is smoothly taking the limit $A \rightarrow 0$ (and $\gamma_{RS}\to0$)  as we enter the high $q_T$ region. We already know from the previous section (see \eq{Ibk-limit}) that in this limit $I_b^0$ goes to 0 smoothly, which means the resummation factor is 0, and each $I_b^k$ goes to the appropriate fixed order limit as well.

The central profiles we choose for the scales are:
\be\label{central}
\left\{\mu_H,\, \nu_H,\,\mu,\, \nu\right\}_\text{central}=\left\{iQ,\, Q,\, \mur(q_T),\, \nur(q_T) \right\}
\ee
$\mu_H$ is chosen as $i Q$, which implements the resummation of enhanced $\pi^2$ terms at each order in $\alpha_s$, which improves the perturbative convergence \cite{Ahrens:2008qu,Ahrens:2008nc}. While implementing the hard function running, we take the absolute value of $U_H$. 

We make six scale variations for $\mu, \nu$ by a factor of 2 about their central values for each of the profiles shown in Fig. \ref{profile_var}:
\begin{gather}
\frac{\mu_L}{2} \leftarrow \mu_L \rightarrow 2\mu_L \nn \\
\frac{\nu_L}{2} \leftarrow \nu_L\rightarrow 2\nu_L 
\label{scalevariations}
\\
\Bigl(\frac{\mu_L}{2}, \frac{\nu_L}{2}\Bigr)\leftarrow  (\mu_L,\nu_L)\rightarrow (2\mu_L, 2\nu_L)\,, \nn
\end{gather}
 to estimate higher order terms as shown in \fig{profile}. In the first two variations in \eq{scalevariations} $\mu_L$ and $\nu_L$ are each varied independently, while in the last one, they are varied simulatneously. The anticorrelated variation ($(\mu_L/2, 2\nu_L)\leftarrow (\mu_L,\nu_L)\rightarrow ( 2\mu_L, \nu_L/2)$) is not included to avoid double counting \cite{Neill:2015roa}.

We take the largest envelope in $\mu,\nu$ and $q_0$ variations as our estimate of the error band.
In principle, we can make variations at either end of the running (i.e. at the high scales $\mu_H$ and $\nu_H$,
or the low scales $\mu_L$ and $\nu_L$). In practice, since the value of $\as$ is larger at lower scales, the scale
variation at the low scale of running yields the largest error band.
There is implicit $q_T$ dependence in $z(q_T)$, $\mur$, and $\nur$.

Now we are ready to give results for our final resummed cross section matched to the fixed-order QCD result. We give an explicit expression of the transverse spectrum which can be written as sum of resummed and nonsingular parts.
We match the resummed and perturbative QCD results such that final result is valid in both small are large $q_T$ regions.
This can be done by adding the resummed result to nonsingular terms  where
all logarithmic singular terms reproduced by EFT results in \eq{sing} are subtracted from the perturbative result
\be
\sigma = \sigma^\text{res} + \sigma^\text{ns}\,,\quad \sigma^\text{ns}(\mu) = \sigma^\text{pert} (\mu) -\sigma^\text{sing}(\mu)
\,, \label{nsdef}
\ee
where the differential in $q_T^2$ and rapidity $y$ is implied. The fixed-order $\sigma^{\text{ns}}$ is evaluated at a scale $\mu$, which we choose to be equal to $\abs{\mu_H}$, the high scale to which the profile \eq{run} goes for large $q_T$.
In this section we give explicit expressions for  $\sigma^\text{res}$ and  $\sigma^\text{sing}$, and \appx{per} gives $ \sigma^\text{pert}$.

The resummed cross section $\sigma^\text{res}$ is given by \eq{thefinalresult}, with the modifications from the scale profiles \eqss{run}{ACprofile}{scalevariations} implemented. Meanwhile the expansion \eq{Fexpansion} of the fixed-order coefficient $\wt F$ in \eq{Ibdef} becomes an expansion in logs of $\mur b_0$ since the $\hat\partial_\chi$ derivatives in \eq{logderivatives} bring down powers of $\ln \mur b_0$ inside the $b$ integrand:
\be
\label{Frunexpansion}
\wt F(b,x_1,x_2,Q;\mur,\nur,\nu_H) = \sum_{n=0}^\infty \sum_{k=0}^{2n} \Bigl(\frac{\as(\mur)}{4\pi}\Bigr)^n \wt F^{(n)}_k \ln^k\mur b_0\,,
\ee
where the coefficients $\wt F_k^{(n)}$ which were given to $\cO(\as^2)$ at NNLL accuracy in \eqss{Ftree}{Foneloop}{Ftwoloop} for $\zeta = 0$ (the pure resummation region) now become:
\begin{align}
\label{Fprofile0}
\wt F^{(0)}_0 &= f_i(x_1,\mur)f_{\bar i}(x_2,\mur) \\[12pt]
\label{Fprofile1}
\wt F^{(1)}_2 &= f_i(x_1,\mur)f_{\bar i}(x_2,\mur) \zeta\zed_S \frac{\Gamma_0}{2} \\
\wt F^{(1)}_1 &= f_i(x_1,\mur)f_{\bar i}(x_2,\mur) \biggl[ \zed_S \Gamma_0 \Bigl( (1\minus\zeta) \ln\frac{\nu_L}{\mu_L} + \zeta\ln \frac{\nu_H}{\abs{\mu_H}} \Bigr) + \zed_f \Gamma_0 \ln\frac{\nu_H^2}{Q^2}  + 2\gamma_f^0\biggr] \nn \\
&\quad  - [2P_{ij}^{(0)}\otimes f_j(x_1,\mur)] f_{\bar i}(x_2,\mur) -  f_i(x_1,\mur) [2P_{\bar i j} \otimes f_j(x_2,\mur)] \nn \\ 
\wt F^{(1)}_0 &= f_i(x_1,\mur) f_{\bar i}(x_2,\mur) c_{\wt S}^1 \nn \\
&\quad + [I_{ij}^{(1)}\!\otimes\! f_j(x_1,\mur)] f_{\bar i}(x_2,\mur) + f_i(x_1,\mur)[ I_{\bar i j}^{(1)}\!\otimes\! f_j(x_2,\mur)]\,.\, \nn
\end{align}
while the two-loop coefficients $\wt F_k^{(2)}$ remain zero at NNLL accuracy.
In the pure fixed order limit $\zeta=1$, the resummation is turned off as $\mur=\nur=|\mu_H|=\nu_H$, and we see the term containing explicit $\mu_L,\nu_L$ in \eq{Fprofile1} vanish.
The integrals $I_{b,k}$ take the fixed order form in \eq{Ibkexact-final} with $\mu_L\to \mu_H$.
In this limit, $\wt F_0^{(0,1)}$ actually does not contribute due to the fact that $I_b^0\to 0$.
Either by putting \eq{Fprofile1} and \eq{Ibkexact-final} together or by inverse transforming \eq{sing} to momentum space,
we obtain the singular part of the fixed-order cross section at $\cO(\as)$ as
\begin{align}\label{sigsing}
\frac{\sigma^\text{sing}}{\sigma_0}&=
 \biggl\{ \!  \delta(q_T^2)+\frac{\as(\mu)}{4\pi}\biggl(\! (c^1_H \plus c^1_{\tS} \plus C_i\pi^2) \delta(q_T^2)- \gamma_f^0 \biggl[\frac{1}{q_T^2}\biggr]_+ \! \! -\Gamma_0\biggl[\frac{\ln(q_T^2/Q^2)}{q_T^2}\biggr]_+ \biggr)\! \biggr\}f_i(x_1) f_{\bar i}(x_2)
\nn\\
& \quad +\frac{\as(\mu)}{4\pi}\biggl\{ \! \delta(q_T^2)\,[ I^{(1)}_{ij}\! \otimes \!  f_{j}](x_1)\, f_{\bar i}(x_2)+\biggl[\frac{1}{q_T^2}\biggr]_+ [P^{(0)}_{ij} \! \otimes\!  f_{j}](x_1)\, f_{\bar i}(x_2)+ x_1\! \leftrightarrow \! x_2 \! \biggr\} .
 \end{align}
The coefficients in the singular piece of the cross section can be found in \appx{fixedorder}.
Together with $\sigma^\text{pert}$ given in \appx{per}, \eq{sigsing} and \eq{nsdef} give the non-singular part of the cross section at $\cO(\as)$, which we add to \eq{thefinalresult} to obtain the final resummed and matched cross section up to NNLL$+\cO(\as)$.

\subsection{Final resummed cross section in momentum space}
\label{ssec:final}

Then, the resummed cross section in momentum space \eq{resummedIb}, using \eqss{VGaussian}{Ibk}{Ibk-cH} to express the integral $I_b$, implementing the profile scales \eq{run}, and matching onto the fixed-order QCD cross section for large $q_T$, is given by the expression:
\be
\label{thefinalfinalresult}
\boxed{
\begin{split}
\frac{d\sigma}{dq_T^2 dy} &= \frac{1}{2}\sigma_0 C_t^2(M_t^2,\mu_T)H(Q^2,\mu_H) U(\mur,\mu_H, \mu_T) C(\mur,\nur,\nu_H) \\
&\quad\times \sum_{n=0}^\infty \sum_{k=0}^{2n} \Bigl(\frac{\as(\mur)}{4\pi}\Bigr)^n \wt F^{(n)}_k(x_1,x_2,Q;\mur,\nur,\nu_H)I_b^k(q_T;\mur,\nur,\nu_H;\alpha,a_0;\beta,b_0) \\
&\quad + \frac{d\sigma_{\text{ns}}(\abs{\mu_H})}{dq_T^2 dy}\,.
\end{split}
}
\ee
This fairly compact expression still has many pieces to it. We provide a roadmap to where to find them all:

The basic cross sections $\sigma_0$ for Higgs production ($gg\to H$) and for DY ($q\bar q\to \gamma^*$)  are given by:
\be
\sigma_0^{\text{Higgs}} =\frac{M_h ^2}{576 \pi v^2 s}\,,\quad \sigma_0^{\text{DY}} =\frac{4 \pi (\alpha_{\text{em}}(Q))^2e_i^2}{3 N_c Q^2 s}\,,
\ee
where the vacuum expectation value $v^2=1/(\sqrt{2} G_F)$ is determined by the Fermi coupling $G_F\approx 1.1664\times 10^{-5}\, \GeV^{-2}$ and $e_i$ is the electric charge of quark flavor $i$.
The hard function $H$ is given by \eqss{Hexp}{Cexp}{Cn}, with the necessary one-loop non-cusp anomalous dimensions for DY and Higgs given by \eqs{gaHexp}{gaHexp-g} respectively and the one-loop constants given by \eq{c1}. The top matching coefficient $C_t^2$ for Higgs production is given by \eq{Ct} while it is set to 1 for DY. 
The $\mu$ (virtuality RG) evolution kernel $U(\mu_L,\mu_H,\mu_T)$ is given in \eq{UVdef}, with the parts of the exponent $K_{\Gamma,\gamma},\eta_\Gamma$ defined and expanded up to NNLL accuracy in \appx{def-K}. These all contain dependence on the cusp anomalous dimension, whose coefficients $\Gamma_n$ for DY and Higgs are given by \eq{Gacuspexp}. 

The factor $C(\mu_L,\nu_L,\nu_H)$, which came from expressing the exponentiated ``conformal'' part of the rapidity evolution kernel $V_\Gamma$ in \eq{Vstar} as a Gaussian in $\ln b$, is defined in \eq{ACUeta}, and expanded to NLL and NNLL accuracy in \appx{Gaussian}.

The coefficients $\wt F^{(n)}_k$ in the expansion \eq{Frunexpansion} of the $\wt F$ defined in \eq{fixedorderfactor} that contains the fixed-order terms in the soft function, TMDPDFs, and ``non-conformal'' part of the rapidity evolution kernel $V_\beta$ defined in \eq{VGammaVbeta} have been given above in \eqs{Fprofile0}{Fprofile1} up to the order to which we shall need them for NNLL accuracy.

\begin{figure}
\centerline{\scalebox{.55}{\includegraphics{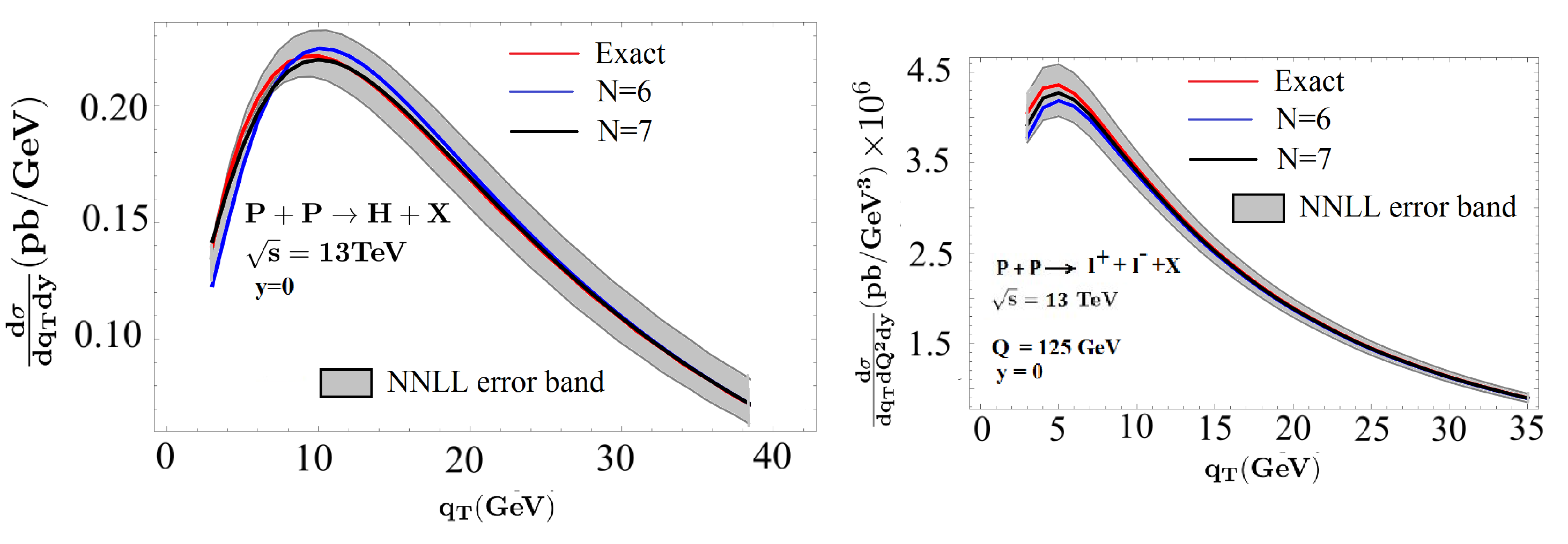}}}
\vskip-0.5cm
\caption[1]{Systematic improvement in the accuracy of Higgs (left) and DY (right) cross sections with increasing number of terms, differential in $y$ (at $y=0$) and $q_T$ (and $Q^2$ for DY, shown at $Q = 125\text{ GeV}$ for comparison). Exact (red) gives resummed cross section without Hermite expansion (\ie numerical $b$ integration). $N=6$ (blue) is the result with six terms in the Hermite expansion, three each for real and imaginary terms. $N=7$ (black) is the result with seven terms, four for real and three for imaginary. Here we plot only the purely resummed result, i.e. with no matching to the fixed order cross section.}
\label{fig:Hcompare} 
\end{figure}

The integrals $I_b^k$ which we defined in \eq{Ibk} are given in final evaluated form in \eq{Ibk-cH}, the calculation of which formed the bulk of this \sec{analytic} and is one of the main results of this paper. That result is in terms of derivatives \eq{dcH} of the integrals $\cH_n$ of Hermite polynomials against the Gaussian in \eq{cH2n} appearing in $I_b^0$ in \eq{IbHermite}, the explicit results for which are given by \eq{Hnresult}. Those integrals depend on the parameters $\alpha,a_0$ and $\beta,b_0$ that we used in the expansion of the function $\Gamma(1-ix)^2$ in terms of Gaussian-weighted Hermite polynomials in \eq{Hermiteexpansion}. For the numerical results in this paper, we used the values
\be
a_0 =2\gamma_E^2+\frac{\pi^2}{6}\approx 2.31129\quad 
\text{and}\quad b_0=\frac{2}{3}\gamma_E^2+\frac{\zeta_3}{3\gamma_E} +\frac{\pi^2}{6}\approx 2.56122
\,, \ee
which were given in \eq{a0b0}, and values of $\alpha,\beta$ given by $\alpha^2-a_0=4$ and $\beta^2-b_0=4$, which were given in \eq{alphabeta}. These choices are by no means unique; other values can also be chosen, which would then give different coefficients in the Hermite expansion. Our choices allowed sufficiently accurate representation of the exact function $\Gamma(1-ix)^2$ with an economical basis of a few terms each for the real and imaginary parts, the coefficients of which we gave in \eq{cn}. \fig{Hcompare} illustrates the agreement of our resummed cross section \eq{thefinalfinalresult} to NNLL accuracy with a total of six or seven (four real, three imaginary) terms in the Hermite expansion vs. the result of numerically integrating the exact $b$ integrand in \eq{resummedIb}. The truncation error lies well within the perturbative NNLL uncertainty band.

\tab{NkLL} gives the orders to which the anomalous dimensions and fixed-order pieces of \eq{thefinalresult} are to be truncated to achieve NLL, NNLL, \emph{etc.} perturbative accuracy in the resummed cross section. The central values of the resummation scales we choose in \eq{thefinalfinalresult} are given in \eq{central}, wherein the central values for $\mu_L=\nu_L$ are given by the solution of \eq{scale} and illustrated in \fig{scaleL}. The perturbative uncertainties are then estimated by performing the variations \eq{scalevariations}.

The final piece $d\sigma_{\text{ns}}/dq_T^2 dy$ is defined by \eq{nsdef} and can be obtained from the fixed-order results in \appx{per}, subtracting off the singular terms \eq{sigsing}.

It is good to remind ourselves at this point that the final formula \eq{thefinalfinalresult} is a threefold expansion as described in \ssec{intro3}: a perturbative expansion in $\as$ and resummed logs, an additional fixed-order expansion of the non-conformal piece $V_\beta$ of the rapidity evolution inside $\wt F$, and an expansion in Hermite polynomials in evaluating each $I_b^k$. It is thus quintessentially a ``formula'' according to a delightful definition we recently encountered: \emph{an expression given by an $=$ sign with a controlled error of known parametric form}.\footnote{A. Manohar, ``The Photon PDF,'' talk at \emph{Lattice QCD} workshop, Santa Fe, NM, Aug. 28--Sep 1, 2017, describing work in \cite{Manohar:2016nzj,Manohar:2017eqh}.}

\section{Remarks on nonperturbative region of $q_T$} 
\label{sec:non_pert}

The final result  \eq{thefinalresult} for the resummed cross section is obtained by doing an integral \eq{Ibdef} over a wide range of $b$. It is a reasonable assumption that the perturbative resummation will hold as long as $1/b > \Lambda_{\text{QCD}} \sim 500\text{ MeV}$. This is roughly the value at which we begin to see the Landau pole in the $b$ space resummation scheme (see \fig{nonpert}).  Beyond this value of $b$, we expect that nonperturbative effects will play a nontrivial role. Although it is not the goal of this paper to include or advance any new method to account for these nonperturbative effects, we will freely make some loose observations here, mainly delineating where we do and do not need to worry about them. While they are certainly important, we will have to leave their incorporation into our results to future work.

\subsection{Remarks on nonperturbative effects}

As we approach such low values of $1/b$, it is no longer possible to match perturbatively the beam functions  onto PDFs as in \eq{beammatching}. Instead we need to retain the complete transverse momentum dependent beam function, which is now a fully nonperturbative function that need to be fit from data.  In the $b$ space resummation scheme, where $\mu \sim 1/b$, giving the result \eq{inverseFT},  the running using the perturbative anomalous dimension will work only for $ 1/b \ll \Lqcd$. A sensible thing to do in this scheme is to freeze the resummation at $ 1/b \sim 2 \Lqcd$ and evaluate the beam function at this frozen low scale which is still perturbative. Interestingly, for the momentum resummation scheme leading to \eq{thefinalresult}, while we still need to replace the PDF's with the full nonperturbative TMDPDFs at low $q_T$, the resummation is always in the perturbative regime since  $\mu \gg \Lqcd$ for any value of $q_T$. 

\begin{figure}
\centerline{\scalebox{.52}{\includegraphics{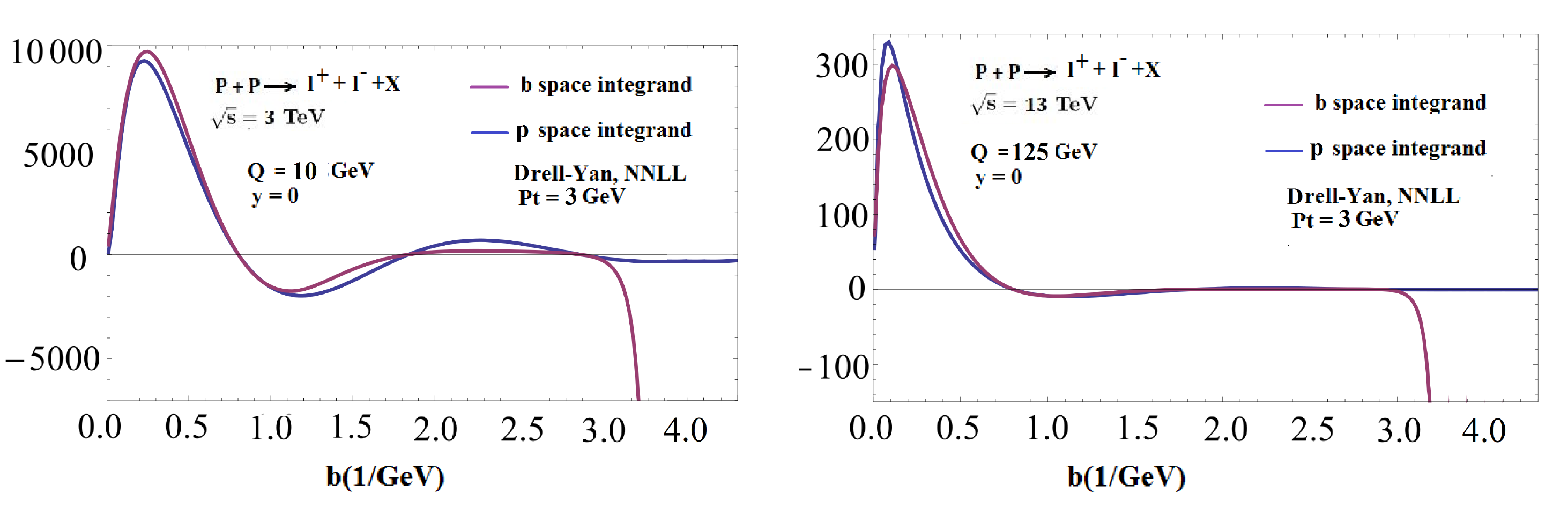}}}
\vskip-0.5cm
\caption[1]{$b$ integrand for different invariant masses, in $b$ space resummation scheme \eq{inverseFT} versus in $p$ space resummation scheme \eq{Ibdef}. The Landau pole in the $b$ space integrand signals the onset of nonperturbative physics. The $p$ space integrand in our scheme vanishes smoothly for large $b$, although this does not mean nonperturbative effects are not important for small $q_T$.}
\label{fig:nonpert} 
\end{figure}

For the case of the soft function, since both resummation schemes use a $b$ dependent value for $\nu$, we need to freeze out the resummation at a perturbative value of $\nu_L \sim 1/b^* > \Lambda_{QCD}$ and fit the soft function at this scale from data. All this seems rather complicated and it might appear that without complete information about the nonperturbative functions, it is not possible to give a prediction for the cross section. However, once again the nature of the resummed perturbative exponent comes to the rescue. Even before we reach a nonperturbative value, the double logarithmic term in $b$ in the exponent completely damps out the integrand. Then the nonperturbative corrections become irrelevant since the region of $b$ space in which they start contributing is heavily suppressed. 

If we consider $b$ space resummation, the damping which is provided by the resummed exponent depends mainly on the hard scale $Q$ and the cusp anomalous dimension. For the Higgs, $Q$ is fixed and the cusp anomalous dimension is large so the damping is always large. Increasing the center-of-mass energy only changes the $x$ value where the PDFs are evaluated, but there is only a mild dependence on this factor. So, we can safely neglect any nonperturbative effects and rely completely on the perturbatively resummed cross section.

For the case of DY, the cusp anomalous dimension is much smaller and the value of $Q$ is variable. So if we go to low $Q$,  nonperturbative effects become important, see \fig{nonpert}. As can be seen from this figure, $b \sim 3\text{ GeV}^{-1}$ is a rough estimate of the value beyond which nonperturbative effects become important where we begin observing the divergence due to the Landau pole in the $b$ space resummation scheme. For low values of $Q$, several different ways of incorporating nonperturbative effects using model functions (which effectively cut off the Landau pole) in $b$ space resummation have been proposed \cite{Collins:2014jpa,Sun:2013dya,DAlesio:2014mrz,Scimemi:2017etj,Qiu:2000hf} . In this paper, we stick to showing results for larger values of $Q$ where these effects are not as important. A detailed discussion of how to handle nonperturbative effects in our hybrid impact parameter-momentum space resummation scheme will be given in future work. We suspect\footnote{Thanks to D. Neill}, among other things, that the nature of our asymptotic $V_\beta$ expansion in \eqs{Vbeta}{VGammaVbeta}  will in fact give clues about the best way to include nonperturbative effects together with our perturbative resummation scheme. This follows from the fact that in the $b$ space resummation scheme, the Landau pole, related to the running of $\alpha_s$  was the indicator of the onset of non-perturbative effects. In our scheme, we have expanded out the running of $\alpha_s$ in the form of the $V_{\beta}$ function and it is natural that the breakdown of this expansion will dictate the onset of nonperturbative effects. (See \cite{Scimemi:2016ffw} for a recent study of nonperturbative power corrections based on renormalon divergences of perturbative expansions of TMD functions.)

\subsection{Remarks on perturbative low $q_T$ limit}
\label{ssec:lowqT}

The $q_T\to 0$ behavior of the perturbative resummed $q_T$ distribution has of course been extensively discussed, and is known to be affected by configurations of multiple large momentum emissions $\vect{k}_T^i$ that cancel vectorially $\sum_i\vec{k}_T^i = \vec{q}_T^{\,i}\sim 0$, leading $d\sigma/dq_T^2$ to go to a constant nonzero value \eg \cite{Parisi:1979se,Becher:2011xn,Ebert:2016gcn,Bizon:2017rah}. We will not add anything to these discussions here, postponing consideration of this regime until we also include nonperturbative effects as remarked above. 

We observe, however, that our scheme in \sec{analytic} applies practically only to the larger $q_T$ regime. Specifically, we are using an approximation for the ratio of gamma functions $F(t) = \Gamma[-t]/\Gamma[1+t]$ appearing in \eq{Mellin-Barnes} in $t$ space.
While the approximation that we are currently using is good enough in the perturbative regime ($q_T \geq 2\text{ GeV}$), below this $q_T$ scale, if we compare it to the CSS resummation then it shows exponential suppression as opposed to a constant behavior.
However, the $q_T$ scale at which this deviation from the exact resummed cross section happens could be systematically lowered by improving our expansion of $F(t)$ (\ie adding more terms to our final formula \eq{IbHermite}).
This is evident from the behavior of the integrand in $t$ space \eq{saddle}. As $q_T$ is lowered, both $A$ and $t_0$ become large. Since $A$ controls the suppression of the integrand which allows us to approximate the ratio of $\Gamma$ functions as a series, as $q_T$ is lowered, this suppression becomes smaller and smaller which will force us to include more and more terms in our expansion in order to maintain the same accuracy. While we can get good accuracy in the perturbative regime $q_T \sim $ 2 GeV with a few terms, below this scale it is impractical to continue using this series approximation. So in principle, to recover the behavior as $q_T\to 0$, we can no longer do an expansion in the Hermite polynomials.

We observe, however, that as $A$ increases at lower $q_T$, the integrand in $b$ space (\eq{Ibk}) is now highly suppressed at large $b$. This means that it would be more practical to now do an expansion of the Bessel function itself (which was not possible at larger $q_T$), rather than go to $t$ space. For $q_T\lesssim 2\text{ GeV}$ we would need a series expansion that approximates $J_0(bq_T)$ well just out to its first peak or so. It should be possible to find such an expansion similar to what we did for large $q_T$ above that still allows us to do the $b$ space integrals analytically and accurately, which by construction would now display constant behaviour at low $q_T$. We defer a presentation of the details of this procedure to future work. 

So the Mellin-Barnes representation is in some sense the dual of the Bessel function, in that at perturbative $q_T$, $t$ space is more amenable to an expansion and hence an analytical result with a few terms, but fails as we move into the low-$q_T$ region where nonperturbative effects kick in, where expanding directly in $b$ space makes more sense.

\section{Comparison with previous formalisms}
\label{sec:compare}

There have been numerous other techniques developed for implementing the resummation for transverse momentum spectra of gauge bosons. We will briefly comment on how ours compares to some of them, though we will not undertake any sort of in-depth comparison here and will not really do justice to any of these other methods. A more detailed discussion of them was given in \cite{Neill:2015roa} as well as \cite{Ebert:2016gcn}.

The earliest one was the CSS formalism  \cite{Collins:1984kg,Collins:2011zzd}, which was applied in, \eg \cite{Bozzi:2010xn,deFlorian:2011xf},  for computing DY and Higgs transverse momentum cross sections. The value of $\mu_L$ is implicitly chosen to be $1/b_0$. There is no explicit independent scale $\nu$, however a comparison of the resummed exponents reveals that the implicit choice for $\nu_L$ is also $1/b_0$. So the central values agree with the $b$ space resummation implemented in \eq{inverseFT} and an earlier paper \cite{Neill:2015roa}.

The difference in the two approaches is two-fold. Firstly, the error analysis using scale variation is different in the absence of an independent scale $\nu$. Varying only $\mu$ scales is likely to underestimate the uncertainty.  Second, in the high $q_T$ region, the matching procedure is different. In the CSS formalism, there is no systematic way of turning off resummation  while matching to the fixed order cross section, so that the predicted results differ in the high $q_T$ region. The Landau pole is handled by implementing a smooth cut-off in $b$ space. However, as we have seen, as long as we are at high $Q$, this does not affect the prediction. An explicit comparison between the two schemes was given in \cite{Neill:2015roa}.

\begin{figure}
\centerline{\scalebox{.55}{\includegraphics{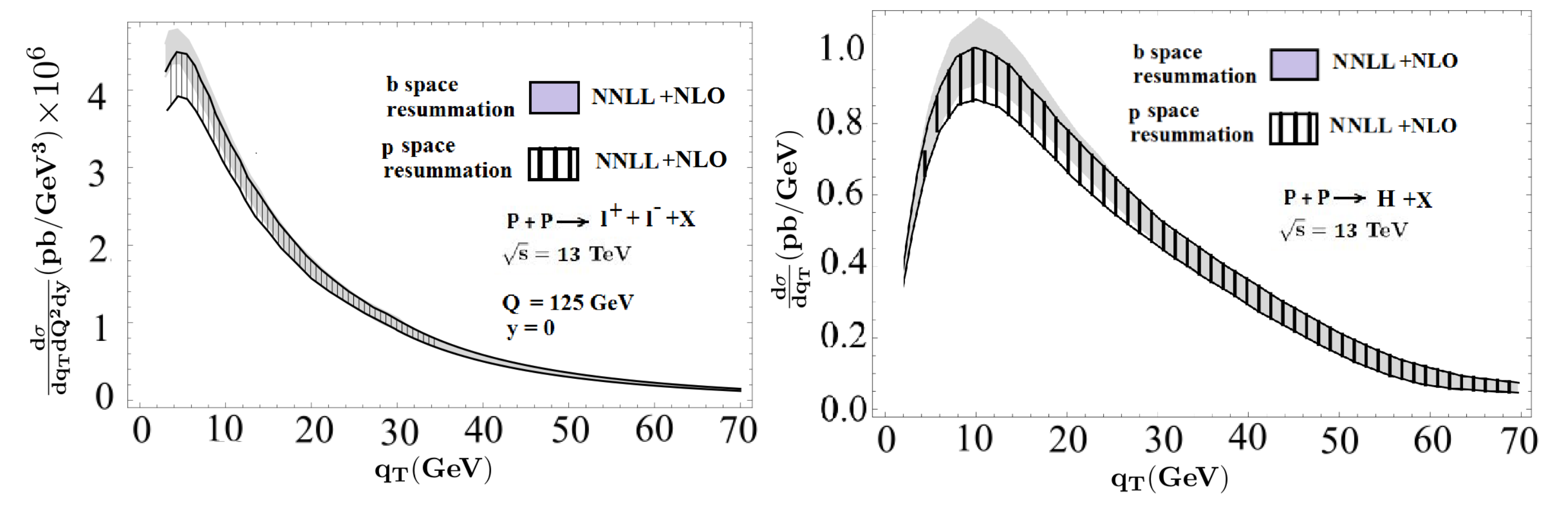}}}
\vskip-0.2cm
\caption[1]{Comparison of NNLL+NLO cross section (resummed cross section matched to O($\alpha_s$) fixed order cross section using profiles) in two schemes, $b$-space resummation \eq{inverseFT} and $p$-space resummation \eq{thefinalresult}. The overlap is a good cross-check of the accuracy of our method, and the improvement in the reliable estimation of uncertainties and computation time in our resummation scheme has been described in the text. The Higgs cross section is differential only in $q_T$.}
\label{fig:nnll} 
\end{figure}

In \fig{nnll}, we compare the $b$ space resummation scheme for the implementation in \cite{Neill:2015roa}  at NNLL+NLO accuracy for the Higgs and DY transverse spectrum with the hybrid $b$-space/momentum-space resummation scheme developed in this paper. This will serve transitively as a comparison also with other $b$ space resummation schemes. We can deduce the following 
\begin{itemize}
\item
The width of the error bands is comparable in the entire region of $q_T$ which is not too surprising since the error analysis in both \cite{Neill:2015roa} and the present paper were based on the same variations \eq{scalevariations} around the respective central values.
\item
In the low $q_T$ region, the central value in our hybrid scheme is lower that the pure $b$ space scheme even though it is within the error band. This is to be expected since the two schemes differ in subleading terms at a given resummation accuracy.
\item
In the high $q_T$ regime, the results agree exactly since in this range, the resummation has been turned off and the cross section is just the one-loop fixed order cross section in both schemes. 
\end{itemize}

Another technique was implemented in \cite{Becher:2011xn,Becher:2012yn} which again follows the CSS formalism with the implicit $\nu$ choice $1/b$, but the $\mu$ choice is made in momentum space choosing $\mu \sim q_T+q_T^*$,  where $q_T^*$ is chosen as 2 GeV for DY and 8 GeV for the Higgs. For the kinematics we chose to illustrate in \ssec{muLscale}, we actually found very similar shifts at low $q_T$ based on our analysis of the scales which minimize the contributions of the residual fixed order logs, which in itself parallels the logic in \cite{Becher:2011xn,Becher:2012yn}, though we do not necessarily adopt the same physical interpretations. The matching procedure is again similar to the CSS case and hence differs from the scheme in \cite{Neill:2015roa} the high $q_T$ regime. Again a detailed discussion of the differences was provided in \cite{Neill:2015roa}.

There also have been methods proposed for setting \emph{all} renormalization scales in momentum space. The  most recent \cite{Ebert:2016gcn} technique has been to solve the RG equations in momentum space directly. In momentum space the beam and soft functions are functions of plus distributions of the form $\left [ 1/q_T^2 \ln^n( q_T^2 /Q^2) \right]_+ $ and these terms are resummed directly in momentum space using  a technique of distributional scale setting. This involves setting the scale under an integral of the plus distribution. The integral turns the plus distribution into ordinary logarithms which can be minimized by choosing a specific momentum scale. However, they also observe that for transverse momentum spectra of gauge bosons, a direct scale choice of $\mu ,\nu \sim q_T$ does not work since this scale choice gives a spurious contribution from highly energetic emissions ($ k_T\gg q_T$) in the phase space and hence a scale that varies with the energy of emissions has to be used so that the region of phase space of energetic emissions is suppressed. This is reflective of the divergence observed in the soft resummation \eq{softNLL} at low values of $b$, which, in our method we chose to cure by adding subleading terms in the cross section through \eq{nuLstar}. Mathematically the solution proposed in \cite{Ebert:2016gcn} is quite elegant. It will be interesting to see its implementation numerically and to compare the results at NNLL.

Another method of obtaining the transverse momentum spectra has been proposed \cite{Monni:2016ktx,Bizon:2017rah} that uses the coherent branching formalism. The cross section is given in terms of a convolution over independent emissions off the initial gluons (or quarks for DY). It then singles out the hardest emission which also sets the scale for $\nu$ which again suppresses the energetic emissions since all other emissions are by construction of lower energy.  This differs mainly from \cite{Ebert:2016gcn} in this scale choice, as in  \cite{Ebert:2016gcn} $\nu \sim k_i$ which follows the energy of each emission instead of just the hardest one.

Amusingly, every proposal we know of so far  (\eg \cite{Ebert:2016gcn,Monni:2016ktx,Bizon:2017rah} and this paper) to implement TMD resummation in momentum space yields a result formally correct at a given order of logarithmic accuracy, but in terms of either an infinite sum or infinite nest of expressions (beyond the perturbative expansion itself) that must be truncated to yield a result that can be evaluated numerically. Refs.~\cite{Ebert:2016gcn,Monni:2016ktx,Bizon:2017rah} obtain their final resummed results in terms of infinite sums over gluon emissions, which \cite{Monni:2016ktx,Bizon:2017rah} implemented in a Monte Carlo routine. Our final result, on the other hand, contains the infinite sum in \eq{IbHermite}, over Hermite polynomials in the basis expansion of the function $\Gamma(t)^2$ arising from the representation \eq{Mellin-Barnes} we used for the Bessel function in the inverse Fourier transform from impact parameter to momentum space. This is of course quite different from sums over gluon emissions. Truncating our formula corresponds to the level of numerical accuracy one attains for the Bessel function and resultant integral, rather than the number of gluons one includes in the emission amplitudes.

All the methods have their pros and cons in terms of the perturbative series obtained, error analysis, and how rigorously or easily nonperturbative effects can be included (see, \eg \cite{Collins:2014jpa}). Again, we leave it to future work to show how we do the latter in our method.

\section{Conclusions}
\label{sec:conclusions}
We took a fresh look at resummation for transverse momentum spectra of gauge bosons in momentum space. In contrast to the classic procedure which chooses both virtuality and rapidity scales $\mu,\nu$ for resummation both in impact parameter space, we proposed a hybrid prescription for resummation, choosing the rapidity renormalization scale $\nu$ with impact parameter dependence and the virtuality scale $\mu$ in momentum space.  We made a choice $\nu_L^* \sim \nu_L(\mu_L b_0)^{-1+p}$ for the low (soft) rapidity scale, and observe that with this choice, the integral over the $b$ space rapidity resummation exponent is convergent. We stress that a well-defined power counting for $\ln(\mu_Lb_0)$ is not possible before we have a stable soft exponent and that only when this exponent is in place, we can treat $\ln(\mu_L b_0) $ as a small log with an appropriate choice of $\mu_L$. We also give a prescription for obtaining the $\mu_L$ scale in momentum space using the analysis of the $b$ space integrand to justify our power counting $\ln(\mu_L b_0) \sim 1 $, which shifts the scale up  from the na\"{i}ve momentum-space choice $\mu_L\sim q_T$.

We then use the idea that restricting the soft exponent in $b$ space to be at most quadratic and thus Gaussian in $\ln (\mu_L b_0)$ allows us to obtain a semi-analytic formula for the cross section.
Using the Mellin-Barnes representation of the Bessel function and the absence of the Landau pole in our resummation formalism, we are able, with certain approximations for the Bessel function appearing in the inverse Fourier transform that are independent of the details of the observable or kinematics, to give a closed-form analytic expression for the cross section at any order of resummation accuracy.

In brief, the main ideas and results of our paper are:
\begin{itemize}
\item Exponentiation of quadratic fixed-order \emph{small} logs of $\mu_L b_0$ from the soft function and rapidity evolution that are formally subleading at a given order of resummation accuracy, but automatically make the $b$ integral in going to momentum space convergent without an additional regulator or cutoff. This is formally achieved by the shifted scale choice $\nu_L \to \nu_L^*$ in \eq{nuLstar}.

\item Division of the rapidity exponent $V$ into an exponentiated part $V_\Gamma(\nu_L^*,\nu_H;\mu_L)$ in \eq{VGaussian} that is quadratic and thus Gaussian in $\mu_L b_0$, and a part $V_\beta$ in \eq{Vbeta} expanded in an asymptotic series, making the $b$ integral \eq{Ibdef} doable.

\item Use of the Mellin-Barnes representation \eq{Mellin-Barnes} of the Bessel function, transformation to the form \eq{Ibimag}, and expansion of the pure function $\Gamma(-c-ix)^2$ appearing therein in a series of Hermite polynomials times Gaussians in \eq{Hermiteexpansion}, which for NNLL accuracy in the cross section can be safely truncated to a few terms each in the real (even) and imaginary (odd) parts, each term of which gives rise to an integral over $b$ (or $x$) which can be done be performed analytically, giving the result \eq{Hnresult}.

\item The above steps give rise to the final resummed cross section in momentum space, \eq{thefinalfinalresult}, in which we can implement scale variations and profiles to reliably estimate theoretical uncertainty and match smoothly onto fixed-order predictions for large $q_T$.
\end{itemize}

The final result \eq{thefinalfinalresult} represents a threefold expansion: \emph{perturbative expansion} in $\as$ and resummed logs in the RG and RRG evolution kernels and fixed-order hard, soft, and collinear functions; \emph{$V_\beta$ expansion}, a fixed-order asymptotic expansion of the non-conformal part of the RRG evolution kernel to ensure a Gaussian rapidity kernel; and \emph{Hermite expansion}, in number of terms in basis expansion of the pure Bessel function in the inverse Fourier transform between $b$ and $q_T$ space. Each of these is systematically improvable. 

We do not even claim that the particular methods, expansions, and strategies we implemented are the fastest or most accurate amongst all similar strategies. It is fast, and it is accurate.  Keeping just a few terms in the Hermite expansion we obtain an error in the cross section at the percent level, much better than the NNLL perturbative accuracy to which we work in this paper, while obtaining the result with a $\sim$five-fold improvement in computation time in our tests. We hope our presentation provides a blueprint and an example to obtaining a faster, more accurate predictions for many TMD observables in momentum space, and is certainly open to further development and improvement.

We applied our results to obtain the transverse spectrum of the Higgs as well as the DY $q_T$ spectrum at NNLL, matched to fixed-order $\cO(\as)$ results at large $q_T$. We give a comparison  with results obtained using the CSS formalism and observe a very good agreement where they should agree, consistent within subleading terms which is observed from the overlapping of the error bands at both NLL and NNLL. We also gave cursory discussions of the relevance of nonperturbative effects in different kinematical regimes, and also of how our method compares with some recently proposed methods of resummation directly in momentum space for all renormalization scales. 

The techniques we have proposed should be applicable to other observables that depend on a transverse momentum or are sensitive to ``soft recoil,'' (\eg \cite{Chiu:2011qc,Chiu:2012ir}), and admit of a factorization of the form \eq{txsec} with a convolution between soft and collinear functions in (2-D) transverse momentum $\vec{q}_T$ describing modes separated in rapidity as in \fig{modes}. When a (semi-)analytic formula can be obtained as we have done, it should drastically cut down computation time and improve our understanding of the physical behavior of the cross section and its computational uncertainties as a function of the scales it depends on. 
We will perform a a more detailed phenomenological study using our expressions with comparisons to data in the near future, and then also apply our techniques to other TMD observables.

\begin{acknowledgments}
We are grateful to D. Neill for many helpful conversations, especially for suggesting the use of an integral representation of the Bessel function to simplify the analytical computation of cross sections, and for a detailed review of a preliminary draft which (we believe) improved its presentation considerably, and to S. Fleming and O.~Z. Labun for collaboration on early stages of this work and on related work. This work was supported by the U.S. Department of Energy through the Office of Science, Office of Nuclear Physics under Contract DE-AC52-06NA25396 and by an Early Career Research Award, through the LANL/LDRD Program, and within the framework of the TMD Topical Collaboration.

\end{acknowledgments}

\appendix

\section{Fixed-order expansions}
\label{app:fixedorder}

\subsection{Evolution kernels}
\label{app:def-K}

The evolution kernels $K_{\Gamma}(\mu_0, \mu)$, $\eta_{\Gamma}(\mu_0, \mu)$, $K_{\gamma} (\mu_0, \mu)$ that appear in the RGE solutions for the hard and soft functions and TMDPDFs were defined in \eqss{UHdef}{US}{Uf}. They can be rewritten in terms of integrals over the coupling $\alpha_s(\mu)$ via the relation
\be
\mu\frac{d}{d\mu} \alpha_s(\mu) = \beta[\alpha_s(\mu)] \Rightarrow \frac{d\mu}{\mu} = \frac{d\as}{\beta(\as)}\,,
\ee
yielding
\begin{align} \label{Keta-def}
K_{\Gamma}(\mu_0, \mu)
&
= \int_{\as(\mu_0)}^{\as(\mu)}\!\frac{\df\as}{\beta(\as)}\,
\Gamma_\cusp(\as) \int_{\as(\mu_0)}^{\as} \frac{\df \as'}{\beta(\as')}
\,,\nn\\
\eta_{\Gamma}(\mu_0, \mu)
&
= \int_{\as(\mu_0)}^{\as(\mu)}\!\frac{\df\as}{\beta(\as)}\, \Gamma_\cusp(\as)
\,, 
\qquad
K_{\gamma}(\mu_0, \mu)
= \int_{\as(\mu_0)}^{\as(\mu)}\!\frac{\df\as}{\beta(\as)}\, \gamma(\as)
\,.\end{align}
Expanding the beta function and anomalous dimensions in powers of $\as$,
\begin{gather}
\label{anomdimexpansions}
\beta(\as) =
- 2 \as \sum_{n=0}^\infty \beta_n\Bigl(\frac{\as}{4\pi}\Bigr)^{n+1}
\,, \\
\Gamma_\cusp(\as) = \sum_{n=0}^\infty \Gamma_n \Bigl(\frac{\as}{4\pi}\Bigr)^{n+1}
\,, \ \gamma(\as) = \sum_{n=0}^\infty \gamma_n \Bigl(\frac{\as}{4\pi}\Bigr)^{n+1}
\,, \nn
\end{gather}
their explicit expressions to NNLL accuracy are (suppressing the superscript $q$ on  $\Ga$),
\begin{align} \label{Keta}
K_\Gamma(\mu_0, \mu) &= -\frac{\Gamma_0}{4\beta_0^2}\,
\biggl\{ \frac{4\pi}{\as(\mu_0)}\, \Bigl(1 - \frac{1}{r} - \ln r\Bigr)
   + \biggl(\frac{\Gamma_1 }{\Gamma_0 } - \frac{\beta_1}{\beta_0}\biggr) (1-r+\ln r)
   + \frac{\beta_1}{2\beta_0} \ln^2 r
\nn\\ & \hspace{-10ex}
+ \frac{\as(\mu_0)}{4\pi}\, \biggl[
  \biggl(\frac{\beta_1^2}{\beta_0^2} \minus \frac{\beta_2}{\beta_0} \biggr) \Bigl(\frac{1 \minus r^2}{2} + \ln r\Bigr)
  + \biggl(\frac{\beta_1\Gamma_1 }{\beta_0 \Gamma_0 } \minus \frac{\beta_1^2}{\beta_0^2} \biggr) (1\minus r+ r\ln r)
  - \biggl(\frac{\Gamma_2 }{\Gamma_0} \minus \frac{\beta_1\Gamma_1}{\beta_0\Gamma_0} \biggr) \frac{(1\minus r)^2}{2}
     \biggr] \biggr\}
\,, \nn\\
\eta_\Gamma(\mu_0, \mu) &=
 - \frac{\Gamma_0}{2\beta_0}\, \biggl[ \ln r
 + \frac{\as(\mu_0)}{4\pi}\, \biggl(\frac{\Gamma_1 }{\Gamma_0 }
 \minus \frac{\beta_1}{\beta_0}\biggr)(r\minus 1)
 + \frac{\as^2(\mu_0)}{16\pi^2} \biggl(
    \frac{\Gamma_2 }{\Gamma_0 } \minus \frac{\beta_1\Gamma_1 }{\beta_0 \Gamma_0 }
      + \frac{\beta_1^2}{\beta_0^2} -\frac{\beta_2}{\beta_0} \biggr) \frac{r^2 \minus 1}{2}
    \biggr]
\,, \nn\\
K_\gamma(\mu_0, \mu) &=
 - \frac{\gamma_0}{2\beta_0}\, \biggl[ \ln r
 + \frac{\as(\mu_0)}{4\pi}\, \biggl(\frac{\gamma_1 }{\gamma_0 }
 - \frac{\beta_1}{\beta_0}\biggr)(r-1) \biggr]
\,.\end{align}
Here, $r = \as(\mu)/\as(\mu_0)$ and the running coupling is given to 3-loop order by the expression
\begin{align} \label{alphas}
\frac{1}{\as(\mu)} &= \frac{X}{\as(\mu_0)}
  +\frac{\beta_1}{4\pi\beta_0}  \ln X
  + \frac{\as(\mu_0)}{16\pi^2} \biggr[
  \frac{\beta_2}{\beta_0} \Bigl(1-\frac{1}{X}\Bigr)
  + \frac{\beta_1^2}{\beta_0^2} \Bigl( \frac{\ln X}{X} +\frac{1}{X} -1\Bigr) \biggl]
\,,\end{align}
where 
\be
X\equiv 1+ \frac{\as(\mu_0)}{2\pi}\beta_0 \ln \frac{\mu}{\mu_0}\,.
\ee

The expressions \eq{Keta} resum to all orders in $\as$ terms in the fixed-order expansion associated with expansion of the running coupling \eq{alphas} in fixed orders in $\as$. Sometimes it is useful, though, to look at the explicit fixed-order expansions to see which terms these are:
\begin{subequations}
\begin{align}
\label{KGammaexp}
K_\Gamma(\mu_0,\mu) &= \quad\ \frac{\as(\mu_0)}{4\pi}\quad\, \biggl[+\frac{1}{2}\Gamma_0\ln^2\frac{\mu}{\mu_0}\biggr] \\
&\quad + \Bigl(\frac{\as(\mu_0)}{4\pi}\Bigr)^2 \biggl[ - \frac{2}{3} \Gamma_0\beta_0 \ln^3\frac{\mu}{\mu_0} + \frac{\Gamma_1}{2} \ln^2\frac{\mu}{\mu_0} \biggr] \nn\\
&\quad + \Bigl(\frac{\as(\mu_0)}{4\pi}\Bigr)^3 \biggl[ + \Gamma_0\beta_0^2 \ln^4\frac{\mu}{\mu_0} -\biggl(\frac{2}{3}\Gamma_0\beta_1 + \frac{4}{3} \Gamma_1 \beta_0\biggr) \ln^3\frac{\mu}{\mu_0}  + \frac{1}{2}\Gamma_2\ln^2\frac{\mu}{\mu_0}\biggr] + \cdots \,, \nn 
\end{align}
exhibiting the towers of leading logs, next-to-leading logs, and next-to-next-to leading logs from left to right. The LL ($\cO(1/\as)$), NLL ($\cO(1)$), and NNLL ($\cO(\as)$) terms in \eq{Keta} automatically sum each infinite tower of logs through functions of the ratios $r$.
Alternatively, the expansion around $\as(\mu)$ is given by:
\begin{align}
\label{KGammaexp2}
K_\Gamma(\mu_0,\mu) &= \quad\ \frac{\as(\mu)}{4\pi}\quad\, \biggl[+\frac{1}{2}\Gamma_0\ln^2\frac{\mu}{\mu_0}\biggr] \\
&\quad + \Bigl(\frac{\as(\mu)}{4\pi}\Bigr)^2 \biggl[ + \frac{1}{3} \Gamma_0\beta_0 \ln^3\frac{\mu}{\mu_0} + \frac{\Gamma_1}{2} \ln^2\frac{\mu}{\mu_0} \biggr] \nn\\
&\quad + \Bigl(\frac{\as(\mu)}{4\pi}\Bigr)^3 \biggl[ + \frac{1}{3}\Gamma_0\beta_0^2 \ln^4\frac{\mu}{\mu_0}  + \biggl(\frac{1}{3}\Gamma_0\beta_1 + \frac{2}{3} \Gamma_1 \beta_0\biggr) \ln^3\frac{\mu}{\mu_0}  + \frac{1}{2}\Gamma_2\ln^2\frac{\mu}{\mu_0}\biggr] + \cdots \,. \nn 
\end{align}
\end{subequations}
Similarly,
\begin{subequations}
\begin{align}
\label{etaexp}
\eta_\Gamma(\mu_0,\mu) &= \quad\ \frac{\as(\mu_0)}{4\pi} \quad\, \biggl[+\Gamma_0 \ln\frac{\mu}{\mu_0}\biggr] \\
&\quad +  \Bigl(\frac{\as(\mu_0)}{4\pi}\Bigr)^2 \biggl[ - \Gamma_0\beta_0 \ln^2\frac{\mu}{\mu_0} + \Gamma_1\ln\frac{\mu}{\mu_0}\biggr] \nn \\
&\quad + \Bigl(\frac{\as(\mu_0)}{4\pi}\Bigr)^3 \biggl[ +\frac{4}{3}\Gamma_0\beta_0^2 \ln^3 \frac{\mu}{\mu_0} - (\Gamma_0 \beta_1 + 2\Gamma_1\beta_0)\ln^2\frac{\mu}{\mu_0} + \Gamma_2\ln\frac{\mu}{\mu_0}\biggr] + \cdots\,. \nn
\end{align}
Since $\eta_\Gamma$ is always multiplied by another large log (\eg $\ln(Q/\mu_L)$ or $\ln(\nu_H/\nu_L)$ in \eq{Utot}), the first tower is again part of the LL series, the second the NLL series, etc. Alternatively, expanded in $\as(\mu)$,
\begin{align}
\label{etaexp2}
\eta_\Gamma(\mu_0,\mu) &= \quad \frac{\as(\mu)}{4\pi} \quad \biggl[+\Gamma_0 \ln\frac{\mu}{\mu_0}\biggr] \\
&\quad +  \Bigl(\frac{\as(\mu)}{4\pi}\Bigr)^2 \biggl[ + \Gamma_0\beta_0 \ln^2\frac{\mu}{\mu_0} + \Gamma_1\ln\frac{\mu}{\mu_0}\biggr] \nn \\
&\quad + \Bigl(\frac{\as(\mu)}{4\pi}\Bigr)^3 \biggl[ +\frac{4}{3}\Gamma_0\beta_0^2 \ln^3 \frac{\mu}{\mu_0} + (\Gamma_0 \beta_1 + 2\Gamma_1\beta_0)\ln^2\frac{\mu}{\mu_0} + \Gamma_2\ln\frac{\mu}{\mu_0}\biggr] + \cdots\,. \nn
\end{align}
\end{subequations}
Finally, $K_\gamma$ is given by the same expansions as \eqs{etaexp}{etaexp2} with $\Gamma_i \to \gamma_i$. 
But $K_\gamma$ is not multiplied by any additional large logs, so in this case the first column of terms in \eqs{etaexp}{etaexp2} for $K_\gamma$ begins at  NLL, the second NNLL, etc.

In our numerical analysis we use the full NNLL expressions for $K_{\Gamma,\gamma},\eta_\Gamma$ in \eq{Keta}, but to be consistent with the value of $\as(\mu)$ used in the NLO PDFs we only use the two-loop truncation of \eq{alphas}, dropping the $\beta_2$ and $\beta_1^2$ terms, to obtain numerical values for $\as(\mu)$. 
Up to three loops, the coefficients of the beta function~\cite{Tarasov:1980au, Larin:1993tp} and cusp anomalous dimension~\cite{Korchemsky:1987wg, Moch:2004pa} in $\overline{\mathrm{MS}}$ are
\begin{align} \label{Gacuspexp}
\beta_0 &= \frac{11}{3}\,C_A -\frac{4}{3}\,T_F\,n_f
\,,\nn\\
\beta_1 &= \frac{34}{3}\,C_A^2  - \Bigl(\frac{20}{3}\,C_A\, + 4 C_F\Bigr)\, T_F\,n_f
\,, \nn\\
\beta_2 &=
\frac{2857}{54}\,C_A^3 + \Bigl(C_F^2 - \frac{205}{18}\,C_F C_A
 - \frac{1415}{54}\,C_A^2 \Bigr)\, 2T_F\,n_f
 + \Bigl(\frac{11}{9}\, C_F + \frac{79}{54}\, C_A \Bigr)\, 4T_F^2\,n_f^2
\,,\nn\\[2ex]
\Gamma_0 &= 4C_i
\,,\nn\\
\Gamma_1 &= 4C_i \Bigl[\Bigl( \frac{67}{9} -\frac{\pi^2}{3} \Bigr)\,C_A  -
   \frac{20}{9}\,T_F\, n_f \Bigr]
\,,\nn\\
\Gamma_2 &= 4C_i \Bigl[
\Bigl(\frac{245}{6} -\frac{134 \pi^2}{27} + \frac{11 \pi ^4}{45}
  + \frac{22 \zeta_3}{3}\Bigr)C_A^2 
  + \Bigl(- \frac{418}{27} + \frac{40 \pi^2}{27}  - \frac{56 \zeta_3}{3} \Bigr)C_A\, T_F\,n_f
\nn\\* & \hspace{8ex}
  + \Bigl(- \frac{55}{3} + 16 \zeta_3 \Bigr) C_F\, T_F\,n_f
  - \frac{16}{27}\,T_F^2\, n_f^2 \Bigr]
\,.\end{align}
where $C_i$ is $C_F$ and $C_A$ for the quark and gluon, respectively.

\subsection{Hard Function}
The hard function $H$ is given as the square of the SCET matching coefficient $C$ arising from matching QCD and SCET amplitudes, $H=\abs{C}^2$. 
The form of the fixed-order expansion of $C$ can be deduced from \eq{Hevolved} and the fixed-order expansion of $U_C$ in \eq{UHdef}, using the evolution kernels expanded in powers of $\as$ in \eqs{KGammaexp}{etaexp}. All logs in the hard coefficient are zero at $\mu_H=iQ$ \cite{Ahrens:2008nc,Ahrens:2008qu}, leaving
\be
\label{cH}
C(Q^2,\mu_H=iQ) = 1 + \sum_{n=1}^\infty \biggl(\frac{\as(iQ)}{2\pi}\biggr)^n c^{n}\,.
\ee
Then using \eqs{Hevolved}{cH} to express the hard coefficient at an arbitrary scale $\mu$,
\be
\begin{split}
C(Q^2,\mu) &= C(Q^2,iQ)U_C(iQ,\mu) \\ 
&= C(Q^2,iQ)\exp\biggl[-\frac{\zed_H}{2} K_\Gamma(iQ,\mu) + K_{\gamma_C}(iQ,\mu)\biggr] \,,
\end{split}
\ee
At an arbitrary scale $\mu$, $C$ then has the expansion,
\be
\label{Cexp}
C(Q^2,\mu) = 1 + \sum_{n=1}^\infty\biggl(\frac{\as(\mu)}{4\pi}\biggr)^n C^{(n)}\,,
\ee
where to $\cO(\as^2)$,
\begin{subequations}
\label{Cn}
\begin{align}
C^{(1)} &= -\zed_H \frac{\Gamma_0}{16} \ln^2\frac{\mu^2}{Q^2} + \frac{\gamma_H^0}{4} \ln\frac{\mu^2}{Q^2} + \frac{c^1}{2} \\
C^{(2)} &=  - \frac{1}{48}\zed_H \Gamma_0\beta_0 \ln^3\frac{\mu^2}{Q^2} + \Bigl(-\zed_H \frac{\Gamma_1}{16} + \frac{\gamma_H^0\beta_0}{8}\Bigr)\ln^2\frac{\mu^2}{Q^2} + \Bigl(\frac{\gamma_H^1}{4} + \frac{c^1}{2}\beta_0\Bigr)\ln\frac{\mu^2}{Q^2} + \frac{c^2}{2}\,, \nn
\end{align}
\end{subequations}
where the one-loop constant for DY is given by
\begin{subequations}
\label{c1}
\be
c^1 = \Bigl(-16 +\frac{\pi^2}{3}\Bigr) C_F
\,,\ee
and for Higgs production by
\be
c^1 =  \frac{\pi^2}{3}C_A 
\,.\ee
\end{subequations}
The 2-loop constant terms can be found in \cite{Idilbi:2006dg, Becher:2006mr,Harlander:2009bw,Pak:2009bx}.

The hard function is then given by 
\be
\label{Hexp}
H(Q^2, \mu) = \abs{C(-Q^2,\mu)}^2 
\ee
The anomalous dimension for the hard function can be written as 
\be
\gamma_H(Q^2,\mu) =  \gamma_C(-Q^2,\mu)+ \text{c.c.}
\ee

In the Higgs production we have the matching coefficient after integrating out the top quark and it can be written as
\be
\label{Ct}
C_t^2(M_t^2,\mu_t=M_t)=\as(M_t)^2 \left[1+ \sum_{n=1}^\infty \biggl(\frac{\as(M_t)}{2\pi}\biggr)^n c_t^n\right]
\,,
\ee
where $c_t^1=5 C_A-3C_F$ and the 2-loop constant is given in \cite{Harlander:2009bw,Pak:2009bx}.
Its anomalous dimension and RGE take the same form as \eqs{hardRGE}{UHdef} except for the cusp part being zero.

The $\overline{\mathrm{MS}}$ non-cusp anomalous dimension $\gamma_H = 2\gamma_C$ for the DY hard function $H$ can be obtained~\cite{Idilbi:2006dg, Becher:2006mr} from the IR divergences of the on-shell massless quark form factor $C(q^2,\mu)$ which are known to three loops~\cite{Moch:2005id}.
Here we write results up to 2 loops
\begin{align} \label{gaHexp}
\gamma_{H\,0} &=2 \gamma_{C\,0} = -12 C_F
\,,\nn\\
\gamma_{H\,1} &=2\gamma_{C\,1}
= -2 C_F 
\Bigl[
  \Bigl(\frac{82}{9} - 52 \zeta_3\Bigr) C_A
+ (3 - 4 \pi^2 + 48 \zeta_3) C_F
+ \Bigl(\frac{65}{9} + \pi^2 \Bigr) \beta_0 \Bigr]
\,.\end{align}

The Higgs production can be obtained from virtual corrections with the virtual top-quark loop providing the effective $ggH$ vertex. The matching coefficient is known at NLO \cite{Harlander:2005rq,Anastasiou:2006hc} and at NNLO \cite{Harlander:2009bw,Pak:2009bx}.
The non-cusp anomalous dimension of the top matching coefficient $C_t^2$ and $gg$ SCET hard function $H$ are given by
\begin{align} \label{gaHexp-g}
\gamma_{t\,0} &=-4\beta_0 &\quad \gamma_{H\,0} &= 0 \,,\\
\gamma_{t\,1}&= -4\beta_1 &\quad \gamma_{H\,1} &=\left( -\frac{236}{9}+8\zeta_3 \right) C_A^2 +\left( -\frac{76}{9} +\frac{2\pi^2}{3}\right) C_A \beta_0\nn
\,.\end{align}

\subsection{Soft Function}

The fixed-order expansion of the soft function $\wt S$ can be deduced from its (R)RG solution \eq{Sevolved} and the fixed order expansions of $U_S$ and $V_S$ in \eqs{US}{VS}. At the scales $\mu_L =\nu_L= 1/b_0$, all the logs in $\wt S$ vanish,
\be
\label{cS}
2\pi \wt S(b,\mu_L=1/b_0,\nu_L=1/b_0) = 1 + \sum_{n=1}^\infty \Bigl(\frac{\as(1/b_0)}{4\pi}\Bigr)^n c_{\wt S}^n\,.
\ee
Evolving it to arbitrary scales $\mu,\nu$ by \eq{Sevolved}, we obtain
\begin{align}
\label{Sevolved2}
\wt S(b;\mu,\nu) &= \wt S(b;1/b_0,1/b_0) V_S(1/b_0,\nu;1/b_0) U_S(1/b_0,\mu;\nu) \\
&=  \wt S(b;1/b_0,1/b_0)  \exp\biggl\{ - \zed_S K_\Gamma(1/b_0,\mu) + \zed_S \eta_\Gamma(1/b_0,\mu) \ln \nu b_0  \nn \\
& \qquad\qquad\qquad + K_{\gamma_S}(1/b_0,\mu)  + \gamma_{RS}[\as(1/b_0)] \ln\nu b_0\biggr\} \,. \nn
\end{align}
We expand the exponent using \eqs{KGammaexp}{etaexp}, the constants in front using \eq{cS}, and the rapidity anomalous dimension in powers of $\as(\mu)$,
using
\begin{align}
\label{gammaRSexp}
\gamma_{RS}[\as(1/b_0)] &= \sum_{n=1}^\infty \Bigl( \frac{\as(1/b_0)}{4\pi}\Bigr)^n \gamma_{RS}^n \\
&= \frac{\as(\mu)}{4\pi} \gamma_{RS}^0 + \Bigl(\frac{\as(\mu)}{4\pi} \Bigr)^2 \Bigl( 2\gamma_{RS}^0 \beta_0 \ln\mu b_0 + \gamma_{RS}^1\Bigr) \nn \\
&\quad + \Bigl(\frac{\as(\mu)}{4\pi} \Bigr)^3 \Bigl[ 4\gamma_{RS}^0 \beta_0^2 \ln^2\mu b_0 + (2\gamma_{RS}^0 \beta_1 + 4\gamma_{RS}^1\beta_0)\ln \mu b_0 + \gamma_{RS}^2 \Bigr] + \cdots \nn
\end{align}
In principle all the higher-order $\beta$-function terms could contribute at the same order as lower-order ones, but we will always evaluate $\mu$ close to $1/b_0$ in a fixed-order soft or beam function or RRG evolution factor, so the higher-order logs are not large, and those terms are genuinely suppressed by powers of $\as$ relative to lower-order terms.

Putting together these pieces, we obtain the expansion of the soft function,
\be
\label{Sfixedorder}
2\pi \wt S(b,\mu,\nu) = 1 + \sum_{n=1}^\infty\biggl(\frac{\as(\mu)}{4\pi}\biggr)^n \wt S^{(n)}\,,
\ee
where to $\cO(\as^2)$,
\begin{subequations}
\label{Sn}
\begin{align}
\wt S^{(1)} &= \zed_S \frac{\Gamma_0}{2} \Bigl(\ln^2 \mu b_0 + 2 \ln \mu b_0 \ln \frac{\nu}{\mu}\Bigr)  + c_{\wt S}^1 \\
\wt S^{(2)} &= \frac{1}{2}\Bigl[\zed_S \frac{\Gamma_0}{2} \Bigl(\ln^2 \mu b_0 + 2 \ln \mu b_0 \ln \frac{\nu}{\mu}\Bigr)  \Bigr]^2 + \zed_S \Gamma_0\beta_0 \biggl( \frac{2}{3}\ln^3 \mu b_0 + \ln^2 \mu b_0 \ln \frac{\nu}{\mu}\biggr)  \\
&\quad + \zed_S \frac{\Gamma_1}{2}  \Bigl(\ln^2 \mu b_0 + 2 \ln \mu b_0 \ln \frac{\nu}{\mu}\Bigr)   + (\gamma_S^1 + 2c_{\wt S}^1\beta_0) \ln\mu b_0 + \gamma_{RS}^1 \ln\nu b_0 + c_{\wt S}^2 \,. \nn
\end{align}
\end{subequations}
Note we used that the one-loop non-cusp anomalous dimensions $\gamma_S^0 = \gamma_{RS}^0 = 0$ vanish. These expansions agree with those given in \cite{Luebbert:2016itl}.
The constant terms are given by
\bea
 c_{\wt S}^1 &=& -C_i \frac{\pi^2}{3}
 \,, \nn \\
 c_{\wt S}^2 &=&C_i^2\frac{\pi^4}{18}
 + C_i \left[  C_A \left( \frac{208}{27} -\frac{2\pi^2}{3} +\frac{\pi^4}{9} \right)
 			+\beta_0 \left( \frac{164}{27} -\frac{5\pi^2}{6}-\frac{14}{3}\zeta_3 \right)    \right]
 \,.\eea

The non-cusp $\mu$- and $\nu$-anomalous dimensions for the TMDPDF and soft function for gluon at 1 loop were calculated in \cite{Chiu:2012ir}.  Their $\mu$-anomalous dimensions are $\gamma_{f\,0} = 2\beta_0$ and $\gamma_{S\,0} =0$.
By replacing $C_A$ by $C_F$ we obtain anomalous dimension of the soft function for the quark from the one for gluon and by using the consistency relation $2 \gamma_f=-\gamma_H-\gamma_S$ we can find anomalous dimension for the TMDPDF although their one loop results for the quark are separately known: $\gamma_{f\,0} = 6 C_F$ and $\gamma_{S\,0} =0$.
At two loops we only need the total $\mu$ anomalous dimension $\gamma_S^1 + 2\gamma_f^1 = -\gamma_H^1$ which was given in \eq{gaHexp}.

The $\nu$-anomalous dimensions up to 2 loops are given by (\eg \cite{Vladimirov:2016dll,Li:2016axz, Li:2016ctv,Luebbert:2016itl,Echevarria:2015byo})
\begin{align}\label{gaBexp}
\gamma_{R\, S\,0} & =-2 \gamma_{R\,f\,0} = 0
\,,\nn\\
 \gamma_{R\, S\,1} & =-2 \gamma_{R\,f\,1} =-2C_i \Big[  \Big(  \frac{64}{9}-28 \zeta_3   \Big) C_A +\frac{56}{9}  \beta_0 \Big] 
\,,\end{align}
where $C_i= C_F, C_A$ for the quark and for the gluon, respectively.

\subsection{TMDPDF}

The TMDPDF matches onto ordinary PDFs via the matching relation
\be
\label{beammatching}
2\pi \tilde f_i^\perp (b,x;\mu,\nu) = \sum_j \int_x^1 \frac{dz}{z} \cI_{ij}(b,z;\mu,\nu) f_j \Bigl(\frac{x}{z},\mu\Bigr)\,.
\ee
The fixed-order expansion of the matching coefficients $\cI$ can be deduced from the (R)RG evolution equation \eq{fevolved} for $\tilde f^\perp$ and from the DGLAP evolution of the PDFs:
\be
\label{DGLAP}
\mu\frac{d}{d\mu} f_i(x,\mu) = \sum_j \int_x^1 \frac{dz}{z}P_{ij}(z,\mu)f_j \Bigl(\frac{x}{z},\mu\Bigr)\,,
\ee
where the splitting function $P_{ij}$ have the perturbative expansion:
\be \label{Pij}
P_{ij}(z,\mu) = \sum_{n=0}^\infty \Bigl(\frac{\as(\mu)}{2\pi}\Bigr)^n P_{ij}^{(n)}(z) \,.
\ee
At the scales $\mu=\mu_L = 1/b_0$ and $\nu = \nu_H = p^\pm$, all logs in the TMDPDF vanish, and it has the form
\be
2\pi f_i^\perp  (b,x;\mu_L = 1/b_0,\nu_H =  p^\pm) = \sum_j\int_x^1\frac{dz}{z} \cI_{ij}(b,z;\mu_L = 1/b_0,\nu_H =  p^\pm) f_j\Bigl(\frac{x}{z},\mu_L=1/b_0\Bigr)\,,
\ee
where
\be
\cI_{ij}(b,z;\mu_L = 1/b_0,\nu_H =  p^\pm)  = \delta_{ij} \delta(1-z) + \frac{\as(1/b_0)}{4\pi} I_{ij}^{(1)}(z) + \Bigl(\frac{\as(1/b_0)}{4\pi}\Bigr)^2 I_{ij}^{(2)}(z) + \cdots
\ee
contains just finite matching functions. Using \eqs{Uf}{Vf} to evolve $f_i^\perp$ to arbitrary scales $\mu,\nu$,
\begin{align}
&\wt f_i^\perp(b,x;\mu,\nu) = \wt f_i^\perp(b,x;1/b_0,p^\pm) V_f (p^\pm,\nu;1/b_0)U_f(1/b_0,\mu;\nu)  \\
&\qquad=\wt f_i^\perp(b,x;1/b_0,p^\pm) \exp\biggl\{ \zed_f \eta_\Gamma(1/b_0,\mu)\ln\frac{\nu}{p^\pm} + K_{\gamma_f}(1/b_0,\mu) + \gamma_{Rf}[\as(1/b_0)] \ln\frac{\nu}{p^\pm}\biggr\} \,, \nn
\end{align}
 and using \eq{DGLAP} to evolve the PDF as well as \eq{gammaRSexp} to expand the rapidity anomalous dimension in powers of $\as(\mu)$, we find that the beam function matching coefficients in \eq{beammatching} have fixed-order expansions taking the form:\footnote{The gluon beam function is decomposed into 2 tensor structures contributing to diagonal part and to off-diagonal parts. Here, we restrict ourselves to the diagonal part because the off-diagonal part begins to contribute at two loops and is not necessary at NNLL accuracy.}
\be
\label{Iijexpansion}
\cI_{ij} =  \delta_{ij}\delta(1-z) + \sum_{n=1}^\infty \Bigl(\frac{\as(\mu)}{4\pi}\Bigr)^n \cI_{ij}^{(n)}\,,
\ee
where to $\cO(\as^2)$,
\begin{subequations}
\label{Iijcoefficients}
\begin{align}
\cI_{ij}^{(1)} &= \delta_{ij}\delta(1-z) \Bigl( \zed_f \Gamma_0 \ln\mu b_0 \ln\frac{\nu}{p^\pm} + \gamma_f^0 \ln\mu b_0\Bigr) - 2P_{ij}^{(0)}(z) \ln\mu b_0 + I_{ij}^{(1)}(z)\,,  \\
\cI_{ij}^{(2)} &= \delta_{ij}\delta(1-z) \biggl[ \frac{1}{2} \Bigl( \zed_f \Gamma_0 \ln\mu b_0 \ln\frac{\nu}{p^\pm} + \gamma_f^0 \ln\mu b_0\Bigr)^2   + \zed_f \Gamma_0\beta_0 \ln^2 \mu b_0 \ln\frac{\nu}{p^\pm} \\
&\qquad\qquad\qquad  + \gamma_f^0 \beta_0 \ln^2\mu b_0  + \zed_f \Gamma_1\ln \mu b_0 \ln\frac{\nu}{p^\pm} + \gamma_f^1 \ln \mu b_0 + \gamma_{Rf}^1 \ln\frac{\nu}{p^\pm} \Bigr]\nn \\
&\quad + \bigl[ I_{ij}^{(1)}(z) - 2P_{ij}^{(0)}(z)\ln \mu b_0\bigr]\Bigl( \zed_f \Gamma_0 \ln\mu b_0 \ln\frac{\nu}{p^\pm} + \gamma_f^0\ln \mu b_0 + 2\beta_0\ln\mu b_0\Bigr) \nn \\
&\quad - 2\ln\mu b_0 \sum_k\int_z^1\frac{dy}{y} I_{ik}^{(1)}(y)P_{kj}^{(0)}\Bigl(\frac{z}{y}\Bigr) + 4\ln^2\mu b_0 \sum_k \int_{z}^1\frac{dy}{y} P_{ik}^{(0)}(y)P_{kj}^{(0)}\Bigl(\frac{z}{y}\Bigr)  \nn \\
&\quad - 4 P_{ij}^{(1)}(z) \ln\mu b_0 + I_{ij}^{(2)}(z)\,.\nn
\end{align}
\end{subequations}

The one-loop constant terms $I^{(1)}_{ij}(z)$~\cite{Chiu:2012ir, Ritzmann:2014mka} are given by
\bea
I^{(1)}_{qq'}(z) &=& I^{(1)}_{\bar q q}(z)=I^{(1)}_{gg}(z)=0
\,,\nn\\
I^{(1)}_{qq}(z) &=&2 C_F (1-z)
\,,\nn\\
I^{(1)}_{qg}(z) &=& I^{(1)}_{\bar q g}(z) =4 T_F z(1-z)
\,,\nn\\
I^{(1)}_{gq}(z) &=&I^{(1)}_{g \bar q}(z)= 2 C_F z
\,.\eea
The two-loop terms $I^{(2)}_{ij}(z)$ can be found from \cite{Luebbert:2016itl} (see also \cite{Gehrmann:2014yya}).
The 1-loop splitting functions defined in \eq{Pij} are given by\footnote{This definition differs by a factor of 2 from the ones used in 
\cite{Gaunt:2014xxa}.}
\bea\label{Pij0}
P^{(0)}_{q_i q_j}(z) &=& 2C_F \delta_{ij}\, \left[ \cL_0(1-z) (1+z^2) +\frac 32 \delta(1-z)\right]
\,,\nn\\
P^{(0)}_{q_i g}(z) &=& P^{(0)}_{\bar q_i g}(z) =2T_F\,  \left[ (1-z)^2 +z^2 \right]
\,,\nn\\
P^{(0)}_{g g}(z) &=& 4 C_A\,  \cL_0(1-z) \frac{(1-z+z^2)^2}{z} +\beta_0 \delta(1-z)
\,,\nn\\
P^{(0)}_{g q_i }(z) &=& P^{(0)}_{g \bar q_i }(z) =2 C_F\, \frac{1+(1-z)^2}{z}
\,. \eea

The anomalous dimensions $\gamma_{Rf}^i$ that we need are given above in \eq{gaBexp}.

\subsection{TMD cross section}

Combining the above fixed-order expansions of the hard and soft functions and TMDPDFs according to the factorization formula \eq{bxsec}, we obtain for the fixed-order expansion up to $\cO(\as^2)$ of the singular pieces of the full QCD cross section (for the $qq$ channel in DY):
\begin{align}\label{sing}
&(2\pi)^3 \widetilde \sigma(b) = f_q(z_1,\mu)f_q(z_2,\mu) \biggl\{ 1 + \frac{\as(\mu)}{4\pi}\Bigl(-\zed_H \frac{\Gamma_0}{2} \ln^2 Qb_0 - \gamma_H^0 \ln Qb_0 + c_H^1 + c_{\wt S}^1\Bigr) \\
&\qquad\quad + \Bigl(\frac{\as(\mu)}{4\pi}\Bigr)^2  \biggl[ \frac{1}{2}\Bigl(-\zed_H \frac{\Gamma_0}{2} \ln^2 Qb_0 - \gamma_H^0 \ln Qb_0\Bigr)^2 - \frac{2}{3}\zed_H\Gamma_0\beta_0 \ln^3 Qb_0  \nn \\
&\qquad\qquad\qquad - \Bigl(\zed_H\frac{\Gamma_1}{2} + \gamma_H^0\beta_0\Bigr)\ln^2 Qb_0 + ( -\gamma_H^1 + \gamma_{RS}^1 + 2c_{\wt S}^1 \beta_0)\ln Qb_0 + c_H^2 + c_{\wt S}^2 \nn \\
&\qquad\qquad\qquad + \Bigl(-\zed_H \frac{\Gamma_0}{2} \ln^2 Qb_0 - \gamma_H^0 \ln Qb_0 + c_H^1 + c_{\wt S}^1\Bigr)2\beta_0\ln\frac{\mu}{Q}\biggr]\biggr\} \nn \\
&\qquad\quad + \frac{\as(\mu)}{4\pi}\sum_j \int_{0}^1\frac{dy}{y}\Bigl(I_{qj}^{(1)}(y) - 2P_{qj}^{(0)}(y)\ln\mu b_0\Bigr) \Bigl[ f_q(z_1,\mu)  f_j\Bigl(\frac{z_2}{y},\mu\Bigr)  + f_j\Bigl(\frac{z_1}{y},\mu\Bigr) f_q(z_2,\mu)\Bigr] \nn \\
&\qquad\qquad \times \biggl\{ 1+ \frac{\as(\mu)}{4\pi} \Bigl(  -\zed_H \frac{\Gamma_0}{2} \ln^2 Qb_0 - \gamma_H^0 \ln Qb_0 + c_H^1 + c_{\wt S}^1\Bigr)\biggr\} \nn \\
&\quad +\Bigl(\frac{\as(\mu)}{4\pi}\Bigr)^2 \sum_{j,k}\int_0^1\frac{dy}{y}\int_y^1\frac{d\xi}{\xi} \Bigl[ -2\ln\mu b_0\,  I_{qk}^{(1)}(\xi) P_{kj}^{(0)}\Bigl(\frac{y}{\xi}\Bigr) + 4\ln^2\mu b_0\, P_{qk}^{(0)}(\xi)P_{kj}^{(0)}\Bigl(\frac{y}{\xi}\Bigr)\Bigr] \nn \\
&\qquad\qquad\qquad\qquad \times \Bigl[\theta(y-z_1)f_j\Bigl(\frac{z_1}{\xi},\mu\Bigr) f_q(z_2,\mu) + f_q(z_1,\mu)\theta(y-z_2)f_j\Bigl(\frac{z_2}{\xi},\mu\Bigr)\Bigr] \nn \\
&\quad +\Bigl(\frac{\as(\mu)}{4\pi}\Bigr)^2  \sum_j \int_{0}^1\frac{dy}{y}\Bigl(I_{qj}^{(2)}(y) - 4P_{qj}^{(1)}(y)\ln\mu b_0\Bigr) \Bigl[ f_q(z_1,\mu)  f_j\Bigl(\frac{z_2}{y},\mu\Bigr)  + f_j\Bigl(\frac{z_1}{y},\mu\Bigr) f_q(z_2,\mu)\Bigr]\biggr\} \,. \nn
\end{align}
The third line of the $\cO(\as^2)$ pieces cancels out the running of $\as(\mu)$ in the $\cO(\as)$ piece on the very first line. All the $P_{ij}^{(0,1)}$ pieces cancel out the evolution of the PDFs $f_q(z_i,\mu)$. The remaining pieces on the first three lines that contain logs are all fixed by the RG evolution of the hard and soft functions and TMDPDFs in \eq{bxsec}.

\subsection{Gaussian rapidity exponent}
\label{app:Gaussian}

The pieces is in the exponentiated rapidity evolution kernel \eq{VGaussian} are given to all orders by \eq{ACUeta}, and to NLL accuracy by:
\bea \label{ACUetaNLL}
A&=&-\zed_S \Gamma_0 \frac{\as(\mu_L)}{4\pi} \left(\frac12+\frac{\as(\mu_L)}{4\pi}\ln \frac{\nu_H}{\nu_L} \right) \\
\chi &=& \exp\left[ \frac{ \ln\nu_H/\nu_L }{1+\frac{\as(\mu_L)}{2\pi}\ln \nu_H/\nu_L }\right] \,,\quad C= e^{A \ln^2\chi} \,, \nn
\eea
and to NNLL accuracy by:
\bea \label{ACUetaNNLL}
A&=&-\zed_S\left[ \Gamma_0 \frac{\as(\mu_L)}{4\pi} +\Gamma_1 \frac{\as^2(\mu_L)}{16\pi^2}\right]
\left(\frac12+\frac{\as(\mu_L)}{4\pi}\ln \frac{\nu_H}{\nu_L} \right) \\
\chi &=& \exp\left[ 
\frac{ \ln\nu_H/\nu_L }{1+\frac{\as(\mu_L)}{2\pi}\ln \nu_H/\nu_L }
+\frac{\gamma_{RS}^{(1)}\frac{\as(\mu_L)}{4\pi} }{2\zed_S\left( \Gamma_0 +\Gamma_1  \frac{\as(\mu_L)}{4\pi}\right)}
\right] \nn \\
C&=& \exp\left[ A \ln^2\chi + \Bigl(\frac{\as(\mu_L)}{4\pi}\Bigr)^2  \gamma_{RS}^{(1)}\ln\frac{\nu_H}{\nu_L} \right]\nn
\eea

\section{A general scheme for soft resummation}
\label{app:Cparameter}

The default prescription for $\nu$ running in our scheme uses the choice (\eq{nuLstar})
\bea
\nu_L^*= \nu_L(\mu_L b_0)^{-1+p} 
\eea
which automatically resums the leading double logarithms of $\ln(\mu_L b_0)$. This scheme also partially resums single and double logarithms of the argument $\mu_Lb_0$ at higher orders in $\alpha_s$. This is the simplest scheme that provides a stable $b$ space kernel that respects the power counting that $l =\ln(\mu_L b_0)$ is small (see \ssec{mom}). This scale choice is by no means unique. There is still a lot of space to play around with the choice of this scale, where each choice would differ from the other in exactly which set of small logs $l$ get included in the exponent.  All of these different schemes, therefore, would only differ from each other in subleading terms, and hence we would expect that each of these would lead to an overlapping error band at any given order in resummation.

In this section, we give a general prescription for the scale choice $\nu_L$, that covers all of these schemes, while still allowing us to obtain an analytical expression. So we still obey the constraint of putting terms at most of quadratic power in l in the exponent. The generalization that we propose is 
\bea
 \tilde \nu_L^*= \nu_L(\mu_L b_0)^{r}\zeta_0^{s}
\eea
where we now expand out both $r$ and $s$ as a power series in $\alpha_s$. $\zeta_0$ is a constant
\be
r = \sum_{i=0}^{\infty} r_i \alpha_s^i\,,\quad  s = \sum_{i=0}^{\infty} s_i \alpha_s^i\,.
\ee
The soft exponent \eq{VGamma} now looks like 
\begin{align}
&V_\Gamma (\nu_L = \tilde \nu_L^*,\nu_H;\mu_L) = \exp\biggl\{ \ln\frac{\nu_H}{ \tilde \nu_L^*} \sum_{n=0}^\infty \Bigl(\frac{\as(\mu_L)}{4\pi}\Bigr)^{n+1}(\zed_S \Gamma_n \ln\mu_L b_0 + \gamma_{RS}^n)\biggr\}\,\\
&\quad=  \exp\biggl\{ \left(\ln\frac{\nu_H}{\nu_L}- \ln(\mu_L b_0) \sum_{i=0}^{\infty} r_i \alpha_s^i -\ln \zeta_0  \sum_{i=0}^{\infty} s_i \alpha_s^i \right) \sum_{n=0}^\infty \Bigl(\frac{\as(\mu_L)}{4\pi}\Bigr)^{n+1}(\zed_S \Gamma_n \ln\mu_L b_0 + \gamma_{RS}^n)\biggr\}\,. \nn
\end{align}
In practice, while resumming to a particular order, we truncate the series in r and s to that order in accuracy. The $r_i$ parameters will control the coefficient of both $l$ and $l^2$ at each order in $\alpha_s$, while the $s_i$ parameters will control the coefficient of $l$. The $s_i$ parameters also induces constant terms in the exponent, which ideally, one would not find in an exponent, however, their effect at each order can be cancelled out by including the corresponding constant terms induced in the fixed order by this scale choice.  We have checked that the effect of several different choices for the parameters $r$ and $s$ produce variations in the resummed cross sections smaller than the inherent perturbative uncertainty already present at each order seen in \fig{NNLL}.

\section{Perturbative QCD results at NLO}\label{app:per}

The QCD results of $q_T$ spectrum for Higgs and for DY are known up to NNLO \cite{deFlorian:1999zd,Glosser:2002gm,Ravindran:2002dc,Ellis:1981nt,Ellis:1981hk,Arnold:1988dp,Gonsalves:1989ar,Catani:2009sm}.
Here we give NLO expression \cite{Kajantie:1978qv,Altarelli:1977kt,Halzen:1978et,Glosser:2002gm} which is used to obtain the nonsingular part defined in \eq{nsdef}
\bea \label{pert}
\frac{d\sigma^\text{pert}}{dq_T^2 dy} = \sigma_0 \frac{\as}{2\pi}\,
\int_{x_1^\text{min}}^1 dx_1\, f_i(x_1) f_j(x_2)\, \frac{ G_{ij}(\hats, \hatt, \hatu)}{x_1-m_T e^y}
\eea
where $G_{ij}$ is a reduced partonic cross section depending on partonic Mandelstam variables.
All variable above are defined as
\bea
m_T&=&\sqrt{q_T^2+Q^2} \nn \,,\\
x_1^\text{min}&=&\frac{ m_T e^y-Q} {\sqrt{s}-m_T e^{-y}}  \,, \qquad
x_2=\frac{m_T e^{-y}}{\sqrt{s}}\frac{ x_1\sqrt{s}-Q^2/m_T e^y}{x_1 \sqrt{s}-m_T  e^y}
\nn\,,\\
\hats &=& x_1 x_2 s  \,, \qquad
\hatt =Q^2-x_2 m_T\sqrt{s} e^y \, , \qquad
\hatu =Q^2-x_1 m_T \sqrt{s} e^{-y}
\eea

For Higgs production, the tree-level and partonic cross sections are given by
\bea
\sigma_0&=&\frac{\pi}{64 s}\left( \frac{\as(\mu)}{3\pi v}\right)^2
\,,\nn\\
G_{gg} &=& C_A \frac{Q^8+\hats^4+\hatt^4+\hatu^4}{ \hats^2 \hatt \hatu}
\,,\nn\\
G_{gq} &=&C_F \frac{\hatt+\hats}{-\hatu \hats}\,, \qquad G_{qg}=G_{gq}|_{\hatt \leftrightarrow \hatu}
\,,\nn\\
G_{q\bar q} &=& 2C_F^2 \frac{\hatt^2+\hatu^2}{\hats^2}
\eea

For DY, they are
\bea
\sigma_0&=&\frac{4\pi\alpha^2}{3sQ^4 N_C}e_i^2
\,,\nn\\
G_{gg} &=& 0
\,,\nn\\
G_{gq} &=&\frac{Q^2}{2\hats} \frac{\hats^2+\hatu^2+2Q^2\hatt^2}{-\hatu \hats}\,, \qquad G_{qg}=G_{gq}|_{\hatt \leftrightarrow \hatu}
\,,\nn\\
G_{q\bar q} &=& C_F \frac{Q^2}{\hats} \frac{ (\hatt-Q^2)^2+(\hatu-Q^2)^2  }{\hatt \hatu}
\eea
where $e_i$ is quark charge $2/3$ and $-1/3 $ for the up- and  down-type quarks and summation over $i$ and $j$ is implicitly implied in \eq{pert}.

\section{Alternative techniques for obtaining analytic resummed result}
\label{app:alternatives}

In this section, we present two other ways in which we can obtain an analytic expression for our resummed soft exponent. One of them involves using again a weighted Hermite basis, now applied to the function $f(t)=\frac{\Gamma(-t) }{\Gamma(1+t)}$. The other uses a more generalized basis, however, is less systematic in terms of determining the expansion coefficients. In the last subsection we consider an alternative expansion applicable to the low $q_T$ regime of the perturbative distribution.

\subsection{Hermite basis with a weight for $\frac{\Gamma(1-ix) }{\Gamma(ix)}$}
The integral in \eq{saddle} can be done numerically but by series expanding $\Gamma(1-ix )/\Gamma(ix)$  directly. We then apply the same strategy that was used in section \ref{HStrategy}, but now use it directly for $f(t)=\frac{\Gamma(-t) }{\Gamma(1+t)}$.

Near $x=0$, we have the Taylor expansion as
\bea \label{Gexpansion2}
\frac{\Gamma(1-ix) }{\Gamma(ix)} &=& \left[i x(1 - a_0\, x^2)- 2\gamma_E  x^2(1-b_0\, x^2) \right]+\cdots
\,,\\ \nn
\tilde a_0&=&2\gamma_E^2\approx 0.81631
\,, \qquad \tilde b_0 = \frac{2\gamma_E^3+\zeta_3}{3\gamma_E} \approx 0.95723
\,.\eea
Note that the values of $\tilde a_0$ and $\tilde b_0$ differ from those for $a_0,b_0$ in \eq{a0b0} by $\pi^2/6$.

Then, the series expansion can be written as
\be
\label{HermiteExpansion2}
\frac{\Gamma(1-ix) }{\Gamma(ix)}   = i x  \,  e^{-\tilde a_0 x^2} \sum_{n=0}^\infty {\tilde c}_{2n} H_{2n}(\tilde\alpha x)
	-2 \gamma_E x^2 \, e^{-\tilde b_0 x^2}\, \sum_{n=0}^\infty {\tilde d}_{2n} H_{2n}(\tilde\beta x)\,.
\ee

\begin{figure}
\centerline{
\includegraphics[width=.5\linewidth]{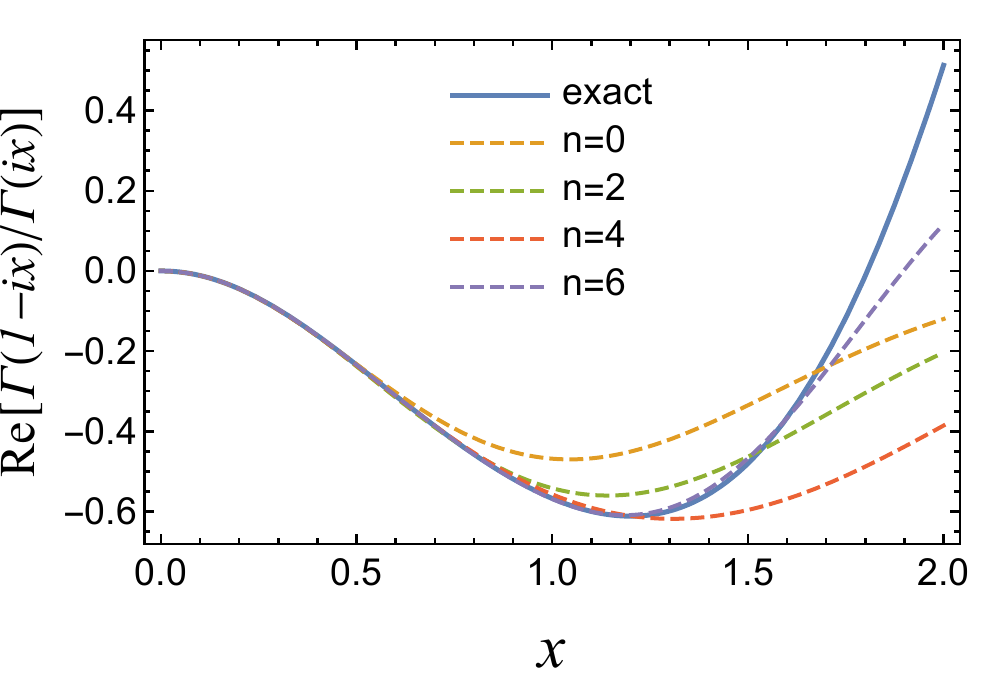}
\hskip 0.5cm
\includegraphics[width=.5\linewidth]{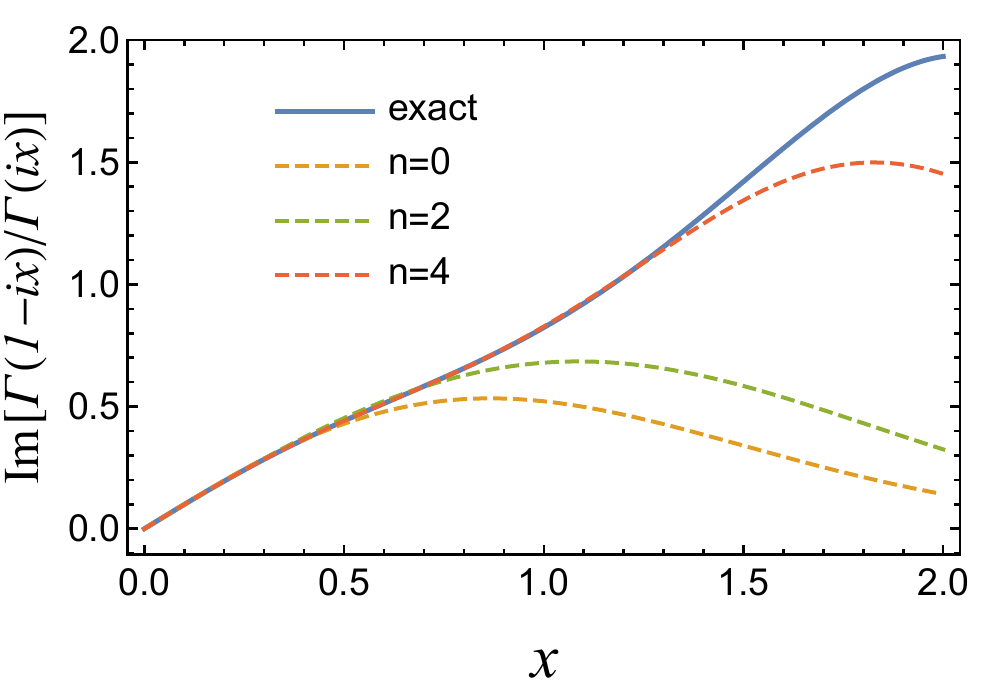}
}
\vskip -0.2cm
\caption[1]{Real and imaginary parts of $\Gamma(1-ix)/\Gamma(ix)$ compared to series expansion in terms of Hermite polynomials up to 4th order.}
\label{fig:Gamma3} 
\end{figure}

So the coefficients in \eq{HermiteExpansion2} are given by
\bea
\label{HermiteCoefficients2}
{\tilde c}_{2n} &=& \frac{\alpha}{\sqrt{\pi}\, 2^{2n} (2n)!} \int_{-\infty}^\infty dx \, \text{Re}\left\{x^{-1}\frac{\Gamma(1-ix)}{\Gamma(ix)}\right\} H_{2n}(\alpha x) e^{-(\alpha^2-a_0) x^2}\,,
\\ \nn
{\tilde d}_{2n} &=& -\frac{\beta}{\gamma_E \sqrt{\pi}\, 2^{2n+1} (2n)!} 
\int_{-\infty}^\infty dx \, \text{Im}\left\{x^{-2}\frac{\Gamma(1-ix)}{\Gamma(ix)} \right\} H_{2n}(\beta x) e^{-(\beta^2-b_0) x^2}\,.
\eea
Empirical tests imply the series converges well for $\tilde\alpha^2-\tilde a_0$ and $\tilde\beta^2-\tilde b_0$ around $3\sim 5$. 
\fig{Gamma2} shows the exact result and series expansion up to 4th order for $\tilde\alpha^2-\tilde a_0=4$ and $\tilde\beta^2-\tilde b_0=4$.
Note that the deviations from the exact results above $x=1.5$ is suppressed by the Gaussian kernel in \eq{saddle} and resulting error in the integral should be smaller than that appearing in \fig{Gamma3}.
   
The coefficients ${\tilde c}_{2n}$ and ${\tilde d}_{2n}$ are given by
\bea
\label{cndn}
{\tilde c}_0& =&1.02257\,,\qquad {\tilde c}_2 = 0.02162\,,\qquad {\tilde c}_4= 0.00168\,, 
\qquad {\tilde c}_6= 3.33\times 10^{-6}\,,
\\ 
{\tilde d}_0& =&1.00941\,,\qquad {\tilde d}_2 = 0.00818\,,\qquad {\tilde d}_4= 0.00042\,,
\qquad {\tilde d}_6= -1.43 \times 10^{-5}\,.
\eea

Now it is straightforward to rewrite the integration in \eq{saddle} in terms of the basis integrations and to obtain the fixed order terms in the similar fashion to \eqs{cH2n}{dcH} in Sec.~\ref{HStrategy}.

\subsection{A tailored basis for expanding $\frac{\Gamma[-t]}{\Gamma[1-t]}$}

While the weighted Hermite polynomial basis presented in the main text is a systematic expansion in an orthogonal basis, we can come up with a basis more closely tailored to the behavior of the function $f(t)=\frac{\Gamma[-t]}{\Gamma[1-t]}$, although it is not as systematic in that it is not orthogonal and there is not a simple formula for the basis coefficients. This basis is:
\be\label{expansion}
f_\text{app} (t)= \sum_n c_n\, f(t; N_n,a_n,b_n)
\ee
\be\label{basis}
f(t; N_n,a_n,b_n)= (t+1)^{N_n}e^{a_n (t+1)^2 +b_n (t+1)}= i^{N_n}(x-x_0)^{N_n} e^{-a_n (x-x_0)^2 +ib_n (x-x_0)}
\,,\ee
where $N_n$ are integers and $a_n$ and $b_n$ are complex constants.
In the 2nd equality, we set $t=c+i x$ and $x_0=i(1+c)$. 
This form has been deliberately chosen in anticipation of our choice of $c = -1$. 

Because we are not aware of a systematic expansion in terms of this basis unlike the weighted Hermite polynomial expansion, the values of $a_n$, $b_n$, and $N_n$ as well as $c_n$ in \eq{expansion} should be determined by fitting to the exact function $f(t)$.
The integration against the evolution kernel is given by
\bea \label{intbasis}
F_{0}(a,b) &=& \frac{1}{i\sqrt{\pi A}} \int_{c-i \infty}^{c+i \infty} dt\,  e^{ \frac{(1+t)^2}{A}-2L (1+t) }\, f(t; 0, a,b)
				= \frac{ e^{-A\frac{\left(b-2L\right)^2}{4(1+a A)}} }{\sqrt{1+a A}}  
\nn\\
F_{N}(a,b) &=& \frac{1}{i\sqrt{\pi A}} \int_{c-i \infty}^{c+i \infty} dt\,  e^{ \frac{(1+t)^2}{A}-2L (1+t) }\, (1+t)^{N}f(t; 0,a,b)
\nn\\
				&=& \left(-\frac{\partial_L}{2}\right)^{N}  F_{0}(a,b)
\nn\\
				&=& \omega_{N}(a,b) \, F_{0}(a,b)
\,,\eea
where $\partial_L=\partial/(\partial L)$ and $\omega_{N}$ is determined by $N$th derivative of $ F_{0}(a,b)$.

The integral $I_b$ is rewritten as
\bea
 I_b = \frac{ 2 C_1}{q_T^2} \sum_{n} c_n \, F_{N_n}(a_n,b_n) 
\,,
\eea
where the derivative on $F_{N}$ can be replaced by $F_0$ multiplied by a coefficient $d_{N,k}$.

\begin{figure}
\centerline{\scalebox{0.55}{\includegraphics{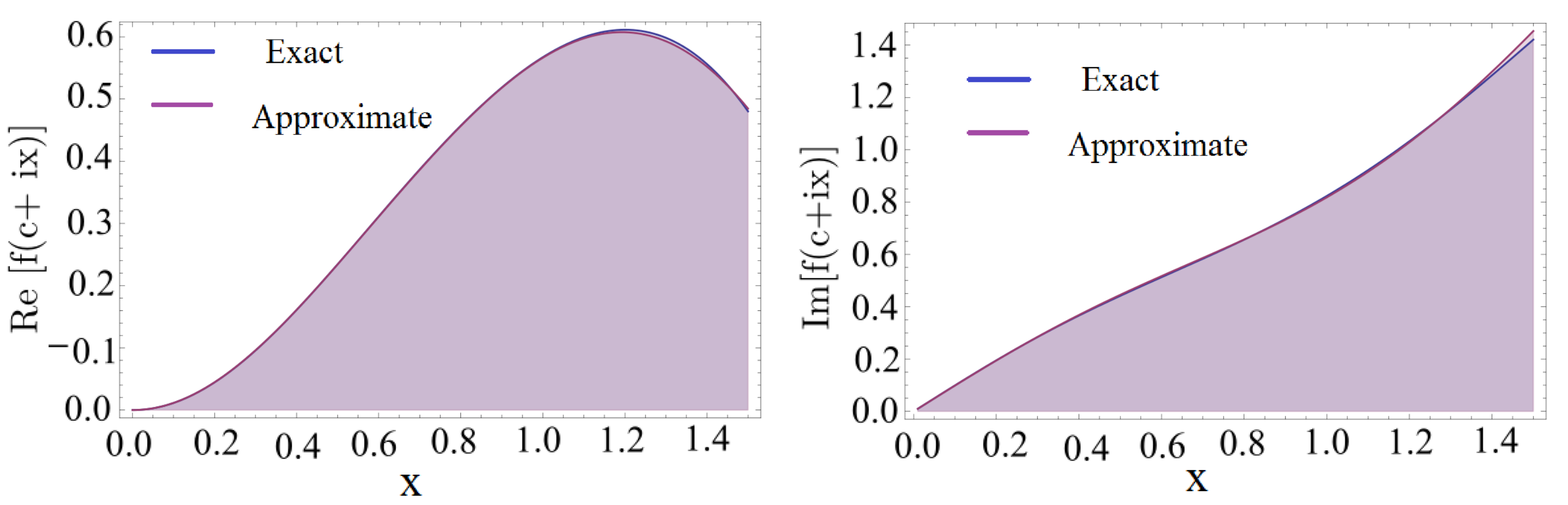}}}
\vskip-0.2cm
\caption[1]{Fit for f(t)}
\label{fit} 
\end{figure}

Here, we show the result of using the basis in \eq{basis} for $N_n=$0 and 2.
The real part of $f_\text{app}(t=-1+ix)$, which we call $f_R(t)$ is even, while the imaginary part $f_I(t) $ is an odd function of $x$. We therefore write 
\bea
f_R(t) &=&  g_1(e^{-g_2 x^2}- \cos[g_3x]) + g_4x^2 e^{-g_5 x^2}
\nn\\
&=&  g_1\left(f(t;0,g_2,0) -\frac 12 [f(t;0,0,g_3)+f(t;0,0,-g_3)]\right) - g_4 f(t;2,g_5,0)
\\ \nn\\
f_I(t) &=& h_1 \sin[h_2 x] +h_3 \sinh (h_4 x)\nn\\   
&=&   \frac{h_1}{2i} \left[ f(t;0,0,h_2)-f(t;0,0,-h_2)\right] +\frac{h_3}{2} \left[ f(t;0,0,-i h_4)-f(t;0,0,ih_4)\right] 
\label{fit}
\eea   
where $t=-1+ix$. In practice we first find the fit in the first equality then, rewrite it in terms of our basis.

The $b$ space integral obtained by replacing $f(t,N_n, a_n,b_n)$ by $F_{N_n}(a_n,b_n)$.
\bea \label{Ibexplicit}
I_b&=&\frac{2C_1}{q_T^2}  \Bigg[
g_1\left(F_0(g_2,0) -\frac 12 [F_0(0,g_3)+F_0(0,-g_3)]\right) - g_4 F_2(g_5,0)
\nn\\ && \qquad \quad
+ \frac{h_1}{2} \left[ F_0(0,h_2)-F_0(0,-h_2)\right] +i\frac{h_3}{2} \left[ F_0(0,-i h_4)-F_0(0,ih_4)\right] 
\Bigg]
\nn\\
&=&
\frac{2C_1}{q_T^2}  e^{-A L^2}
\Bigg[ \frac{g_1 e^{\frac{g_2 A^2L^2}{1+g_2 A}}}{\sqrt{1+g_2 A}}
+\frac{g_4 e^{\frac{g_5 A^2 L^2}{1+g_5A}}}{(1+g_5 A)^{3/2} }\left( \frac{A}{2}-\frac{A^2L^2}{1+ g_5 A}\right) 
\nn\\ && \qquad\quad
-g_1 e^{-\frac{ g_3^2 A}{4}}\cosh[g_3 AL]+h_1e^{-\frac{ h_2^2 A}{4}}\sinh[ h_2 A L]+h_3e^{\frac{ h_4^2 A}{4}} \sin[h_4 A L ]  \Bigg]
\eea
where we have defined $L= \ln \tfrac{2 \Omega}{q_T}$.

The value of the parameters $ g_i, h_i$ shifts as we shift the contour via the value of $c$, so that the final result is independent of the contour chosen. In this paper we have made the following choice for the contour and hence the corresponding parameters $c= -1, g_1=0.5532, g_2=1.77, g_3=2.465, g_4=0.4582, g_5=2.42, h_1 =0.0525, h_2 =4.09, h_3= 0.98, h_4= 0.793$. It is to be stressed that once the contour is fixed, these parameters are also fixed and hence can be used for any observable in any kinematical regime. This is because the fitting is only done for the ratio of Gamma functions $f(t)$ which, in no way involves the details of the specific observable or its kinematics. The only condition as we specified earlier that $A$ be a small number to ensure adequate suppression.  

Let us check if our functions \eq{fit} satisfies the constraints in \eq{dfdt}
\bea
f_\text{app}(-1) &=& g_1(1-1) =0
\nn \\
\frac{d}{dt} f_\text{app}(-1) &=& h_1h_2+h_3 h_4  =0.992
\nn \\
\frac{d^2}{dt^2} f_\text{app}(-1) &=& -g_1 (2 g_2 - g_3^2) + 2 g_4  = 2.31944
\eea
This agrees with \eq{dfdt} better than 1\% that is acceptable at NNLL accuracy.

\section{Mathematical Proofs}

\subsection{A proof of the Mellin-Barnes identity for the Bessel function}\label{app:proof}

Here we present a short proof of the key identity \eq{Mellin-Barnes} we use in the $b$-space integral against the Bessel function:
\bea
\label{J0identity}
J_0(z) = \frac{1}{2 \pi i}\int_{c-i \infty}^{c+i \infty} dt \frac{\Gamma[-t]}{\Gamma[1+t]} \left(\frac{1}{2} z \right)^{2t}  
\,.\eea
This identity can be found, \eg in \cite{NIST:DLMF}, \S 10.9.22, which is given as valid for $J_\alpha(z)$ for $\alpha>0$. We briefly verify that it works also for $\alpha=0$.

For convenience let us change the integration variable in \eq{J0identity} from $t\to -t$. Then we have
\be
\label{J0minust}
J_0(z) = \frac{1}{2 \pi i}\int_{c-i \infty}^{c+i \infty} dt \frac{\Gamma(t)}{\Gamma(1-t)} \left(\frac{1}{2} z \right)^{-2t}  \,,
\ee
where now the contour lies to the \emph{right} of all the poles of $\Gamma(t)$, \ie $c>0$. The contour can be closed in the left half plane out at $t\to -\infty$. The value of the integrand falls rapidly in this limit, and the circular part of the contour contributes zero to the integral. Deforming the contour, we pick up the residues at all the poles $t = -n$ of the Gamma function $\Gamma(t)$, which are
\be
\Res\Gamma(t=-n) = \frac{(-1)^n}{n!}\,.
\ee
Then the integral in \eq{J0minust} has the value
\be
J_0(z) =  \sum_{n=0}^\infty \frac{(-1)^n}{n!} \frac{1}{\Gamma(1+n)} \Bigl(\frac{z}{2}\Bigr)^{2n}\,,
\ee
which is precisely the series representation of the Bessel function $J_0$, proving the identity.

\subsection{Integral of complex Gaussian}

\begin{figure}
\label{fig:contour}
\begin{center}
\includegraphics[width=.75\textwidth]{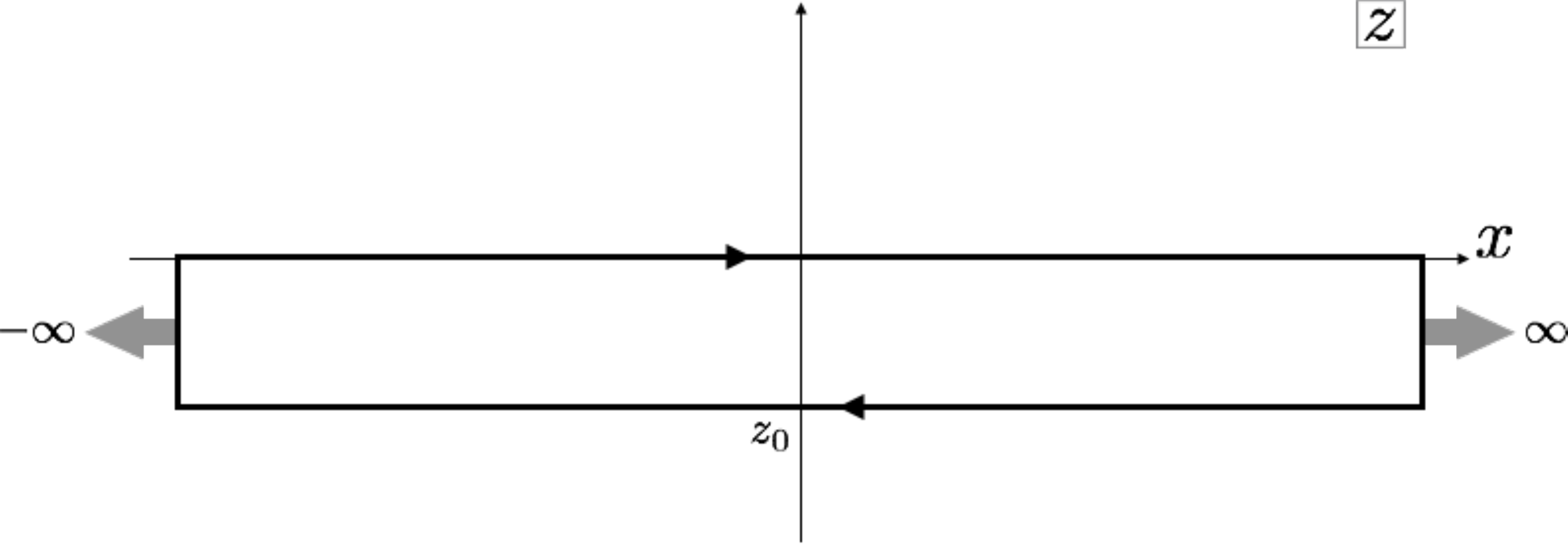}
\end{center}
\vspace{-2em}
\caption{Contour of integration in \eq{contour}. The vertical segments at $x\to\pm\infty$ contribute zero to the integral as $e^{-x^2}\to 0$ there.}
\end{figure}

We can evaluate the integral over a complex Gaussian using the contour integral:
\be
\label{contour}
0 = \int_\cC dz\, e^{-z^2} = \int_{-\infty}^\infty dx \, e^{-x^2} + \int_{\infty}^{-\infty} dx \, e^{-(x - i z_0)^2} = \sqrt{\pi} - \int_{-\infty}^{\infty} dx \, e^{-(x - i z_0)^2} \,,
\ee
where $\cC$ is the contour shown in \fig{contour}.

\subsection{Proof of Gaussian integral of Hermite polynomials}
\label{app:integral}

\begin{figure}
\label{fig:sum}
\begin{center}
\includegraphics[width=.5\textwidth]{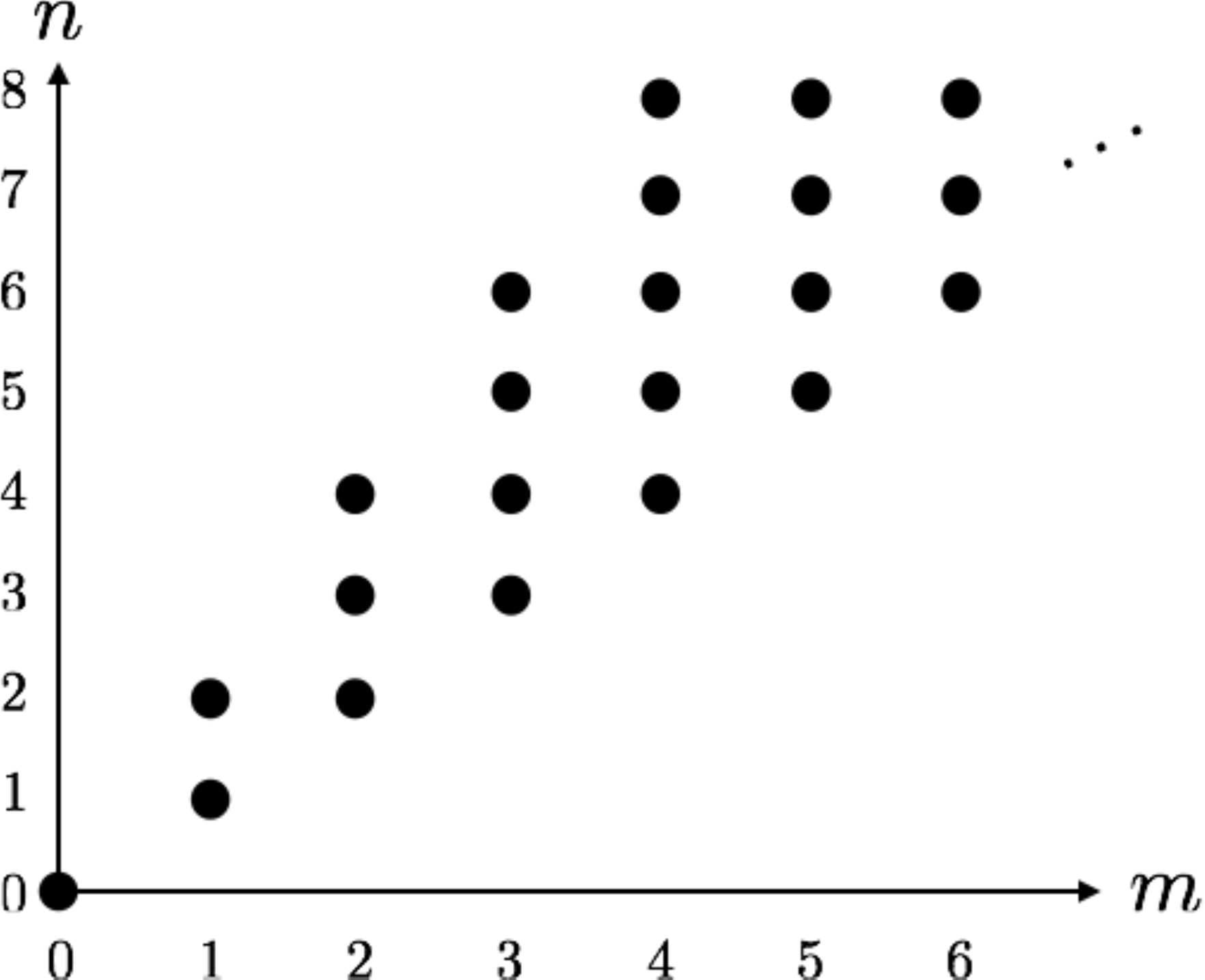}
\end{center}
\vspace{-2em}
\caption{Terms in double sum \eq{summn}.}
\end{figure}

Here we prove the result \eq{Hnresult} for the integrals $\cH_n$ given in \eq{HGaussian}. Starting from the form of the result \eq{Ht}, we have
\be
\cH = \cH_0 \sum_{m=0}^\infty \frac{t^m}{m!}\biggl(\frac{-2\alpha z_0}{1+a_0 A}\biggr)^m \biggl( 1 - t \frac{A\alpha^2 - 1 - a_0 A}{2\alpha z_0}\biggr)^m\,,
\ee
We need to identify the coefficient of each single power $t^n$ of $t$ in order to read off the coefficients $\cH_n$ in \eq{Hnseries}. Using the binomial theorem,
\be
\cH = \cH_0 \sum_{m=0}^\infty \sum_{k=0}^m \frac{t^m}{m!}\binom{m}{k}\biggl(\frac{-2\alpha z_0}{1+a_0 A}\biggr)^m\biggl( - t \frac{A\alpha^2 - 1 - a_0 A}{2\alpha z_0}\biggr)^k \,.
\ee
Using
\be
\binom{m}{k} = \frac{m!}{k!(m-k)!}\,,
\ee
and reindexing the $k$ integral using $n\equiv m+k$, we obtain
\be
\label{summn}
\cH = \cH_0 \sum_{m=0}^\infty \sum_{n=m}^{2m} \frac{(-t)^n}{(n-m)!(2m-n)!}\biggl(\frac{2\alpha z_0}{1+a_0 A}\biggr)^m\biggl(  \frac{A\alpha^2 - 1 - a_0 A}{2\alpha z_0}\biggr)^{n-m} \,,
\ee
This is almost in the form \eq{Hnseries} where we can read off the coefficient of $t^n$, but the order of summation needs to be flipped. As illustrated by \fig{sum}, the following sums are equivalent:
\be
\sum_{m=0}^\infty \sum_{n=m}^{2m} = \sum_{n=0}^\infty \sum_{m=\ceil*{n/2}}^n \,.
\ee
Thus,
\be
\cH = \cH_0 \sum_{n=0}^\infty \sum_{m=\ceil*{n/2}}^{n} \frac{(-t)^n}{(n-m)!(2m-n)!}\biggl(\frac{2\alpha z_0}{1+a_0 A}\biggr)^m\biggl(  \frac{A\alpha^2 - 1 - a_0 A}{2\alpha z_0}\biggr)^{n-m} \,,
\ee
Now we can read off the coefficient of $t^n$ in the series in \eq{Hnseries}, and obtain
\be
\cH_n = \cH_0(-1)^n n! \sum_{m=\ceil*{n/2}}^{n} \frac{1}{(n-m)!(2m-n)!}\biggl(\frac{2\alpha z_0}{1+a_0 A}\biggr)^m\biggl(  \frac{A\alpha^2 - 1 - a_0 A}{2\alpha z_0}\biggr)^{n-m}\,.
\ee
For convenience, we reindex the sum over $m$ by taking $m\to n-m$, and obtain
\be
\cH_n = \cH_0(-1)^n n! \sum_{m=0}^{\floor*{n/2}} \frac{1}{(m)!(n-2m)!}\biggl(\frac{2\alpha z_0}{1+a_0 A}\biggr)^{n-m}\biggl(  \frac{A\alpha^2 - 1 - a_0 A}{2\alpha z_0}\biggr)^{m}\,,
\ee
which after a rearrangement of factors gives the claimed result \eq{Hnresult}.

Explicitly, the first several $\cH_n$ given by \eq{Hnresult} are
\begin{align}
\label{Hnexplicit}
\cH_0 &= e^{-\frac{A}{1+a_0 A}(L-i\pi/2)^2} \frac{1}{\sqrt{1+a_0 A}}  \\
\cH_1 &= - \frac{2z_0\alpha}{1+a_0 A}\cH_0 \nn \\
\cH_2 &=  \frac{\cH_0}{(1+a_0 A)^2} [ 4\alpha^2 z_0^2 + 2(A(\alpha^2-a_0) - 1 )(1+a_0 A)]  \nn \\
\cH_3 &= \frac{\cH_1}{(1+a_0 A)^2} \biggl[ 4 \alpha^2 z_0^2 + 6(A(\alpha^2-a_0) - 1)(1+a_0 A)\biggr]  \nn \\
\cH_4 &= \frac{\cH_0}{(1+a_0 A)^4} \biggl[ 16 \alpha^4 z_0^4 + 48 \alpha^2 z_0^2 (A(\alpha^2 \minus a_0) - 1)(1\plus a_0 A)   + 12(A(\alpha^2\minus a_0) - 1)^2 (1\plus a_0 A)^2\biggr]  \nn \\
\cH_5 &= \frac{\cH_1}{(1+a_0 A)^4} \biggl[ 16 \alpha^4 z_0^4 + 80 \alpha^2 z_0^2 (A(\alpha^2\minus a_0) - 1)(1\plus a_0 A) + 60 (A(\alpha^2\minus a_0) - 1)^2(1\plus a_0 A)^2\biggr] \nn \\
\cH_6 &= \frac{\cH_0}{(1+a_0 A)^6} \biggl[ 64 \alpha^6 z_0^4 + 480 \alpha^4 z_0^4 (A(\alpha^2\minus a_0) - 1)(1\plus a_0 A) \nn\\
&\qquad + 720 (A(\alpha^2\minus a_0) - 1)^2(1\plus a_0 A)^2\alpha^2 z_0^2+ 120 (A(\alpha^2\minus a_0) - 1)^3(1\plus a_0 A)^3\biggr]\nn
\end{align}

\subsection{Recursion relation for $\mathcal{H}_n$ derivative}
\label{app:Hderivative}

Here we prove the recursion relation \eq{recursion} for derivatives of the integrals $\cH_n$ of Hermite polynomials in \eq{cH2n}. We can prove the relation either from this integral \eq{cH2n} directly, or from the final result \eq{Hnresult} of the integration. We present both computations here.

\subsubsection{First proof}

Beginning from the integral definition of $\cH_n$ in \eq{cH2n}, we obtain
\be
\partial_L \cH_n(\alpha,a_0) = \frac{1}{\sqrt{\pi A}} e^{-A(L-i\pi/2)^2} \!\! \int_{-\infty}^\infty \!\! dx\,H_n(\alpha x) e^{-a_0 x^2 - \frac{1}{A}(x+z_0)^2 } \biggl[ -2A\Bigl(L \minus \frac{i\pi}{2}\Bigr) - 2i(x\plus z_0)\biggr] \,,
\ee
where we used $z_0 = A(\frac{\pi}{2} + iL)$. The first term in brackets then cancels with the $z_0$ term, and we have simply
\be
\partial_L \cH_n(\alpha,a_0) = \frac{1}{\sqrt{\pi A}} e^{-A(L-i\pi/2)^2} \!\! \int_{-\infty}^\infty \!\! dx\, (-2ix) H_n(\alpha x) e^{-a_0 x^2 - \frac{1}{A}(x+z_0)^2 }\,.
\ee
Now, we note that the factor $-2x$ can be expressed as a derivative on the Gaussian:
\be
\frac{d}{dx}e^{-a_0 x^2 - \frac{1}{A}(x+z_0)^2 } = \biggl[ -\frac{2}{A}(1+a_0 A) x - \frac{2z_0}{A}\biggr]e^{-a_0 x^2 - \frac{1}{A}(x+z_0)^2 }\,,
\ee
so
\begin{align}
\partial_L \cH_n(\alpha,a_0) &= \frac{iA}{1+a_0A} \frac{1}{\sqrt{\pi A}}  e^{-A(L-i\pi/2)^2} \!\! \int_{-\infty}^\infty \!\! dx\, H_n(\alpha x) \biggl( \frac{d}{dx} + \frac{2z_0}{A}\biggr)e^{-a_0 x^2 - \frac{1}{A}(x+z_0)^2 } \\
&= \frac{2iz_0}{1+a_0A} \cH_n - \frac{iA}{1+a_0A}  \frac{1}{\sqrt{\pi A}} e^{-A(L-i\pi/2)^2} \!\! \int_{-\infty}^\infty \!\! dx\, \alpha H_n'(\alpha x) e^{-a_0 x^2 - \frac{1}{A}(x+z_0)^2 } \,,\nn
\end{align}
where in the last line we integrated the $d/dx$ term by parts, using that the boundary terms are zero at $x\to \pm \infty$. Now we can use the known recursion relation for derivatives of Hermite polynomials:
\be
H_n'(x) = 2n H_{n-1}(x)\,,
\ee
and obtain
\be
\partial_L\cH_n = \frac{2iz_0}{1+a_0A}\cH_n - \frac{2in\alpha A}{1+a_0 A} \cH_{n-1}\,,
\ee
which proves \eq{recursion}.

\subsubsection{Second proof}
Using the all orders result for $\mathcal{H}_n$ in \eq{Hnresult}.
We then compute 
\begin{align}
 &\partial_L\cH_n(\alpha,a_0) = \partial_L \Big\{\cH_0(\alpha,a_0) \Big\}\frac{(-1)^n n!}{(1+a_0 A)^n}\sum_{m=0}^{\floor*{n/2}}\frac{\bigl\{ [ A(\alpha^2 - a_0)-1] (1+a_0 A)\bigr\}^m (2\alpha z_0)^{n-2m}}{m!(n-2m)!} \nn \\
&\quad +  \cH_0(\alpha,a_0)\frac{(-1)^n n!}{(1+a_0 A)^n}\sum_{m=0}^{\floor*{n/2}}\frac{1}{m!}\frac{1}{(n-2m)!} \Bigl\{ [ A(\alpha^2 - a_0)-1] (1+a_0 A)\Bigr\}^m \partial_L \Big\{ (2\alpha z_0)^{n-2m}\Big\}\nn\\
&= \frac{ i 2z_0}{1+a_0 A}\cH_n(\alpha,a_0)\\
&\quad +  \cH_0(\alpha,a_0)\frac{(-1)^n (n-1)!}{(1+a_0 A)^n}\sum_{m=0}^{\floor*{(n-1)/2}}\frac{i 2An\alpha \bigl\{ [ A(\alpha^2 - a_0)-1] (1+a_0 A)\bigr\}^m  (2\alpha z_0)^{n-1-2m} }{m!(n-1-2m)!} \nn
\end{align}
where we have used 
\bea
 \partial_L \Big\{\cH_0(\alpha,a_0) \Big\} =  \frac{ 2i z_0}{1+a_0 A}\cH_0(\alpha,a_0)
\eea
For the second term in $\partial_L \cH_n(\alpha,a_0)$, we can make the following observations 
\begin{itemize}
\item 
For even values of n, the term $m =n/2$ does not contribute to the derivative.
\item
For odd values of n, $\floor*{n/2}$ is the same as $\floor*{(n-1)/2}$.
\end{itemize} 
We can then immediately write down 
\bea
 \partial_L\cH_n(\alpha,a_0)= \frac{2 i z_0}{1+a_0 A}\cH_n(\alpha,a_0) -\frac{ 2i A n \alpha }{1+a_0 A}\cH_{n-1}(\alpha,a_0)
\eea
which proves \eq{recursion}.


\bibliography{DY}
\end{document}